\documentclass{aa}
\usepackage{graphicx}
\usepackage{txfonts}
\usepackage{hyperref}
\hypersetup{colorlinks=true, citecolor=blue}
\usepackage{numprint}

\begin{document}

\title{The ALPINE-ALMA [C~\textsc{ii}] survey: dust attenuation curves at $z=4.4$--$5.5$}
\author{Médéric Boquien\inst{1} \and Véronique Buat\inst{2,3} \and Denis Burgarella\inst{2} \and Sandro Bardelli\inst{4} \and Matthieu Béthermin\inst{2} \and Andreas Faisst \inst{5} \and Michele Ginolfi\inst{6} \and Nimish Hathi\inst{7} \and Gareth Jones\inst{8,9} \and Anton Koekemoer\inst{7} \and Brian Lemaux\inst{10} \and Desika Narayanan\inst{11,12} \and Michael Romano\inst{13,14} \and Daniel Schaerer\inst{15} \and Daniela Vergani\inst{4} \and Giovanni Zamorani\inst{4} \and Elena Zucca\inst{4}}
\institute{Centro de Astronomía (CITEVA), Universidad de Antofagasta, Avenida Angamos 601, Antofagasta, Chile\\\email{mederic.boquien@uantof.cl}\and
           Aix Marseille Univ., CNRS, LAM, Marseille, France\and
           Institut Universitaire de France (IUF), Paris, France\and
           INAF -- Osservatorio di Astrofisica e Scienza dello Spazio di Bologna, via Gobetti 93/3, 40129, Bologna, Italy\and
           IPAC, California Institute of Technology 1200 E California Boulevard, Pasadena, CA 91125, USA\and
           European Southern Observatory, Karl-Schwarzschild-Strasse 2, 85748, Garching, Germany\and
           Space Telescope Science Institute, 3700 San Martin Dr., Baltimore, MD 21218, USA\and
           Cavendish Laboratory, University of Cambridge, 19 J. J. Thomson Ave., Cambridge CB3 0HE, UK\and
           Kavli Institute for Cosmology, University of Cambridge, Madingley Road, Cambridge CB3 0HA, UK\and
           Gemini Observatory, NSF's NOIRLab, 670 N. A'ohoku Place, Hilo, Hawai'i, 96720, USA\and
           Department of Astronomy, University of Florida, 211 Bryant Space Science Center, Gainesville, FL 32611, USA\and
           Cosmic Dawn Center (DAWN), Niels Bohr Institute, University of Copenhagen, Juliane Maries vej 30, DK-2100 Copenhagen, Denmark\and
           Dipartimento di Fisica e Astronomia, Universit\`a di Padova, Vicolo dell'Osservatorio 3, I-35122, Padova, Italy\and
           INAF -- Osservatorio Astronomico di Padova, Vicolo dell'Osservatorio 5, I-35122, Padova, Italy\and
           Observatoire de Genève, Université de Genève, 51 Ch. des Maillettes, CH-1290 Versoix, Switzerland}
\date{}
\abstract
{There is now ample evidence that dust is already present in abundance at high redshift. However, given the faintness of distant galaxies in the optical and the near-infrared, datasets are still limited and how the dust affects the emerging radiation of galaxies at very high redshift is not yet fully understood.}
{Using the ALMA Large Program to INvestigate [C~\textsc{ii}] at Early times (ALPINE), our objective is to quantify the dust attenuation properties in galaxies at $z=4.4$--$5.5$, and in particular the shape of their attenuation curve.}
{Using the CIGALE Spectral Energy Distribution (SED) code, we model the stellar populations and their interaction with the dust in order to measure some of the physical properties of the ALPINE sample. We select a subsample of 23 main-sequence galaxies requiring a detection in at least 6 bands in the rest-frame UV and optical, as well as in the dust continuum around 158~$\mu$m and/or the [C~\textsc{ii}] fine structure line, whose inclusion is important to improve the constraints on the physical properties, while having a reasonably small uncertainty on the slope of the attenuation curves.}
{We find that the attenuation curves span a broad range of properties, from curves that are much steeper than the SMC extinction curve, to shallower than the starburst attenuation curve. The shape of the attenuation curves strongly depends on the V-band attenuation. Galaxies with the lowest attenuation also present the steepest curves. The steepness of such curves is probably the consequence of the combination of the intrinsic physical properties of the dust, the relative distribution of stars and dust in the interstellar medium, and the differential reddening, with an important fraction of the dust concentrated in star-forming regions. The broad range of attenuation curves found at $z \sim 5$ shows that no single attenuation curve is appropriate for main sequence galaxies and that assuming a fixed curve can lead to large errors, for instance in the interpretation and use of the IRX-$\beta$ diagram, if SED modeling is not feasible.}
{Great caution should be exercised when correcting high redshift galaxies for the presence of dust using the UV slope $\beta$ as it can affect the estimation of both star-formation rate and stellar mass even at low V-band attenuation due to the steepness of the attenuation curve. However, when SED modeling can be used, the impact of the choice of the attenuation curve on the star formation rate and the stellar mass is limited.}
\keywords{Galaxies: high-redshift}
\maketitle
  
\section{Introduction\label{sec:introduction}}

Dust is one of the key components of the interstellar medium (ISM) of galaxies. Not only does it dim and redden the far-ultraviolet (FUV) to near-infrared (NIR) radiation emerging from galaxies, which directly affects our ability to measure their physical properties, it also plays a key role in the matter cycle, acting as a catalyst in the formation of the molecular hydrogen that feeds star formation. While the formation of population III stars in a dust-free and metal-free context is not fully understood, it is nevertheless clear that dust formed in large quantities early in the history of the universe \citep[e.g.,][]{riechers2013a, michalowski2015a, watson2015a, burgarella2020a, peroux2020a}.

In the absence of dust, the spectral emission of a normal star-forming galaxy is dominated by stellar populations of different ages with superimposed nebular emission, mainly in the form of recombination lines as well as continuum. The interaction with dust has a dramatic effect, both dimming and reddening the emission from stars and ionized gas. This negatively impacts our ability to measure star formation as energetic photons produced by massive young stars are far more easily attenuated than longer wavelength photons and even a small quantity of dust can lead to an important attenuation in the ultraviolet (UV). In the case of particularly dust-rich galaxies, it can render their detection in the rest-frame UV especially difficult. However, as the FUV emission vanishes due to dust attenuation, this dust re-emits the absorbed energy in the mid-infrared (MIR) and far-infrared (FIR), which can in turn be exploited to trace star formation. Except for the most extreme of cases (e.g., when the dust content is negligible or conversely when almost all of the UV photons are absorbed by dust), an attenuation correction must be carried out to retrieve star formation. One of the most direct ways is to simply apply a hybrid star formation rate (SFR) estimator combining the rest-frame UV with the IR \citep[e.g.,][]{hao2011a, boquien2016a}. The obvious downside is that this requires observations of the dust emission that are costly and difficult to obtain, and even more so at increasing redshifts, where they tend to be limited to vanishingly small samples. With the rest-frame UV emission being relatively easy to obtain from the ground from $z\sim2$ and beyond, techniques have been developed to relate the UV slope ($\beta$) to the UV attenuation (the IRX-$\beta$ relation). While this initially appeared to work remarkably well in the case of starburst galaxies \citep{meurer1999a}, there is now ample evidence that there is no tight universal relation between the UV slope and the attenuation \cite[e.g.,][]{buat2005a, seibert2005a, howell2010a, casey2014a}. In fact this relation relies on two strong underlying assumptions: the intrinsic UV slope of the stellar populations in the absence of dust and the exact shape of the attenuation curve. Numerous studies have analyzed their respective impact in an attempt to understand why and when such relations fail and build more reliable ones \cite[e.g.,][and many others]{kong2004a, boquien2009a, boquien2012a, popping2017a}. In particular the recent study of \cite{salim2019a} found that the diversity of attenuation curves is a strong driver of the scatter around the IRX-$\beta$ relation. This finding, which is consistent with simulations \citep{narayanan2018a, liang2021a}, is especially important in that we can observe a broad variety of attenuation curves at all redshifts \citep[e.g.,][]{salmon2016a, buat2018a, salim2018a}. With the shape of the attenuation curve being strongly dependent on the relative geometry of stars, ionized gas, and dust \citep{salim2020a}, from the disturbed morphologies observed at higher redshifts, we can only expect important variations there too \citep[e.g.,][]{faisst2017a}. Yet, due to the great difficulty to measure them and the sparsity of the data available, our knowledge of attenuation curves beyond $z=4$ remains limited. In effect, most observational studies on the attenuation properties of distant galaxies tend to concentrate on redshifts between 2 and 4 \citep[e.g.,][]{noll2009b, buat2012a, buat2019a, reddy2012a, reddy2015a, shivaei2015a, alvarez2016a, salmon2016a,  fudamoto2017a, fudamoto2020b, lofaro2017a, alvarez2019a, reddy2018a, koprowski2020a}. There is only a handful of examples at higher redshift \citep[][]{capak2015a, scoville2015a, bouwens2016a, barisic2017a, koprowski2018a}. Because of the inherent limits of the observations, studies based on numerical simulations of galaxies at very high redshift \citep[e.g.,][]{mancini2016a, cullen2017a, dimascia2021a} are an important source of information. However, they lead to contrasted results, finding both flat \citep{cullen2017a} and steep \citep{mancini2016a} attenuation curves.

With the rapid build-up of dust at high redshift and the range of new facilities that will come online through the 2020s to peer into distant galaxies with unprecedented detail, it is timely we shed a new light on dust attenuation in the early universe. To reach this goal, we exploit the ALPINE \citep[ALMA Large Program to INvestigate {[}C~\textsc{ii}{]} at Early times,][]{lefevre2020a} survey along with ancillary multi-wavelength data in combination with the CIGALE Spectral Energy Distribution (SED) modeling code \citep{burgarella2005a, noll2009a, boquien2019a}.

The article is organized as follows. We briefly present ALPINE and the sample selection in Sect.~\ref{sec:ALPINE}. We describe our SED modeling to measure the physical properties of the galaxies in our sample in Sect.~\ref{sec:modeling}. We present the results in Sect.~\ref{sec:results}, which we discuss in Sect.~\ref{sec:discussion}, before concluding in Sect.~\ref{sec:conclusion}. Throughout this article we assume a \cite{chabrier2003a} initial mass function (IMF) and a flat $\Lambda$-Cold Dark Matter cosmology with $H_0=70$~km~s$^{-1}$, $\Omega_m=0.3$, and $\Omega_\Lambda=0.7$.

\section{Sample\label{sec:ALPINE}}

\subsection{The ALPINE survey}

ALPINE is a large survey of 118 high-redshift ($4.4 \lesssim z \lesssim 5.9$) galaxies observed with ALMA \citep{lefevre2020a}, targeting the [C~\textsc{ii}] line at 158~$\mu$m. In a nutshell, the ALPINE sample in based on the COSMOS \citep{scoville2007a} and ECDFS \citep{giacconi2002a} surveys. Galaxies were selected from the UV ($L_{UV}>0.6L^*$) and sources hosting a type I Active Galactic Nucleus were specifically excluded. Finally, to ensure proper targeting of the {[}C~\textsc{ii}{]} line with ALMA, all the objects had to have a secure redshift. The resulting sample is representative of star-forming galaxies down to an SFR of 1~M$_\odot$~yr$^{-1}$ around a redshift of 5. As we will see in Sect.~\ref{sssec:MS}, our ALPINE subsample follows well the main sequence at that redshift. The survey and the complementary data have been described in great detail in \cite{lefevre2020a}, \cite{bethermin2020a} and \cite{faisst2020a} and we refer to these articles for more information on the sample, the observations, and data processing.

Being the largest targeted survey of high-redshift galaxies with ALMA, ALPINE is a real treasure trove to study outflows \citep{ginolfi2020a}, mergers and morpho-kinematic diversity \citep{lefevre2020a, jones2020a, jones2021a, romano2021a}, Ly-$\alpha$ emitters \citep{cassata2020a}, [C~\textsc{ii}] as an SFR tracer \citep{schaerer2020a}, [C~\textsc{ii}] sizes \citep{fujimoto2020a}, molecular gas \citep{dessauges2020a}, [C~\textsc{ii}] luminosity functions \citep{yan2020a, gruppioni2020a, loiacono2021a}, the main sequence \citep{khusanova2021a}, the dust mass budget \citep{pozzi2021a}, and even the study of serendipitously discovered objects \citep{gruppioni2020a, romano2020a, loiacono2021a}. Of particular interest to this article, \cite{fudamoto2020a} delved into the attenuation properties of ALPINE galaxies. In a nutshell, they examined the relation between IRX, the stellar mass (M$_\star$), and $\beta$. They found that both through stacking analysis and for the galaxies detected in the dust continuum, the attenuation curve is steeper than for galaxies at lower redshift, but with an overall smaller attenuation. This study focused on the exploitation of rest-frame UV-optical data modeled with the LePhare code \citep{arnouts1999a, ilbert2006a} on one side, and on the ALMA observations on the other side, independently from one another. However, given the high quality of the ALPINE data, an alternative approach is to model the full multi-wavelength emission simultaneously with the ALMA observations. This allows to obtain more reliable estimates in the physical properties of galaxies, albeit with a smaller number. We present in more detail this approach in Sect.~\ref{sec:modeling}.

\subsection{Available data and sample selection}

For this work, in addition to the ALMA data, we consider all the bands presented in \cite{faisst2020a}, which range from $\sim 360$~nm to 8~$\mu$m in the observed frame. However, as described in Sect.~\ref{ssec:bands}, requirements on a minimum signal-to-noise ratio (SNR) means that for each object only a fraction of the available bands are used in the fits.

We have divided the data into two tiers to which further cuts are applied in Sect.~\ref{ssec:bands} and \ref{ssec:final-selection}. Tier 1 is initially comprised of the 21 galaxies that have been detected both in the continuum around 158~$\mu$m rest-frame and in the [C~\textsc{ii}] line. Tier 2 is made of the 53 galaxies that are not detected in the continuum but their [C~\textsc{ii}] emission is. We exclude a total of 43 galaxies from our final sample. Two of them are detected in the dust continuum around 158~$\mu$m but not in the [C~\textsc{ii}] line. This deficit can be suggestive of the presence of an obscured active nucleus \citep[e.g.,][]{sargsyan2012a, sargsyan2014a} that would negatively affect the reliability of our measurements. In addition, the uncertainty on the spectroscopic redshift of these two objects is such that the [C~\textsc{ii}] line may have fallen outside the ALMA spectral window. The other 41 objects are detected neither in [C~\textsc{ii}] nor in the continuum and are deemed too unreliable for this analysis as the lack of FIR detection severely hampers our ability to constrain many of the physical properties that are critical for constraining the attenuation curves, as we will see in Sect.~\ref{ssec:reliability}. We find that many of these objects are very blue, which along their non-detection by ALMA, suggests that they are very dust poor. Others unfortunately lack data at short wavelength, not allowing to constrain their UV properties. However this in turn suggests that most of these objects have a low mass, making them too faint to be detected at a sufficient level in the UV-optical.

\section{Multi-wavelength modeling\label{sec:modeling}}

\subsection{CIGALE\label{ssec:cigale}}

To measure the physical properties of the sample, we carry out a multi-wavelength modeling with CIGALE \citep{burgarella2005a, noll2009a, boquien2019a}. Briefly, CIGALE is an energy balance SED modeling code including flexible star formation histories, stellar populations, and dust in absorption and emission, and which relies on Bayesian inference to estimate the physical properties of galaxies. The idea behind energy balance is that the luminosity absorbed by the dust in the UV-to-NIR is re-emitted self-consistently in the MIR and FIR. Modeling and fitting the full SED simultaneously is particularly useful to break degeneracies. For instance a galaxy can be optically red either because it is very dusty or because it does not contain young stellar populations. In the former case the dust emission will be higher than in the latter case. Energy balance allows us to determine the actual nature of the object. At the same time the Bayesian inference technique is essential to estimate reliably the physical properties.

Another key feature is that CIGALE seamlessly handles upper limits, which allows to obtain constraints in the case of non-detections. It is based on the method presented in \cite{sawicki2012a}. See in particular Sect.~4.3 of \cite{boquien2019a} for its implementation in CIGALE. The handling of upper limits is especially important for this work as the dust continuum is undetected in over 2/3 of the sample. At the same time, CIGALE can also fit intensive (that is, quantities that are independent from the size of the object, such as the age) and extensive (that is, quantities that are dependent on the size of the object, such as the luminosity) physical properties rather than just fluxes. We use these features to exploit both the dust continuum and the [C~\textsc{ii}] line ALMA observations. For the 21 galaxies for which the 158~$\mu$m continuum emission is measured, the dust luminosity is a strong constraint on energy balance and we fit this extensive property along with the panchromatic fluxes. Rather than estimating the dust luminosity ourselves, we adopt the meticulously derived values of \cite{bethermin2020a} who used a combination of models with empirical fits. For tier 2 galaxies, only [C~\textsc{ii}] is detected. This is not a quantity that can be computed in \textsc{cigale} models. Indeed, given the important variations of the luminosity of this fine-structure line that are not well-understood, modeling reliably the [C~\textsc{ii}] emission of galaxies is especially difficult. We therefore fit the SFR after conversion from [C~\textsc{ii}], which provides us with a more indirect and less reliable constraint on the energy balance compared to the dust luminosity. For this, we use the carefully curated statistical relation between the [C~\textsc{ii}] luminosity and the total SED-based SFR (obscured and unobscured) presented in \cite{schaerer2020a}, log(L([C~\textsc{ii}])/L$_\odot$)=7.05+0.96$\times$log(SFR/M$_\odot$ yr$^{-1}$), which is based on models with LePhare. In addition to the SFR and the multi-wavelength fluxes, we also include upper limits on the dust luminosity in the fit.

We use the following strategy to model our sample of galaxies. First we assume a delayed Star Formation History (SFH): $\mathrm{SFR}\propto t\times\exp(-t/\tau)$, with $t$ the time elapsed since the beginning of star formation, and $\tau$ the time at which the SFR reaches its maximum value. Assuming that the first star was born approximately 100~Myr after the big bang, we fix $t$ to 900~Myr for galaxies at $z \sim 5.5$ and 1300~Myr for galaxies at $z \sim 4.5$. Based on a simple physical argument, this allows fixing a parameter that is otherwise poorly constrained, which can have negative effects on the determination of other physical properties. We allow $\tau$ to vary from 100~Myr (star formation peaks early) to 1300~Myr (the SFR always increases over the timespan considered) in steps of 200~Myr. In practice the assumption of an inflexible functional form, such as that of a pure delayed SFH, can bias the measure of some key physical properties of galaxies \citep{ciesla2017a, lower2020a} and lead to overly rigid relations between SFR and M$_\star$ with little dispersion. To solve this, we add a second component to take into account the most recent episode of star formation. For simplicity we assume that this episode can be modeled as a constant SFR over a duration of 10, 25, 50, 75, or 100~Myr and whose strength is parametrized by the fraction of the stellar mass formed during this episode, with respect to the total stellar mass formed over the lifetime of the galaxy. We sample the mass fraction with 21 linearly spaced values from 0\% (no burst) to 10\% (strong burst). One characteristic to note is that at high redshift, it has been argued that that the SFR should be rising on average, as indicated by the cosmic SFR density evolution \citep{madau2014a}. With the adopted parametrization, this is possible through two mechanisms. First, when $\tau$ is larger than the age, the SFR is always increasing over the considered period. Then, there is the possibility of a fairly strong episode of star formation even when there is a decrease of the SFR on the middle term. Overall, this parametrization allows us to have a variety of SFH which yields a dispersion of $\pm0.3$~dex around the main sequence as we will see in Sect.~\ref{sssec:MS}. A consequence of the flexibility is that the uncertainties on the physical properties are larger but probably more realistic. This is due to degeneracies that are otherwise not taken into account with smooth and rigid SFH. For instance the spectrum of a galaxy with a stronger but older episode of star formation will be fairly similar to that of a galaxy with a weaker but more recent episode of star formation.

\begin{figure}[!htbp]
  \includegraphics[width=\columnwidth]{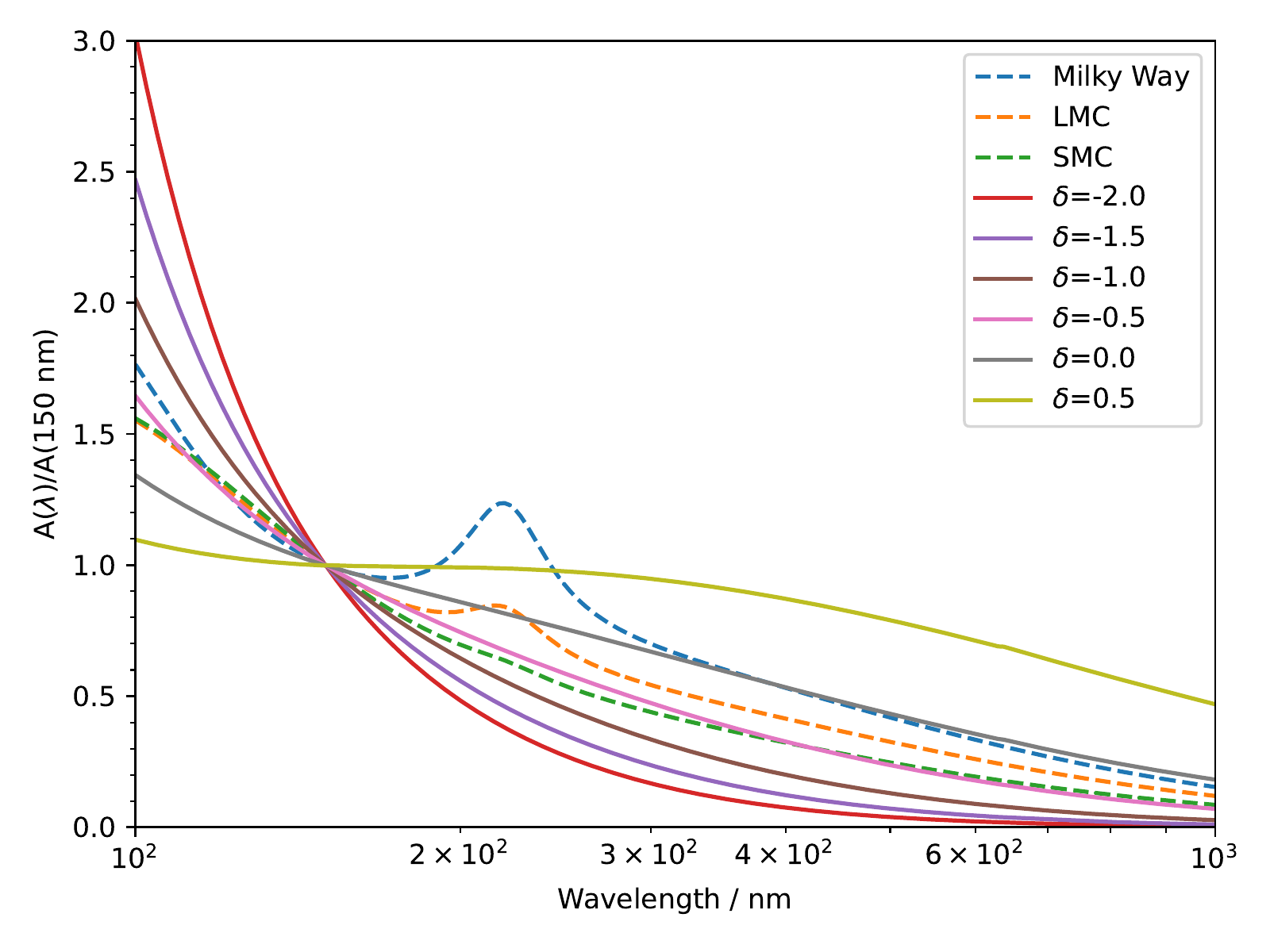}
  \caption{Modified starburst attenuation curves normalized to 150~nm for different values of $\delta$ (solid lines). A starburst curve corresponds to $\delta=0$. The models include 101 curves ranging from $\delta=-2$ to $\delta=0.5$. For comparison we also plot the extinction curves of the Milky Way, the Large Magellanic Cloud, and the Small Magellanic Cloud (dashed lines).\label{fig:curves}}
\end{figure}

With the SFH defined, we compute the stellar spectrum using the \cite{bruzual2003a} single stellar populations, assuming a \cite{chabrier2003a} IMF, and a metallicity $Z=0.008$ as a compromise between metallicities of $Z=0.02$ (solar) and $Z=0.004$ adopted by \cite{faisst2020a}. This choice is motivated by the fact that the galaxies in our subsample are reasonably massive for this redshift (M$_\star\gtrsim10^{10}$~M$_\odot$) and at such masses, \cite{faisst2016a} found $12+\log O/H=8.32^{+0.40}_{-0.74}$ for galaxies at $z\sim5$. Given the range of masses reaching over $10^{11}$~M$_\odot$ and that there is an important dispersion making it difficult to assign a specific metallicity to a given galaxy, a subsolar metallicity of $Z=0.008$ appears reasonable. This being said, the choice of the metallicity only has a very small influence on the estimates of the attenuation curve and it does not affect our conclusions. We then add a nebular component that includes the continuum (free-free, free-bound, and 2-photon processes) as well as hydrogen and metal recombination lines, which are scaled based on the production rate of Lyman continuum photons. This is important because at high redshift, the equivalent width of emission lines can become very large and thus account for a substantial fraction of the total flux in passbands \cite[e.g.,][]{sargent1970a,atek2011a,yuan2019a}. We have adopted an ionization parameter $\log U=-2$ but adopting a lower value of $\log U=-3$ gives qualitatively identical results. The next step concerns the attenuation. As mentioned earlier, there is now established evidence of ample variations in the attenuation curves of galaxies. To account for this, we have chosen a bump-free modified \cite{calzetti1994a,calzetti2000a} prescription \citep[][see Sect.~\ref{ssec:curve} for the specific motivation for this choice]{noll2009a}. In short, it considers a starburst curve extended by the \cite{leitherer2002a} curve under 180~nm. The resulting curve is then modified by multiplying it with a power law with a free index $\delta$ ($\propto \lambda^\delta$) ranging from $-2.0$ (steeper) to 0.5 (shallower) and that is applied both on the stellar and the gas emission without differential reddeningso that the effective attenuation curve is directly tied to $\delta$. For indication, $\delta=0$ corresponds to a bona fide starburst attenuation curve and is also similar to a Milky Way extinction curve in the optical, $\delta=-0.3$ is akin to a Large Magellanic cloud extinction curve (avoiding the bump), and $\delta=-0.5$ to a Small Magellanic Cloud extinction curve. We show these curves in Fig.~\ref{fig:curves}.
Even though curves steeper than that of the Small Magellanic Cloud may seem unusual, they can be found both in simulations \citep[e.g.,][]{trayford2020a} and observationally \citep[e.g.,][]{salim2018a}. Finally we include the intergalactic medium (IGM) of \cite{meiksin2006a}, which has a strong effect at these redshifts on the radiation shortward of Ly-$\alpha$ and we redshift the spectra. We would like to emphasize that even though this set of models has been optimized for the measurement of attenuation properties, we have ensured that the SFH is flexible enough to cover variations we could expect from main sequence galaxies. Additional analysis using models with either lower or higher degrees of flexibility in their SFH has shown that they ultimately lead to qualitatively similar results.

Sampling this parameter space, we fit \numprint{14847000} models to each galaxy. For the entire sample, we compute and fit (including to a synthetic catalog, see Sect.~\ref{ssec:reliability}) a total of \numprint{727503000} models to all the galaxies in approximately 15 hours on a 48-core dual AMD EPYC 7451 server. We give the modules and their corresponding parameters in Table~\ref{tab:cigale}.
\begin{table*}
  \centering
  \begin{tabular}{lll}
  \hline\hline
  Module&Parameter&Value\\\hline
  \texttt{sfhdelayed}&\texttt{age\_main} (Myr) &\texttt{900} ($z\sim5.5$) or \texttt{1300} ($z\sim4.5$)\\
                     &\texttt{tau\_main} (Myr) &\texttt{100, 300, 500, 700, 900, 1100, 1300}\\
                     &\texttt{age\_burst} (Myr)&\texttt{10, 25, 50, 75, 100}\\
                     &\texttt{tau\_burst} (Myr)&\texttt{1e10}\\
                     &\texttt{f\_burst}        &21 linearly spaced values from \texttt{0.0} to \texttt{0.1}\\\hline
  \texttt{bc03}&\texttt{imf}                    &\texttt{1} \citep{chabrier2003a}\\
               &\texttt{metallicity} (Z$_\odot$)&\texttt{0.008}\\\hline
  \texttt{nebular}&\texttt{logU}                      &\texttt{-2}\\
                  &\texttt{f\_esc}                    &\texttt{0}\\
                  &\texttt{f\_dust}                   &\texttt{0}\\
                  &\texttt{lines\_width} (km~s$^{-1}$)&\texttt{300}\\\hline
  \texttt{dustatt\_calzleit}&\texttt{E\_BVs\_young}                        &100 linearly spaced values from \texttt{0.005} to \texttt{0.500} mag\\
                                       &\texttt{E\_BVs\_old\_factor}        &\texttt{1.0}\\
                                       &\texttt{powerlaw\_slope}           &101 linearly spaced values from \texttt{-2.0} to \texttt{0.5}\\
                                       &\texttt{filters}                   &\texttt{FUV, V\_B90, 150, 550}\\\hline
  \texttt{restframe\_parameters}       &\texttt{beta\_calz94}              &\texttt{true}\\
                                       &\texttt{D4000}                     &\texttt{false}\\
                                       &\texttt{IRX}                       &\texttt{true}\\
                                       &\texttt{EW\_lines}                 &---\\
                                       &\texttt{luminosity\_filters}       &---\\
                                       &\texttt{colours\_filters}          &---\\\hline
  \texttt{redshifting}                 &\texttt{redshift}                  &---\\\hline
  \hline
  \end{tabular}
  \caption{Modules and their corresponding parameters used by CIGALE for building a grid of \numprint{727503000} models (\numprint{14847000} models at a given redshift) and estimate the physical properties of the ALPINE sample. We refer to Appendix C of \cite{boquien2019a} for a detailed definition of each parameter.\label{tab:cigale}}
\end{table*}

\begin{figure*}[!htbp]
  \includegraphics[width=0.495\textwidth]{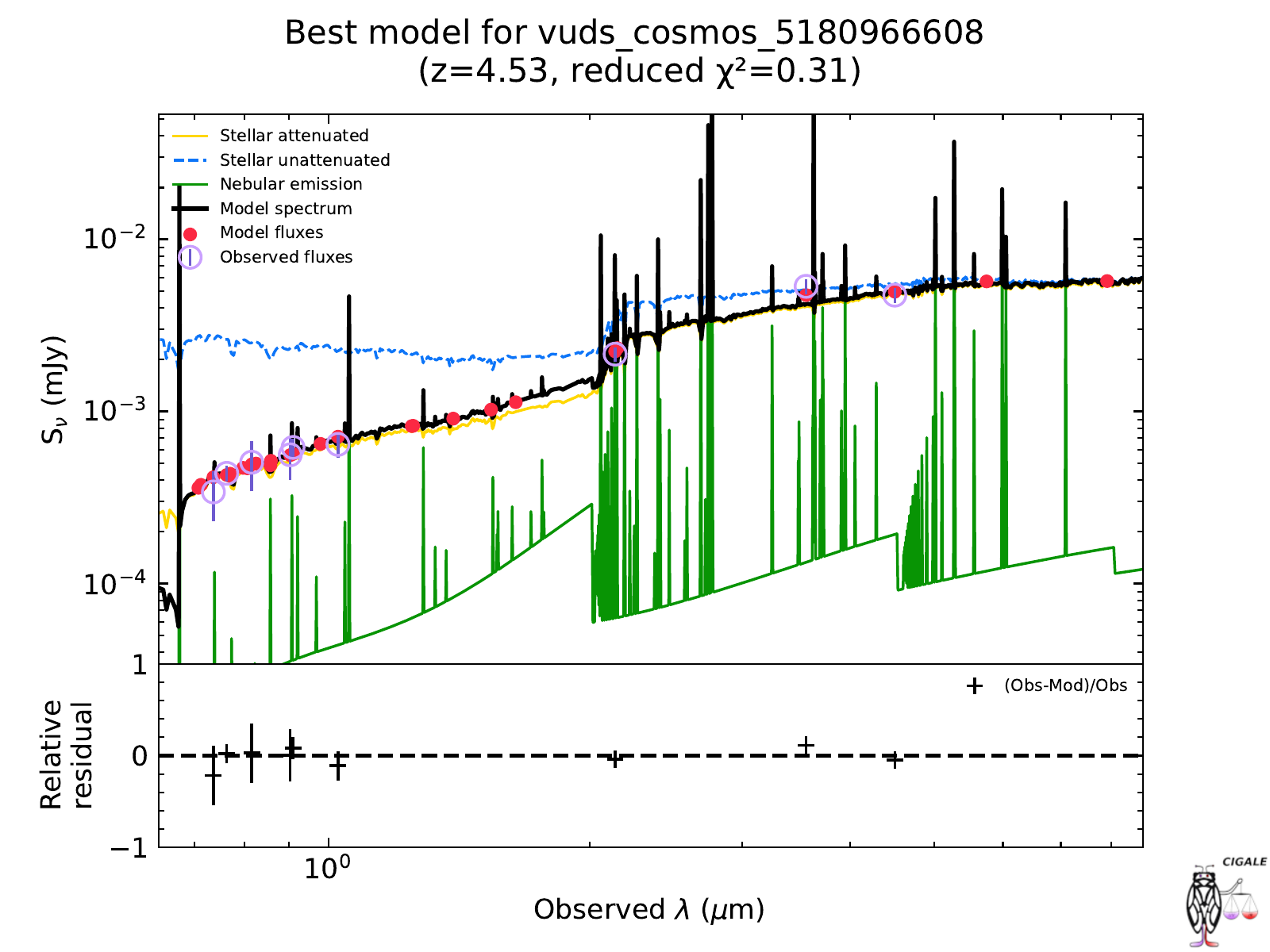}
  \includegraphics[width=0.495\textwidth]{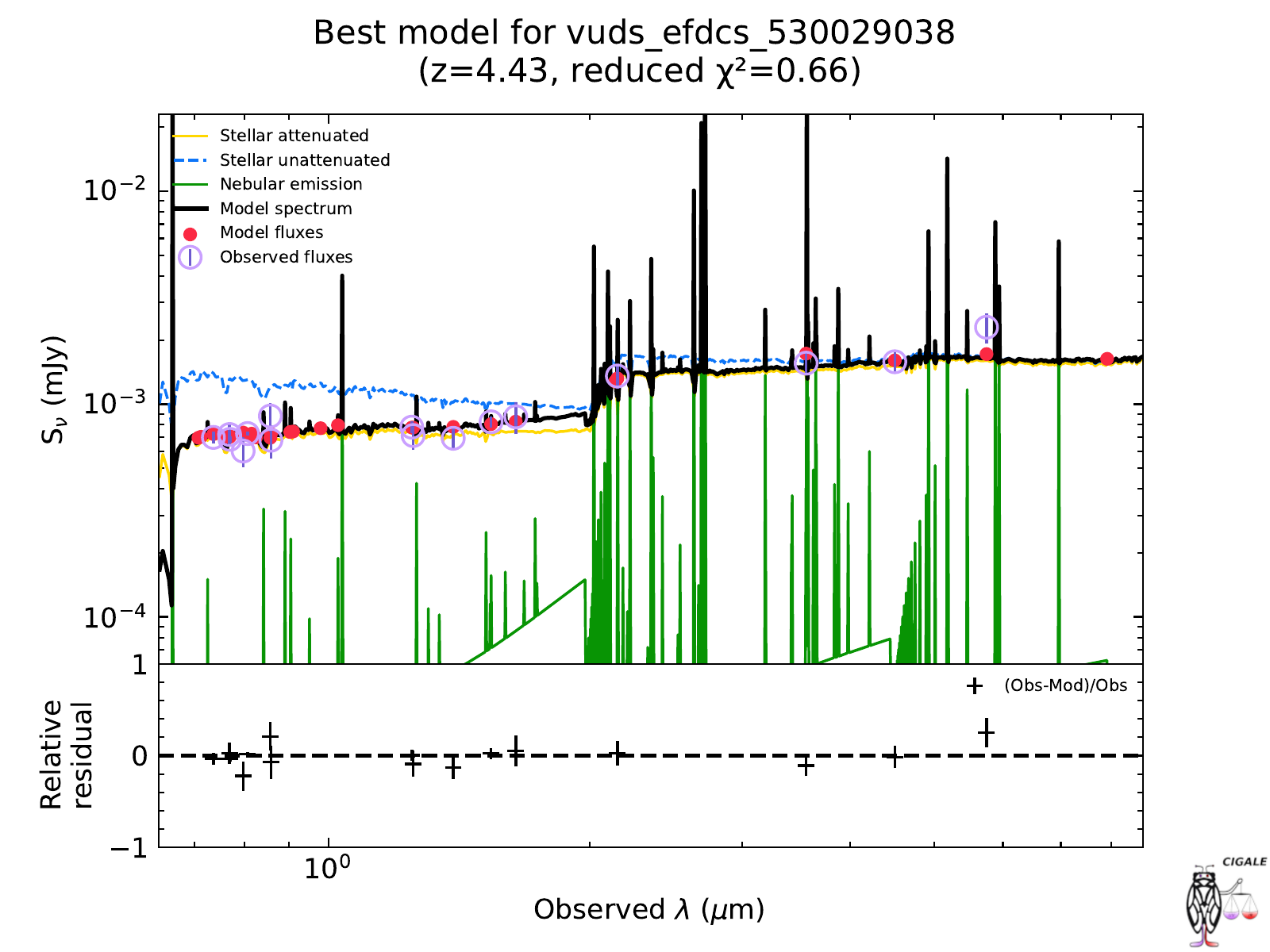}
  \caption{Best-fits of galaxies vuds\_cosmos\_5180966608 at $z=4.53$ (left) and vuds\_efdcs\_530029038 at $z=4.43$ (right). The galaxy spectra are shown in black, with the dust-attenuated stellar populations in yellow and the ionized gas in green. For comparison, the dust-free stellar populations are shown in blue. The model fluxes integrated in passbands are shown as red circles and the observations and 1-$\sigma$ uncertainties are indicated with purple circles with vertical lines. Finally the relative residuals are shown in the bottom panels of each plot.\label{fig:best-fit}}
\end{figure*}

Finally, the physical properties and their uncertainties are estimated by computing the likelihood-weighted mean and standard deviation of the parameters of all the models fitted to the data. The attenuations are determined for each model computing the dust-free and dust-attenuated luminosities through the GALEX FUV and the Bessel V bands. The IRX parameter is taken as the decimal logarithm of the dust luminosity to the rest-frame UV luminosity in the GALEX FUV filter. The UV slope $\beta$ is computed by fitting a powerlaw on the restframe UV spectrum of each model using the wavelength windows defined in \cite{calzetti1994a}. The dust-free UV slope $\beta_0$ is computed in the same way. All the other parameters either correspond to input parameters ($\delta$) or to values derived from the models (SFR and M$_\star$).

\subsection{Band selection, adjustment of uncertainties\label{ssec:bands}}

We construct the photometric table using the compilation of \cite{faisst2020a} in the rest-frame UV-optical, which is based on the COSMOS15 \citep{laigle2016a} and 3D-HST \citep{brammer2012a, skelton2014a} catalogs, which in turn were based on the HST imaging obtained by CANDELS \citep{grogin2011a, koekemoer2011a}. From this large dataset, we eliminate bands contaminated by the Ly-$\alpha$ line due to its uncertain escape fraction. We also remove bands at shorter wavelengths as the intervening IGM renders the models somewhat less reliable in this range and because there is evidence that the attenuation curve flattens at shorter wavelengths at low redshift \citep{buat2002a, leitherer2002a} as well as for more distant galaxies \cite{reddy2016a}, which would only add to uncertainties.

An important aspect to note is that in order to avoid over-fitting due to underestimated flux uncertainties, we follow the strategy of \cite{faisst2020a} that was adopted for the LePhare modeling. We scale the uncertainties upward by 10\% and we add in quadrature 0.01~mag for ground-based images, 0.05~mag for intermediate bands, 0.1~mag for Spitzer 3.6~$\mu$m and 4.5~$\mu$m, and 0.3~mag for Spitzer 5.8~$\mu$m and 8.0~$\mu$m. At the same time, we disable the additional uncertainty of 10\% that is included in CIGALE by default.

To prevent diluting the fit with a large number of low quality fluxes, we make a further band down selection object by object, selecting only bands detected at a signal-to-noise ratio of at least 3. We limit the selection to objects with a least six non-ALMA observations so a sufficient number of bands is available to constrain the attenuation curve. This reduces the sample to 20 tier 1 and 32 tier 2 galaxies.

Overall we include Subaru/Suprime-Cam ($i^\prime$, $z^\prime$, and $z^{\prime\prime}$, IA709, IA738, IA767, IA827, IA856, NB711, and NB816), Subaru/Hyper-Suprime-Cam ($y$), Spitzer IRAC (3.6~$\mu$m, 4.5~$\mu$m, 5.8~$\mu$m, and 8.0~$\mu$m), CFHT/WIRCAM (H$^w$ and K$_s^w$), CFHT/VIRCAM (Y, J, H, and K$_s$), HST/ACS (F755W and F814W), HST/WFC3 (F125W, F140W, and F160W), MPG-ESO/WFI (I), and VLT/ISAAC (J, H, and K$_s$) bands.

\subsection{Final sample selection\label{ssec:final-selection}}

Finally, we only consider galaxies with a maximum absolute uncertainty on the slope $\delta$ of 0.45. The reason for doing so is that the inclusion of galaxies with poor constraints on $\delta$ may affect our conclusions. The choice of this value is somewhat subjective. However, an uncertainty of 0.45 on $\delta$ keeps the estimated uncertainty well under a quarter of the full range. Naturally, this criterion tends to exclude galaxies that are not detected in the dust continuum, and by extension galaxies at low attenuation with a flat attenuation curve (see Sect.~\ref{ssec:impact}). This being said, we find that this choice is appropriate to discriminate with confidence between different regimes of attenuation curves. 

Unsurprisingly, this final selection criterion has a strong effect on the tier 2 sample in particular, which is reduced to 9 galaxies. The reason is that only having upper limits on the dust continuum emission limits our ability to constrain the shape of the attenuation curve. The final tier 1 sample suffers from a much more modest reduction in its size, down to 14 galaxies. Out of the 23 galaxies of our final sample, 4 are located beyond a redshift of 5.

As an illustration, we show examples of typical fits of tier 1 galaxies in Fig.~\ref{fig:best-fit}. The complete set of best-fits is presented in Fig.~\ref{fig:best-fits-1} and \ref{fig:best-fits-2} for the rest of tier 1 and tier 2 galaxies. The physical properties (IRX, $\beta$, $\delta$, AFUV, $\mathrm{A_V}$, SFR, and M$_\star$) and the corresponding uncertainties for these 23 galaxies are presented in Table~\ref{tab:properties}. For clarity, all these physical properties are directly estimated from CIGALE. IRX is computed as the ratio of the total dust luminosity to the luminosity in the GALEX FUV band and $\beta$ is determined by fitting the UV continuum in the spectral windows given in Table 2 of \cite{calzetti1994a}.

\subsection{Reliability of the estimation of physical properties\label{ssec:reliability}}

\begin{figure*}[!htbp]
 \centering
 \includegraphics[width=0.4\textwidth]{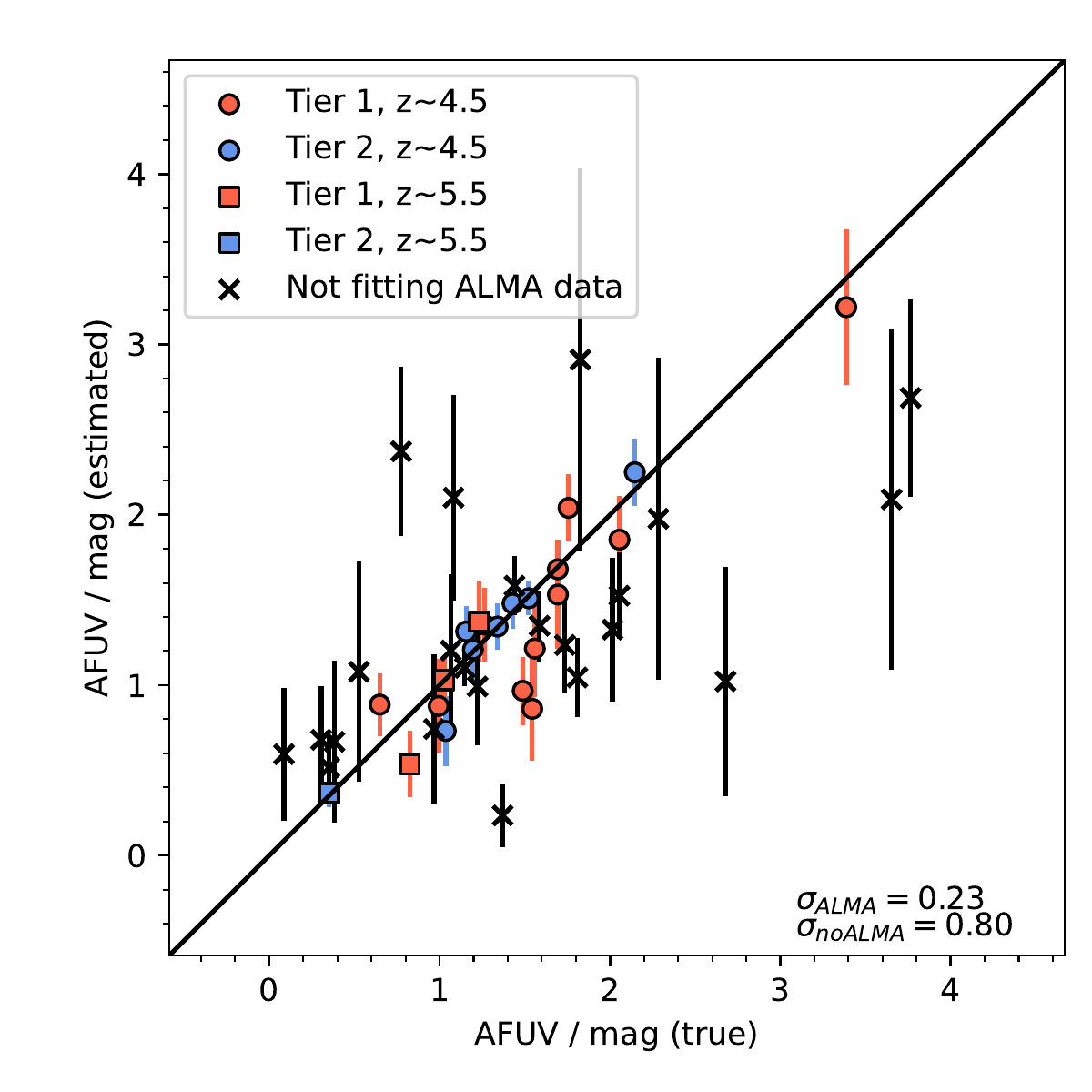}
 \includegraphics[width=0.4\textwidth]{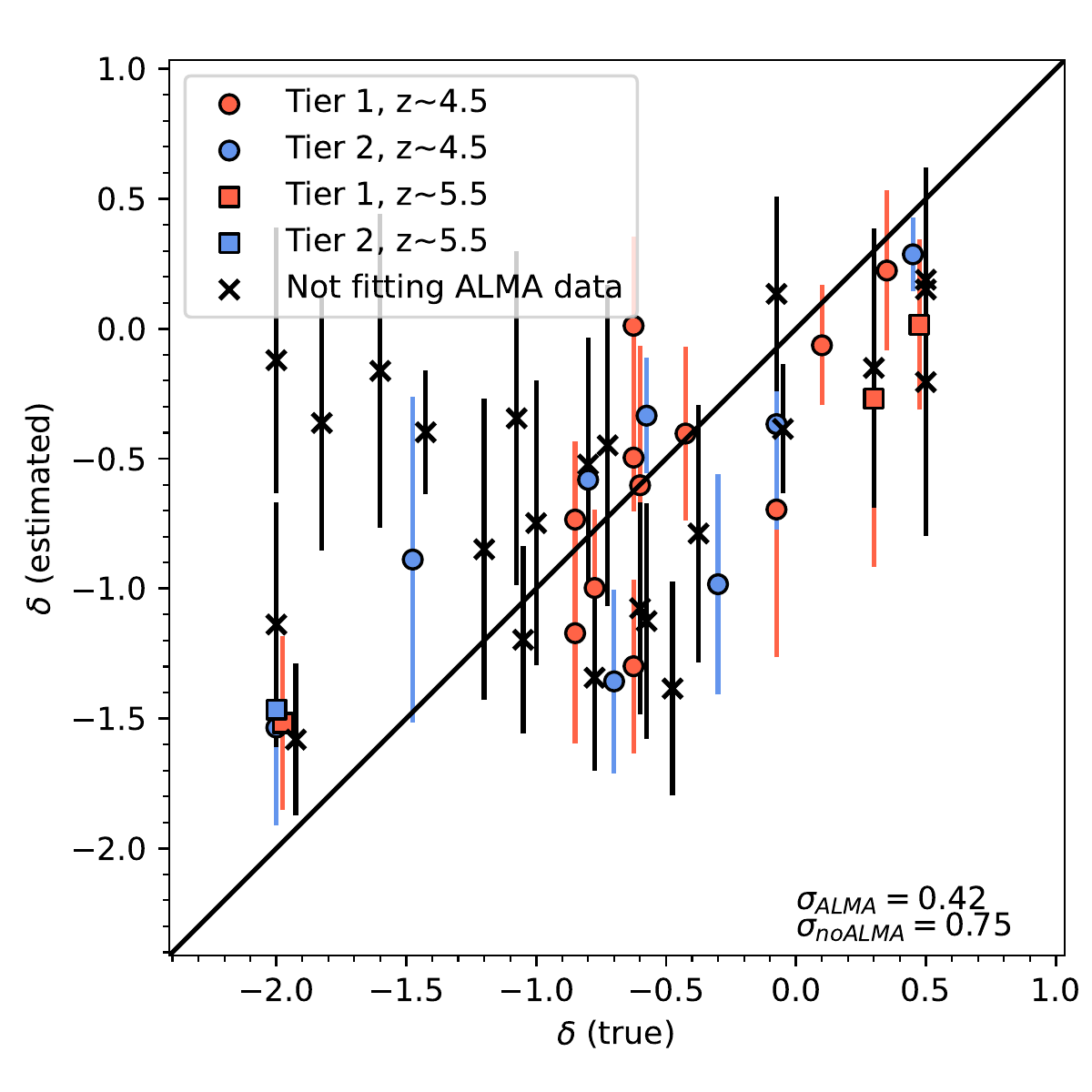}\\
 \includegraphics[width=0.4\textwidth]{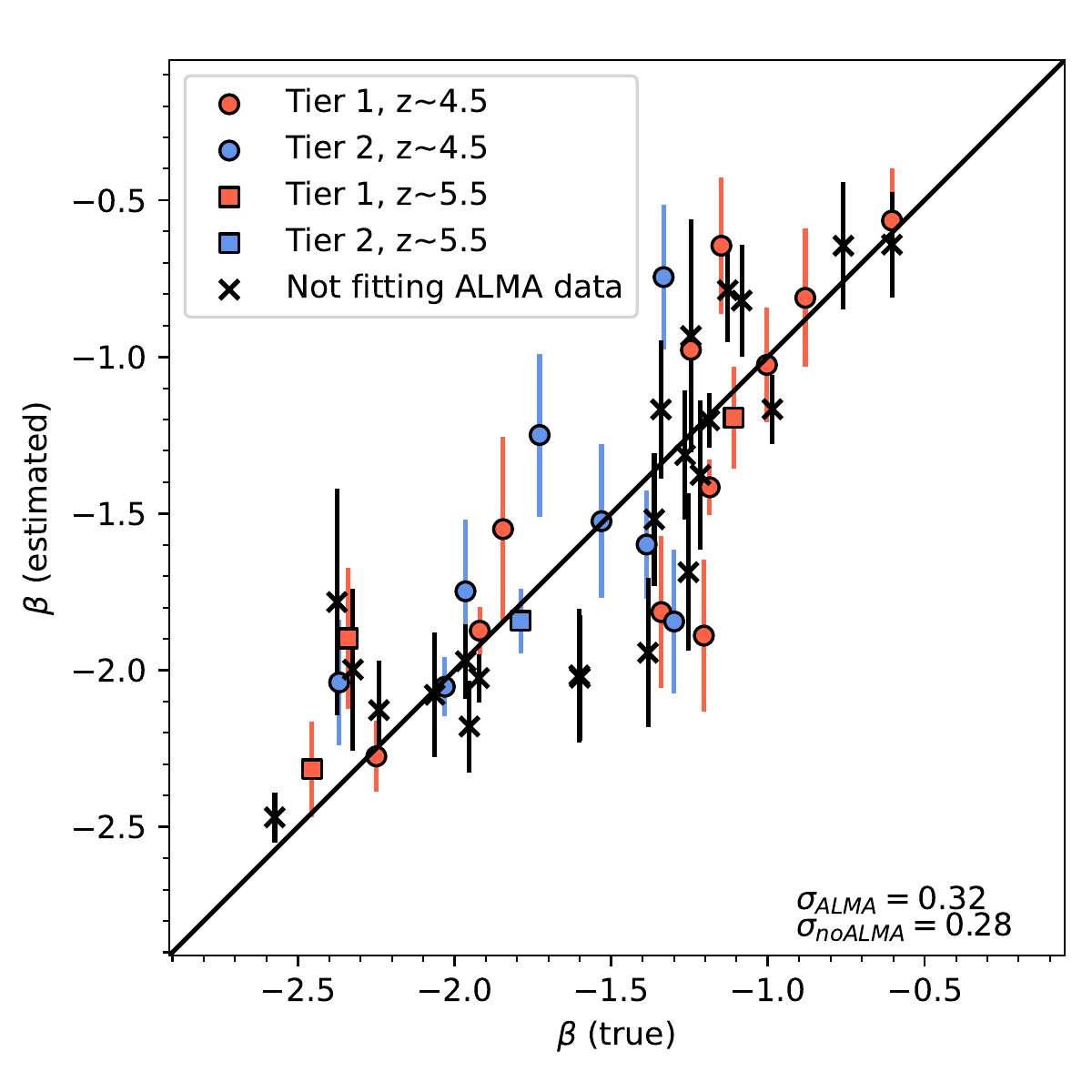}
 \includegraphics[width=0.4\textwidth]{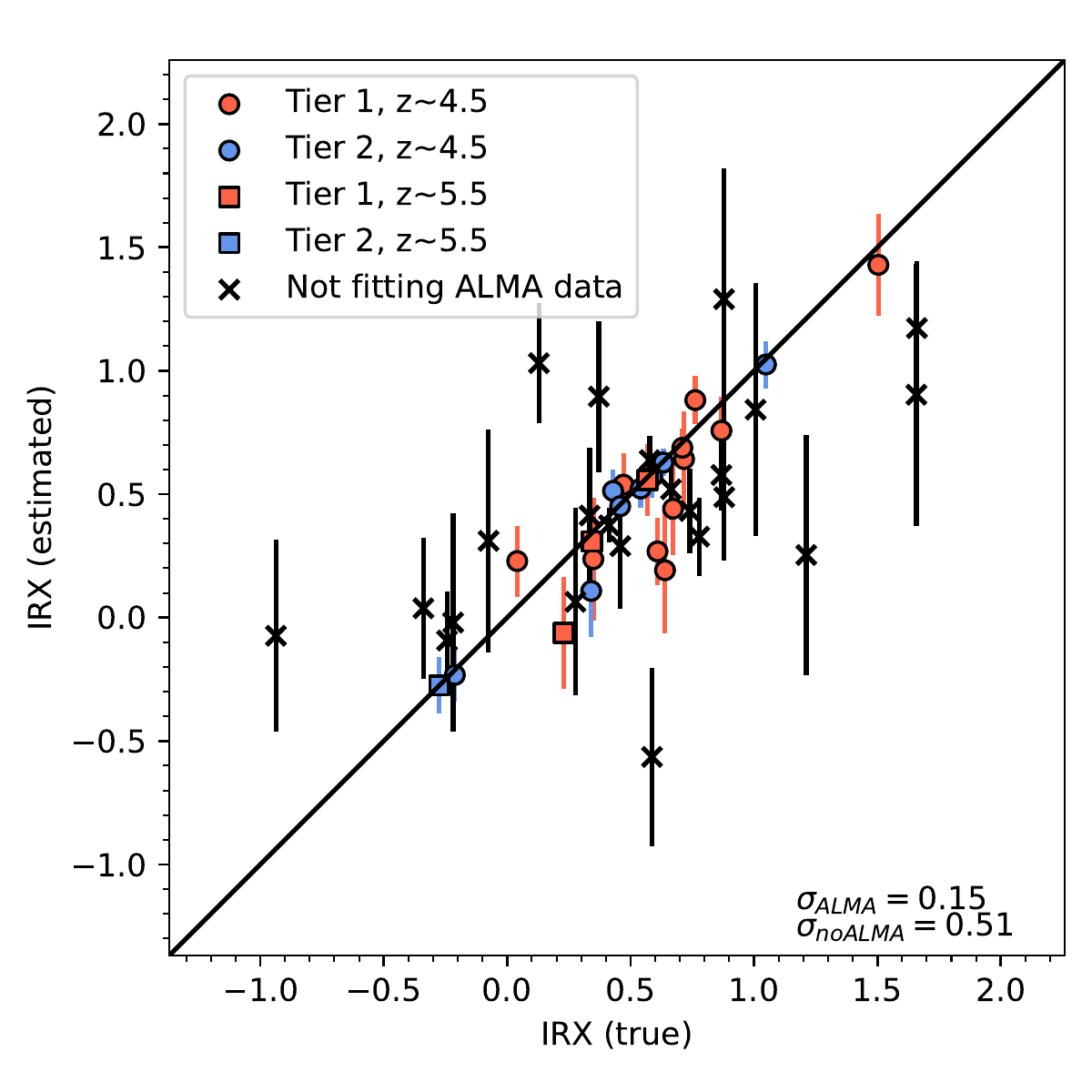}\\
 \includegraphics[width=0.4\textwidth]{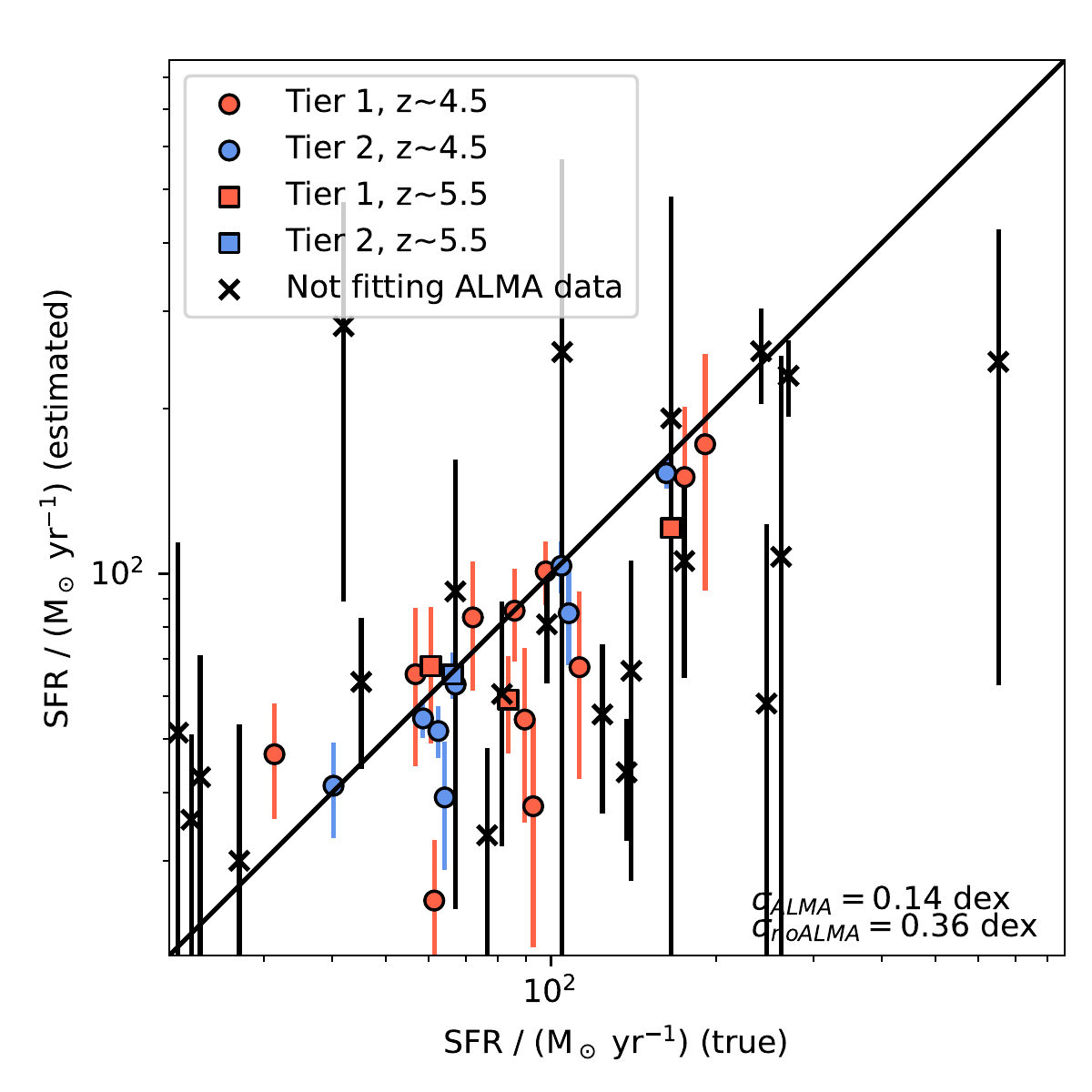}
 \includegraphics[width=0.4\textwidth]{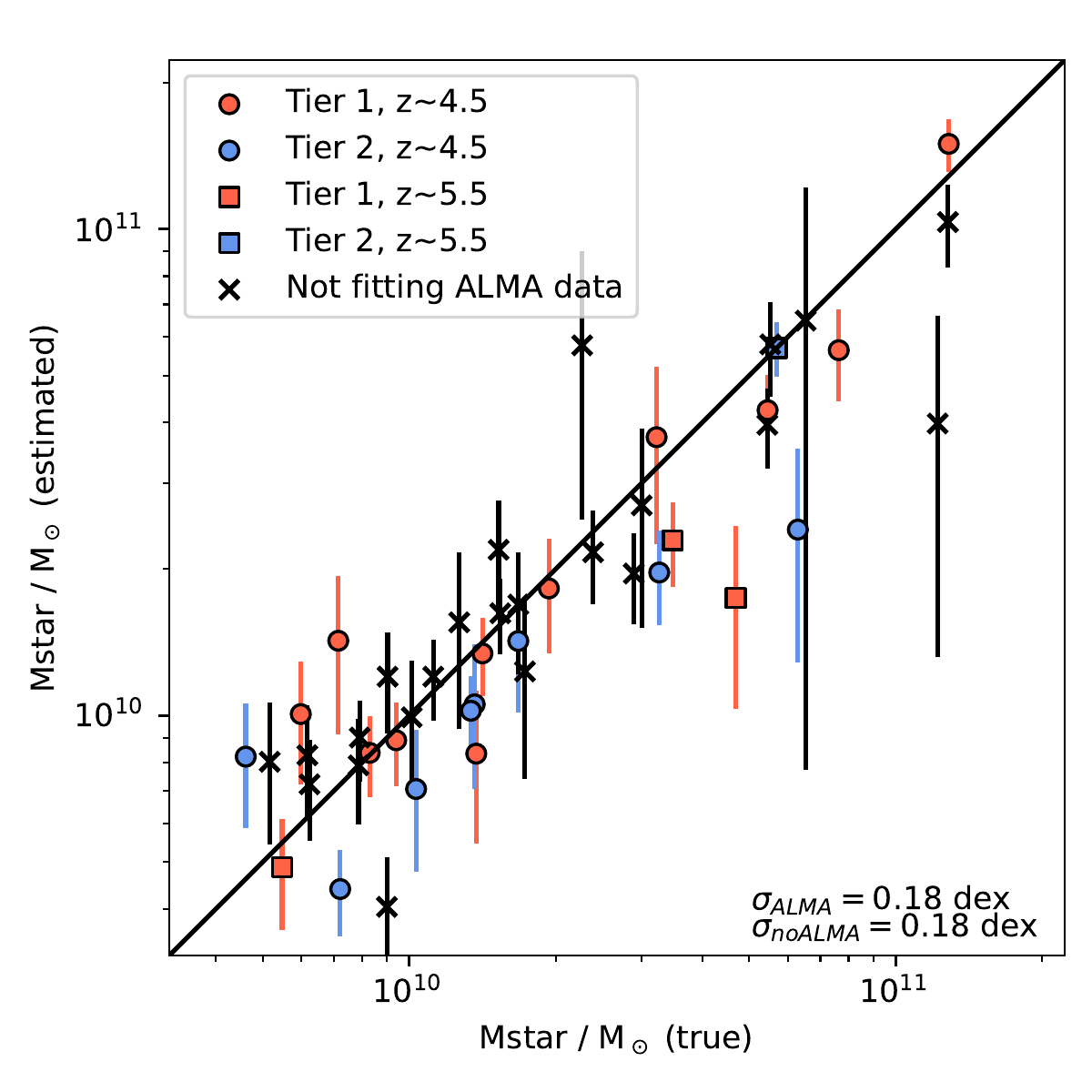}
 \caption{Comparison of the true ($x$ axis) and estimated values ($y$ axis) for AFUV, $\delta$, $\beta$, IRX, SFR, and M$_\star$, from top-left to bottom-right. The solid black line indicates the one-to-one relation. The red (blue) symbols represent galaxies detected in the continuum at 158~$\mu$m rest-frame (detected only in the [C~\textsc{ii}] line). The black crosses represent the same objects when the ALMA data ([C~\textsc{ii}] and/or 158~$\mu$m continuum) are not taken into account in the fit, essentially limiting it to rest-frame UV and optical observations. The inclusion of ALMA data visibly improves the estimates of many of the physical properties. For a physical property as difficult to estimate as the slope of the attenuation curve $\delta$, the inclusion of ALMA data is decisive. Even though the scatter remains important, in particular for the steepest (most negative $\delta$) curves, this is a considerable progress compared to the situation without ALMA observations for which constraints are much more lax. The standard deviation of the physical properties estimated with and without ALMA data is indicated in the bottom-right corner of each panel.\label{fig:mocks}}
\end{figure*}

One important improvement over the modeling presented in \cite{faisst2020a, fudamoto2020a} is the inclusion of ALMA observations (both the [C~\textsc{ii}] line and the total infrared luminosity). First, the ALMA data may bring decisive information to break degeneracies, for instance between the SFH and the attenuation. In order to assess the capacity of the modeling to estimate reliably the physical properties with and without ALMA data, we follow the strategy of \cite{giovannoli2011a}. We first build a synthetic catalog from the best-fits. For each object, random perturbations are then injected into the best-fit fluxes. These perturbations are drawn from a Gaussian distribution with a standard deviation corresponding to the uncertainty given in the input catalog. Finally, we fit this synthetic catalog and we compare the Bayesian estimates of the physical properties to the exact values obtained from the original best-fits. We show the results in Fig.~\ref{fig:mocks}.
It appears that AFUV, $\delta$, $\beta$, IRX, SFR, and M$_\star$ are all well determined for objects from tiers 1 and 2.

If we fit the same objects, this time not including ALMA observations, we obtain a starkly different picture as is testified by the strongly increased standard deviation for most physical properties. The estimates for most physical properties appear much more uncertain. This is true in particular for $\delta$, which is left nearly unconstrained, with little dependence between the true and estimated values, especially for the steepest slopes. For other physical properties the scatter is visibly larger in most cases. This is notably the case of AFUV, IRX, and the SFR. This demonstrates once again the importance of observing the emission of the dust and [C~\textsc{ii}] (taken indirectly into account through the SFR) in order to constrain the physical properties of galaxies.

We must emphasize that using synthetic catalogs, our main aim here is to verify that with the current set of observations and models, the code is able to retrieve the physical properties self-consistently. For the physical properties it cannot retrieve, then we have the assurance that we cannot trust these estimates. For the physical properties that are well-retrieved, we can trust these estimates to the extent that the choice of the models is appropriate. So, even though such tests are not foolproof they are efficient at uncovering limitations in the estimation of the physical properties.

\section{Results\label{sec:results}}

\subsection{General properties: SFR, M$_\star$, IRX, and $\beta$}

As mentioned earlier, \cite{fudamoto2020a} explored the relation between IRX, M$_\star$, and $\beta$ in the ALPINE sample. The major difference in our case is that we compute all the physical properties of each galaxy through self-consistent SED modeling from rest-frame FUV to 158~$\mu$m. We examine here the relations between these quantities, also including the SFR.

\subsubsection{SFR-M$_\star$ relation\label{sssec:MS}}

\begin{figure}[!htbp]
 \includegraphics[width=\columnwidth]{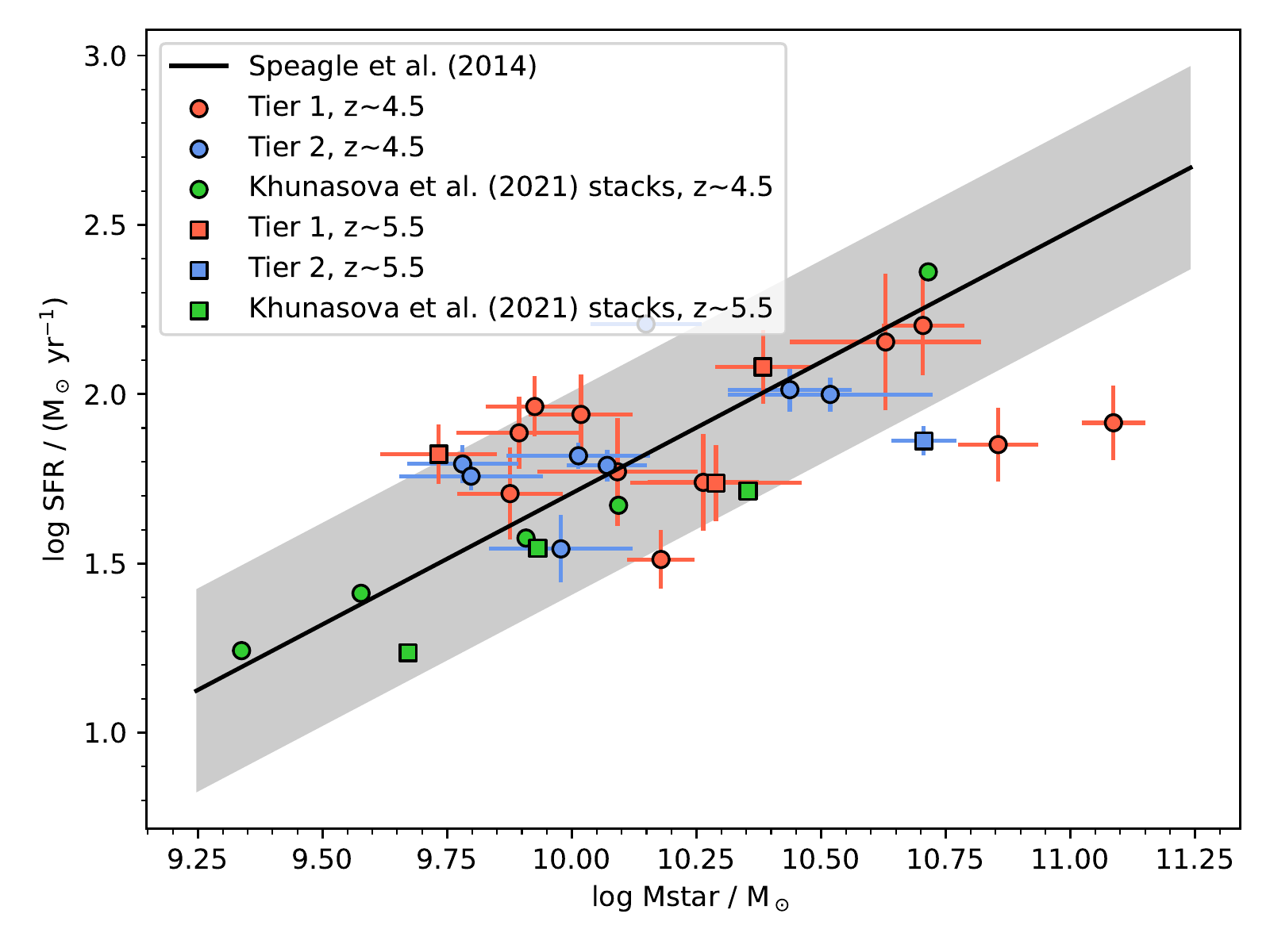}
 \caption{Relation between the SFR and M$_\star$ as estimated by CIGALE for our sample. The symbols are identical to those of Fig.~\ref{fig:mocks}, with the addition of the stacks of \cite{khusanova2021a} in green. The black line corresponds to the relation derived by \cite{speagle2014a} at $z\sim5$, converted from a \cite{kroupa2001a} to a \cite{chabrier2003a} IMF. The shaded area corresponds to a typical $\pm0.3$~dex scatter in SFR. We see that it forms a well defined sequence over approximately a dex in both SFR and M$\star$ and matches the relation inferred by \cite{speagle2014a}. This is expected as the original ALPINE sample was constructed to contain normal star-forming galaxies.\label{fig:MS}}
\end{figure}

As the SFR and M$_\star$ are tightly related, the latter being directly dependent on the integral of the former, the scaling relation between these two quantities is one of our main tools to understand the evolution of galaxies across redshifts \citep[e.g.,][and many others]{brinchmann2004a,noeske2007a}. It is found that most massive galaxies form more stars but in a somewhat sublinear fashion and that at fixed mass, galaxies at lower redshift tend to form stars at a lower rate. We examine in Fig.~\ref{fig:MS} the relation between SFR and M$_\star$ followed by our sample.
We see that it forms a well defined sequence over approximately a dex in both SFR and M$_\star$ and matches the relation inferred by \cite{speagle2014a}. This is expected as the original ALPINE sample was constructed to contain normal star-forming galaxies nd we observe a good agreement with the stacks of \cite{khusanova2021a}.

A fair question, however, would be how the SFR compares with that obtained from simple estimators that are often used at high redshift. To answer this question, we have converted the rest-frame FUV emission estimated with CIGALE and the dust luminosity to SFR using the conversion factors presented in \cite{kennicutt2012a}. After conversion to the \cite{chabrier2003a} IMF using the coefficients of \cite{madau2014a}, we find that the sum of the FUV and dust SFR show no offset compared to the SFR determined with CIGALE with the average logarithmic difference equal to $0.00 \pm 0.05$~dex.

\subsubsection{IRX-$\beta$ relation\label{sssec:irx-beta}}

As we have described in Sect.~\ref{sec:introduction}, the relation between IRX and $\beta$ is an important tool to quantify the attenuation of galaxies. However, it is now well established that there can be important deviations away from the original relation \citep{meurer1999a}, which was derived for UV-bright starburst galaxies. Three physical properties play a key role in defining the locus of a galaxy in the IRX-$\beta$ plane: the intrinsic UV slope $\beta_0$ in the absence of dust, which depends on the SFH as well as on the intrinsic properties of the stellar populations, and the shape and the amplitude of the attenuation curve, which has the effect of reddening the stellar spectrum. The more dust, the redder the UV slope $\beta$ is at fixed attenuation curve slope. Similarly, the steeper the slope of the attenuation curve, the redder the UV slope $\beta$ is at fixed amplitude of the attenuation. At the same time, a redder $\beta_0$ will also yield a redder $\beta$. In the end, interpreting IRX-$\beta$ diagrams to estimate the attenuation always bears the uncertainty of an incorrect assumption on these parameters.

\begin{figure}[!htbp]
 \includegraphics[width=\columnwidth]{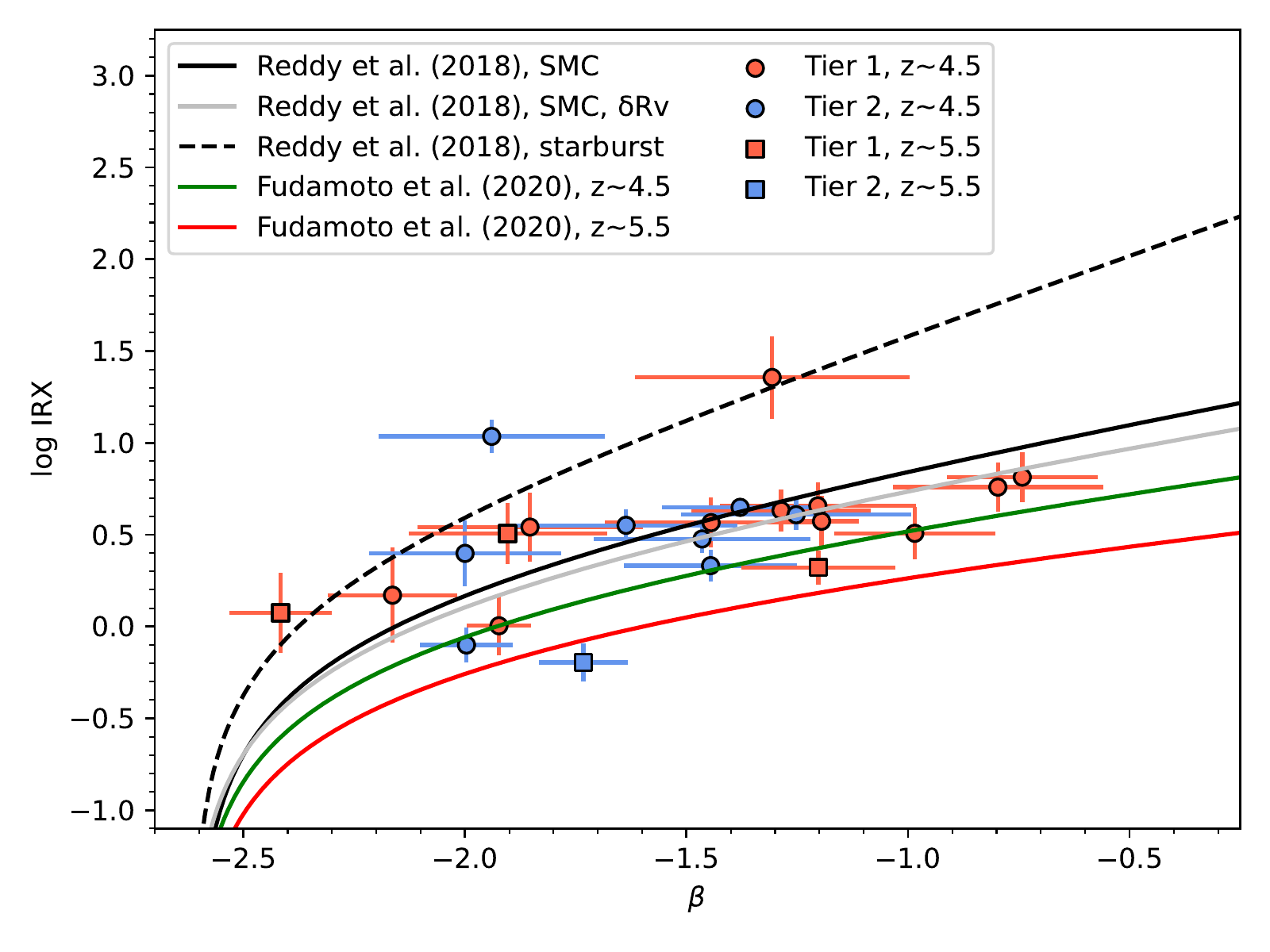}
 \caption{Relation between IRX and $\beta$ as estimated by CIGALE in our sample. The symbols are identical to those of Fig.~\ref{fig:mocks}. The lines correspond to different previously published relations. The dashed black, solid black and gray lines respectively correspond to the starburst, unmodified SMC, and SMC with a modified Rv attenuation curves, as published in \cite{reddy2018a}. The green and red lines correspond to relations derived by \cite{fudamoto2020a} for stacked ALPINE galaxies at $z\sim4.5$ and $z\sim5.5$, respectively. The reasonably small range in the intrinsic UV slope $\beta_0$ and the distribution of the data with respect to various relations suggest that the dispersion is driven by variations of the attenuation curves across the sample.\label{fig:irx-beta}}
\end{figure}

In Fig.~\ref{fig:irx-beta}, we plot the IRX-$\beta$ relation for the galaxies in our sample.
We see that the sample covers the full range, from attenuation curves shallower than the starburst attenuation curve, to steeper than the SMC extinction curve. The relations of \cite{fudamoto2020a} computed from ALPINE stacks at $z\sim4.5$ and $z\sim5.5$ form an approximate lower limit for our sample.

As mentioned above, the assumption on $\beta_0$ can affect the interpretation of this diagram. Both \cite{reddy2018a} and \cite{fudamoto2020a} assume $\beta_0=-2.62$, which is bluer than classical relations probably from differences in the stellar models as \cite{reddy2018a} used the BPASS models \citep{eldridge2009a}, which include a refined treatment of binary evolution. For comparison, \cite{meurer1999a} assumes $\beta_0=-2.23$. In addition, adopting a fixed $\beta_0$, there is the implicit assumption that all galaxies have the same intrinsic color. To assess whether setting $\beta_0=-2.62$ is sound, we examine the CIGALE estimates of $\beta_0$. We find that over our final sample of 23 galaxies, $\left<\beta_0\right>=-2.54\pm0.07$. Even though it is slightly redder than the adopted value of \cite{reddy2018a}, the difference remains small. Another possible point of comparison is with numerical simulations. \cite{mancini2016a} find $\left<\beta_0\right>\lesssim-2.5$ for galaxies at $z=5-8$. For synthetic objects at $z \sim 5$, \cite{cullen2017a} find $\left<\beta_0\right>=-2.52$. Interestingly, these values appear consistent even though these studies lead to very different attenuation curves. Overall, this shows that, at least from the point of view of UV colors, our stellar populations are consistent with both observations and simulations. In contrast, adopting the starburst relation of \cite{reddy2018a}, to reproduce the observations $\beta_0$ would need to range from $-3.0$ to $-1.5$. Such a broad range seems unlikely given the selection of the sample and the previous considerations. By extension, this means that the dispersion we find is probably driven by variations of the attenuation curve across our sample. Indeed, we find that even though a few points are compatible with the starburst relation, this is excluded for a substantial part of the sample beyond the 3-$\sigma$ level. Conversely, many points are compatible with the SMC curve at a 2-$\sigma$, with just a few galaxies beyond 3-$\sigma$. We will examine the variations of the attenuation curve in more detail in Sect.~\ref{ssec:curve}.

\subsubsection{IRX-M$_\star$ relation}

\begin{figure}[!htbp]
 \includegraphics[width=\columnwidth]{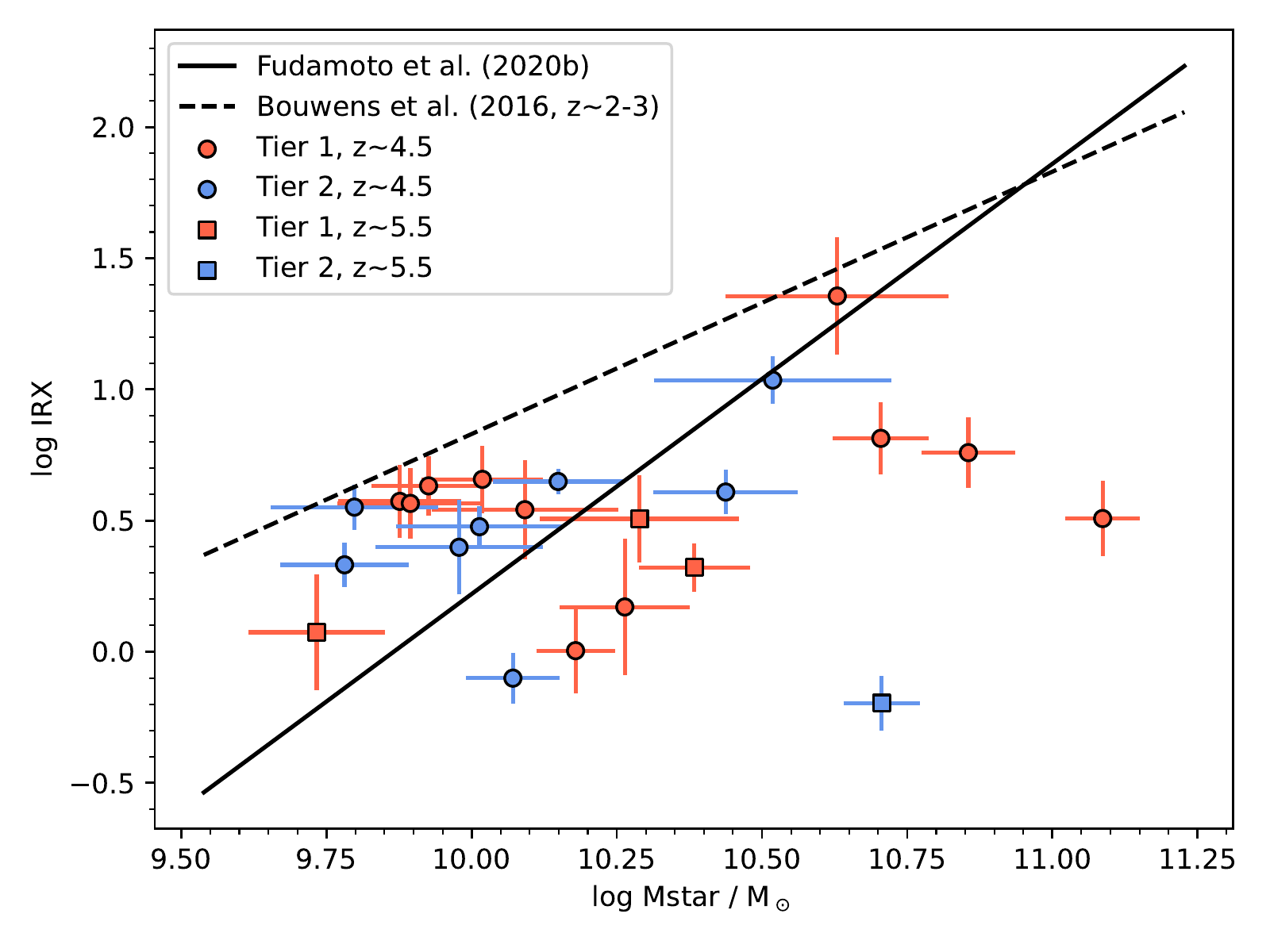}
 \caption{Relation between IRX and M$_\star$ as estimated by CIGALE in our sample. The symbols are identical to those of Fig.~\ref{fig:mocks}. The solid and dashed lines correspond to the relations of \cite{fudamoto2020b} at $z\sim3$ and the relation of \cite{bouwens2016a}, respectively. Similarly to \cite{fudamoto2020a}, we find that our sample of ALPINE galaxies also follows the relation of \cite{fudamoto2020b}, which was computed using an unbiased sample, while falling clearly below the relation of \cite{bouwens2016a}.\label{fig:irx-Mstar}}
\end{figure}

Measuring the slope $\beta$ can be challenging and uncertain for distant galaxies, at least relative to the measure of the stellar mass. Consequently, there have been a number of studies establishing a relation between IRX and M$_\star$ as an alternate way to estimate the dust attenuation in galaxies \citep[e.g.,][]{xu2007a, buat2009a, heinis2014a, bouwens2016a, bogdanoska2020a, fudamoto2020b}. In their study of ALPINE galaxies, \cite{fudamoto2020a} found that there is an evolution of the normalization of the relation with redshift. This is to be noted as, at lower redshifts, the relation shows little change. This was shown at $z \sim 2-3$ by \cite{bouwens2016a} who compiled the relations of \cite{reddy2010a, whitaker2014a, alvarez2016a}. We show in Fig.~\ref{fig:irx-Mstar} the relation between IRX and M$_\star$, which has been determined using CIGALE while taking into account constraints from ALMA observations.
Similarly to \cite{fudamoto2020a}, we find that our sample of ALPINE galaxies also follows the relation of \cite{fudamoto2020b}, which was computed using an unbiased sample, while falling clearly below the relation of \cite{bouwens2016a}. The small number of $z \sim 5.5$ galaxies is not sufficient to comment in detail on an evolution from $z \sim 4.5$. However, in their study which included stacks, \cite{fudamoto2020a} noted a further decrease in the normalization at higher redshift. Overall, we find that the addition of ALMA constraints and the smaller sample does not affect the conclusions of \cite{fudamoto2020a}.

\subsection{Shape of the attenuation curve\label{ssec:curve}}

The combination of multi-wavelengths data with ALMA observations provides us with a unique opportunity to investigate the shape of attenuation curves at $z \sim 5$. As we saw in Sect.~\ref{sssec:irx-beta}, the locus of ALPINE galaxies on the IRX-$\beta$ diagram strongly suggests that there is a diversity of attenuation curves and that no single law can reproduce all of the observations.

\subsubsection{Parametrization of the attenuation curve}

How we quantify the shape of the attenuation curve is important if we want to study its variations. In CIGALE, we have modeled the attenuation adopting a starburst curve that is modified by a power law of index $\delta$ \citep[][see Sect.~\ref{ssec:cigale}]{noll2009a}. The strongest advantage of this parametrization is that $\delta$ is a direct indication of the steepness of the effective attenuation curve. In addition, because it can be used as a screen, it is easy to apply on observations to correct for the presence of dust. We emphasize, however, that the starburst curve itself does not account or assume any particular geometry. Rather, the effect of the geometry is statistically factored into it. A strong limitation of this curve, however, is that it does not explicitly take into account the additional reddening affecting star-forming regions, which can affect the steepness of the curve. In other words, the modified starburst curve empirically reproduces an effective attenuation curve but only gives limited information on the physics behind it.

The model of \cite{charlot2000a} was specifically made to account for the differential attenuation between stars located in star forming regions (Birth Clouds, BC) and those in the ISM. The attenuation curve of each of these two components is described by a power law, with its own index, and the amount of dust associated with each component can also be variable. To make matters more complex, photons from stars in BC are attenuated both by the dust in the BC and the ISM, whereas stars not belonging to BC are only attenuated by the ISM component. This model is consistent with results from radiation transfer modeling \citep{chevallard2013a, buat2018a} and has been remarkably successful to reproduce the observations and provides for a natural explanation for the shape of the starburst attenuation curve, and deviations around it. As a consequence, it is frequently used in models. The main characteristic, however, is that there is only an indirect link between the shape of the attenuation curve of each component and the shape of the effective attenuation curve as would be measured from the observations. That is, the model of \cite{charlot2000a} does not directly give an attenuation curve, unlike the modified \cite{calzetti1994a, calzetti2000a} curve \citep{noll2009a}. Rather, even with fixed model parameters, the shape of the attenuation curve depends on the properties of the stellar populations, and therefore, on the SFH.

This being said, even though the \cite{charlot2000a} model reproduces better the observations in the NIR \citep{lofaro2017a} and models based on radiation transfer computations \citep{chevallard2013a}, the level of flexibility of both curves is similar in the UV and optical domains and the \cite{charlot2000a} model shines in a more theoretical setting such as numerical simulations and SED modeling. For this study, we have therefore decided to adopt the starburst curve for the modeling with CIGALE, as it allows for a simpler measurement of the shape of the attenuation curve through the $\delta$ index. We will however use the \cite{charlot2000a} parametrization in Sect.~\ref{sec:discussion} to inform on the physical interpretation of the results.

\subsubsection{Relation between $\delta$ and $\mathrm{A_V}$}

\begin{figure}[!htbp]
 \includegraphics[width=\columnwidth]{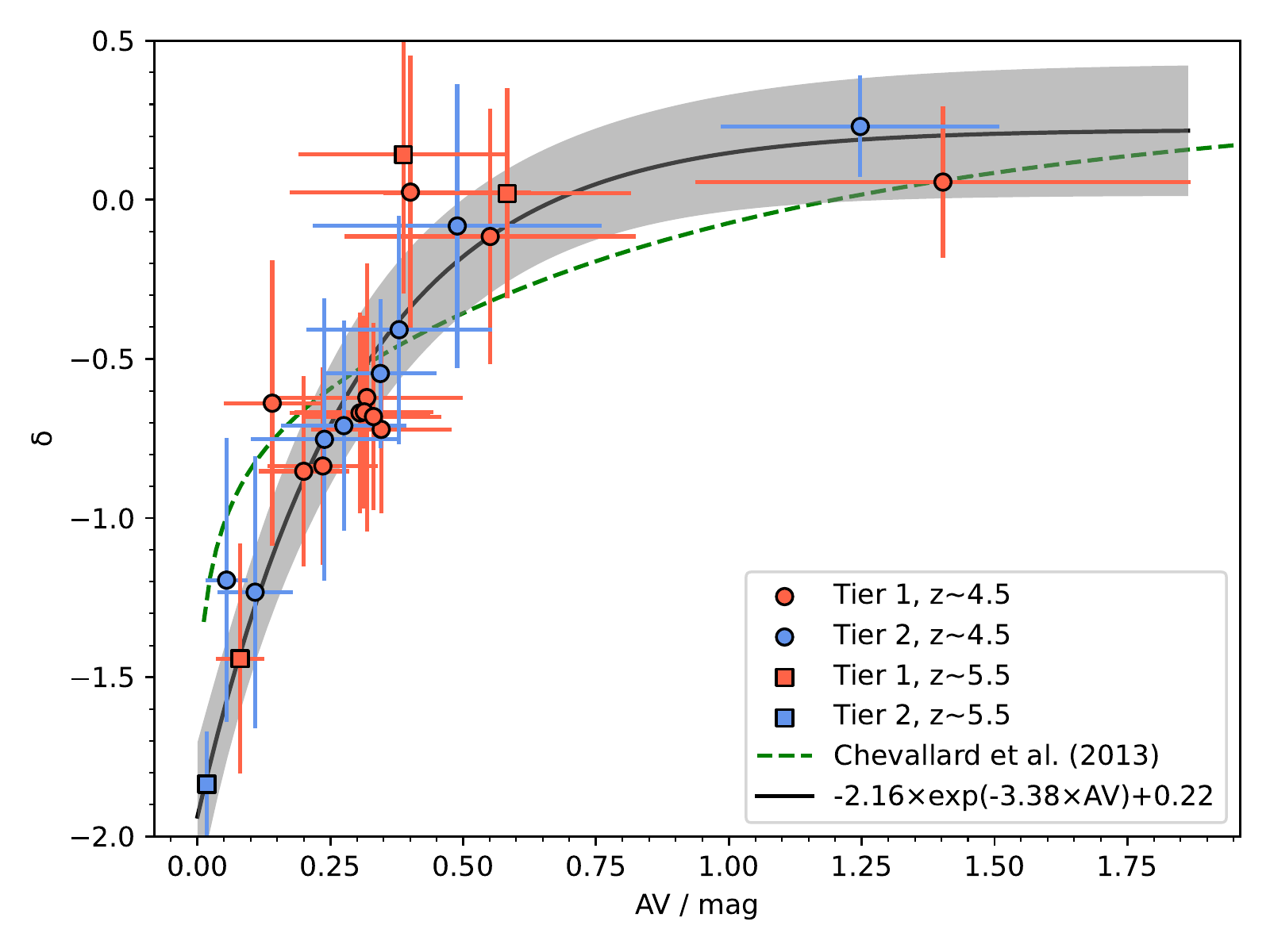}
 \caption{Index $\delta$ of the power law modifying the starburst attenuation curve versus the V band attenuation. The symbols are identical to those of Fig.~\ref{fig:mocks}. The black line corresponds to the fit of a function of the form $a\times\exp\left(b\times\mathrm{A_V}\right)+c$ using an orthogonal distance regression method, with the gray area representing the 1--$\sigma$ dispersion. The dashed green line represents the model of \cite{chevallard2013a} as described in Sect.~\ref{ssec:discussion-dust-properties}. We see that there is a clear relation between the V-band attenuation and $\delta$, with higher attenuation galaxies having a shallower attenuation curve. Such a variation is expected. It has been found in models and simulations alike.\label{fig:delta-AV}}
\end{figure}

To have a first glimpse of the attenuation in ALPINE galaxies, we show in Fig.~\ref{fig:delta-AV} the range of attenuation curves in our sample in terms of amplitude and slope as parametrized through the index $\delta$.
An interesting property of our sample is the generally moderate attenuation of its members. We find that only two galaxies have a V-band attenuation larger than 1 magnitude. Even though the parent sample targets main-sequence galaxies selected in the UV, it is not surprising that some galaxies at higher attenuation are included, in particular among the more massive objects. These two galaxies are among the more massive of the samples with a stellar mass of $3.30\times10^{10}$~M$_\odot$ and $4.26\times10^{10}$~M$_\odot$, respectively. It is to be noted that by virtue of our selection, the objects must be detected with ALMA, which means that at fixed mass, our sample is naturally biased towards more dust-rich galaxies compared to the full ALPINE sample. This highlights both the sensitivity of the survey and the low dust content of main-sequence galaxies at $z\sim5$. Another interesting finding is the large range of $\delta$ covered by the sample, from $-1.84$ to $0.23$. We remind that a starburst curve corresponds to $\delta=0$ and an SMC-like curve to $\delta\simeq-0.5$. This confirms our initial assessment from the inspection of the IRX-$\beta$ diagram that the use of a single attenuation curve is not appropriate at $z \sim 5$. It is important to note that, as we see in Fig.~\ref{fig:delta-AV}, $\delta$ is not randomly distributed. Indeed, we find that there is a clear relation between the V-band attenuation and $\delta$, with higher attenuation galaxies having a shallower attenuation curve. Such a variation is expected. It has been found in models and simulations \citep[e.g.,][]{witt2000a, chevallard2013a, seon2016a, narayanan2018b, trayford2020a, lin2021a} as well as observationally \citep[e.g.,][]{salmon2016a, salim2018a}. We will discuss this relation and its physical origin in Sect.~\ref{sec:discussion}.

\section{Discussion\label{sec:discussion}}

\subsection{Impact of degeneracies}

\begin{figure}[!htbp]
 \includegraphics[width=\columnwidth]{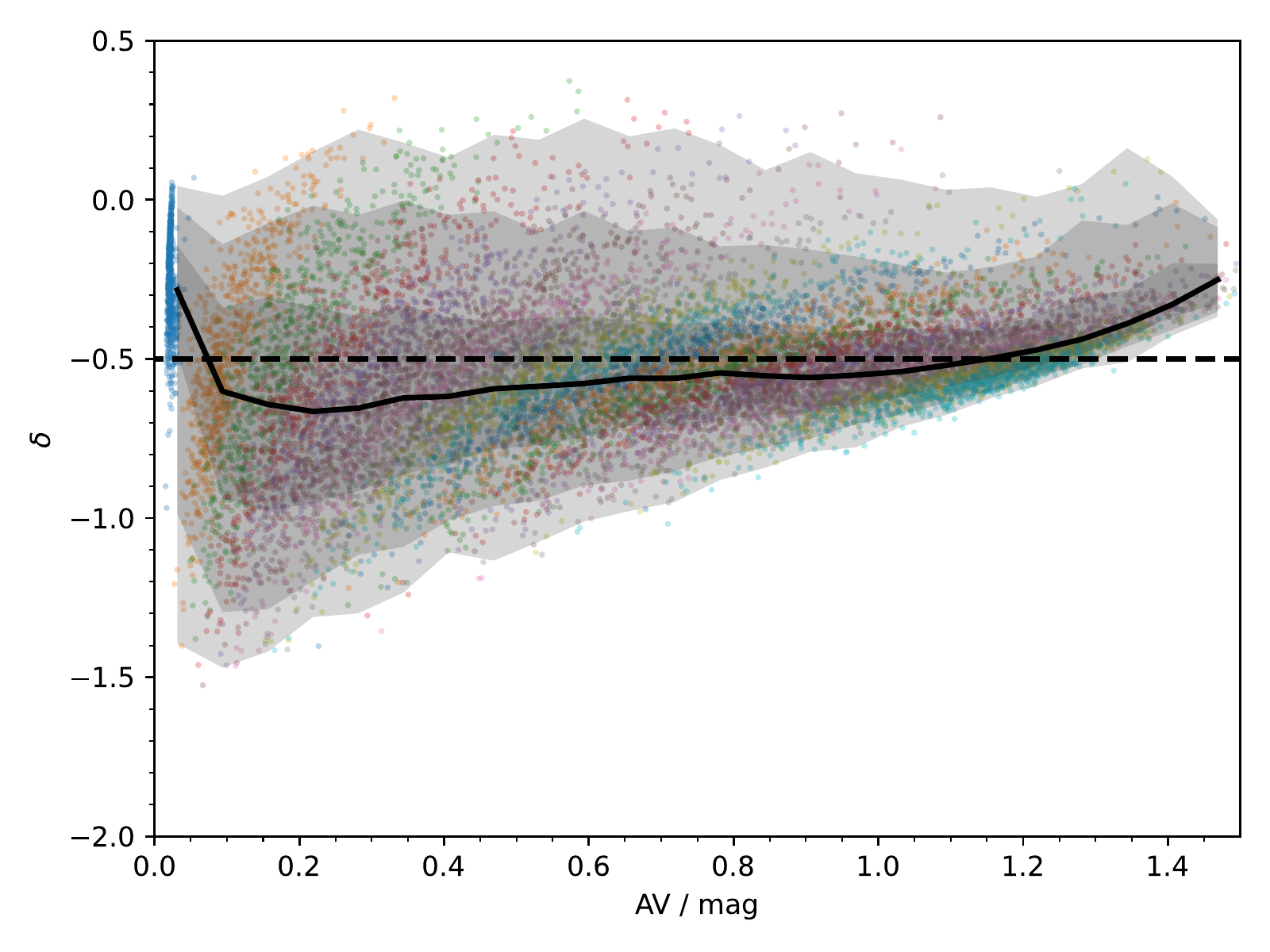}
 \caption{Estimation of $\delta$ versus the V-band attenuation for a set of synthetic galaxies with fixed $\delta=-0.5$ (dashed black horizontal line) and a variable attenuation. Each value of the true reddening is plotted in a different color. The shaded areas represent the 1-, 2-, and 3-$\sigma$ dispersion on $\delta$, with the solid line indicating the median. The effect of the degeneracy is clearly visible for galaxies of the same true reddening. However, considering a range of attenuations, its effect is little different from a dispersion on $\delta$ with only a weak trend towards lower $\delta$ at low $\mathrm{A_V}$. In addition, very few objects have an extremely steep $\delta$, making it unlikely that they are generated due to degeneracies that would smear the results for a single attenuation curve. \label{fig:delta-AV-degeneracy}}
\end{figure}

An important aspect to keep in mind in the rest of the article, however, is that $\delta$ and $\mathrm{A_V}$ are intrinsically degenerate with one another \citep[see for instance the recent study of][]{qin2022a}. The strength of this degeneracy is directly linked to the extent of the multi-wavelength coverage and the uncertainties on the observations. A fair question is therefore whether the relation we see in Fig.~\ref{fig:delta-AV} would be driven by degeneracies rather than physics. In other words, would we obtain this relation if all galaxies followed the same attenuation curve but covered a range of $\mathrm{A_V}$? To answer this question, we have generated a synthetic catalog from the best-fit models of our sample of 23 galaxies but setting a fixed $\delta=-0.5$ and an E(B-V) reddening 0.005~mag to 0.5~mag in 20 steps. The set of bands and the relative uncertainties are taken to be identical to the observations of each galaxy, as are the redshifts. This is especially important as the results may vary significantly based on the bands used and their SNR, as mentioned above. This therefore allows us to reproduce the observed configuration as closely as possible. For each galaxy and each reddening, we generate 50 synthetic galaxies, injecting to each flux and to the dust luminosity a random noise drawn from a normal distribution corresponding to the uncertainty in each band. The physical properties of each of these 23000 galaxies are then evaluated in the same way as ALPINE galaxies, running \textsc{cigale} with the parameters given in Table~\ref{tab:cigale}. We show the results in Fig.~\ref{fig:delta-AV-degeneracy}.
We see that the effect of the degeneracy between $\delta$ and $\mathrm{A_V}$ is clearly visible for galaxies with the same dust properties (points with the same color). However, considering a range of measurement attenuations, its effect is little different from a dispersion on $\delta$ with only a weak trend towards lower $\delta$ at low $\mathrm{A_V}$ as can be seen from the median. This variation is much smaller than the dynamical range of the relation. In addition, very few objects have an extremely steep $\delta$, making it unlikely that they are generated due to degeneracies that would smear the results for a single attenuation curve. This being said, it is possible that considering a narrower range of $\mathrm{A_V}$, therefore considering objects that are very similar in $\mathrm{A_V}$, the local (as opposed to global) trends may be driven by the degeneracy. From these considerations, we think that the degeneracy stretches and amplifies the relation between $\delta$ and $\mathrm{A_V}$ to some degree. In reality it may be somewhat weaker but it is nevertheless real. It is necessary to remember this limitation with interpreting the results.

\subsection{Dust properties and the $\delta$-$\mathrm{A_V}$ relation\label{ssec:discussion-dust-properties}}

The broad diversity of attenuation curves in our homogeneous sample of 23 main-sequence galaxies is striking, going from curves shallower than starburst curves to steeper than SMC curves. At first sight it may be surprising as these curves are often perceived as two extremes. However, the physical properties of the dust, the properties of the stellar populations, and the relative distribution of stars and dust all affect the shape of the attenuation curves, ultimately leading to the $\delta$-$\mathrm{A_V}$ relation presented in Fig.~\ref{fig:delta-AV}.

\begin{figure}[!htbp]
 \includegraphics[width=\columnwidth]{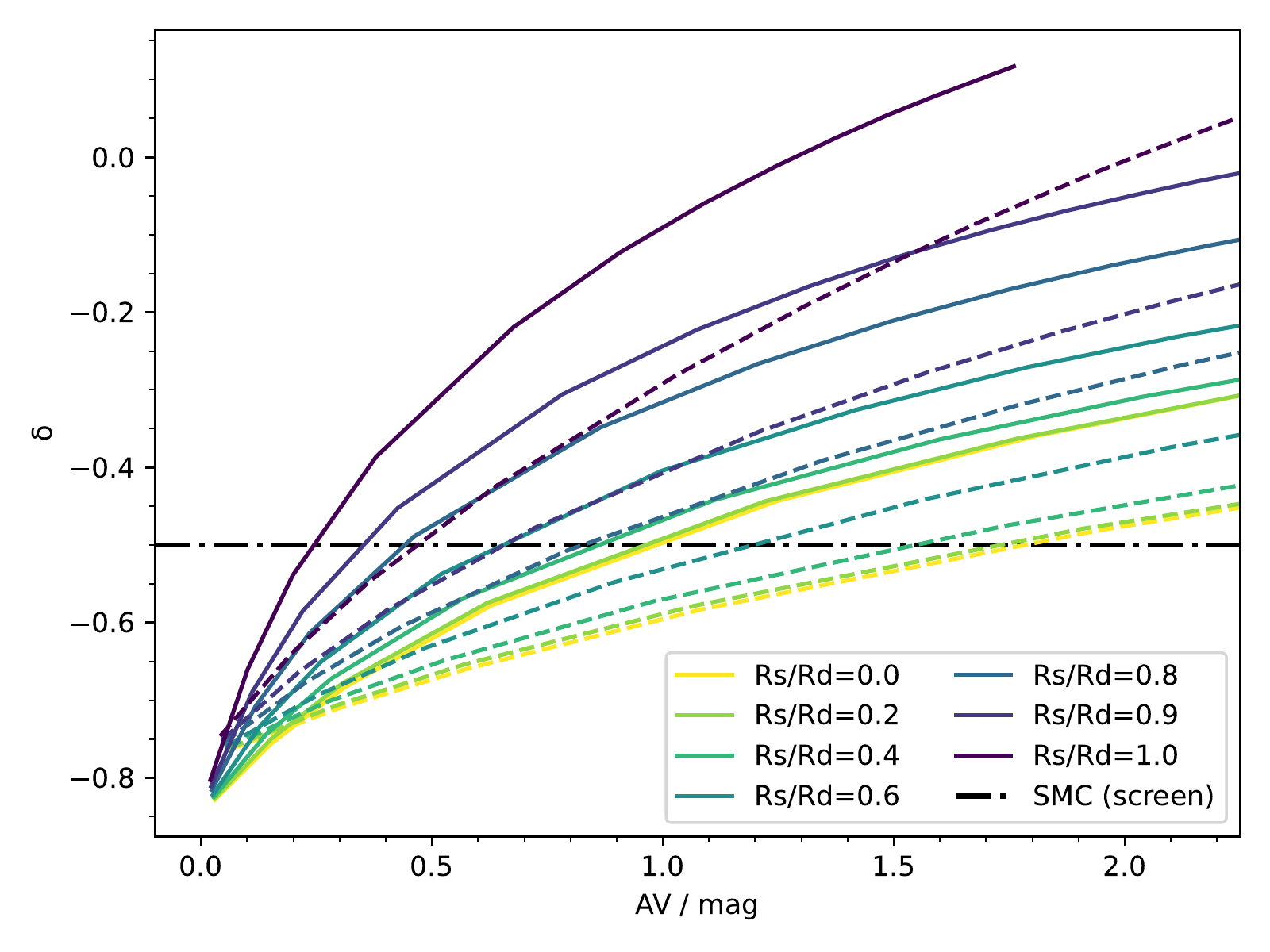}
 \caption{Index $\delta$ of the power law modifying the starburst attenuation curve versus the V band attenuation for the clumpy ISM models of \cite{seon2016a} with a Mach number of 4. Each colored line represents a model with a varying V-band optical depth and for a fixed distribution of stars and dust, parametrized by the ratio of their radii \citep[Rs/Rd, for additional detail see][]{seon2016a}. The solid lines represent a dust composition following the SMC dust model of \cite{weingartner2001a} and the dashed lines that of \cite{witt2000a}. The dash-dotted horizontal line indicates the slope of an SMC curve in the case of a pure screen geometry. We see that at fixed attenuation, the more extended the stellar distribution is compared to the dust, the flatter the attenuation curve is.\label{fig:seon}}
\end{figure}

A first aspect to consider is the relative distribution between stars and dust. Indeed, extinction curves are only valid in the case of a simple screen geometry. In detailed radiation transfer simulations of a clumpy ISM, \cite{seon2016a} showed that the attenuation curve can take a broad range of shapes depending on the properties of the dust and its distribution. The steepest curves are found at low column density, with the variation of the scattering albedo of the dust grains with wavelength playing a key role \citep{inoue2006a, seon2016a}. To illustrate the steepening of the curve at low attenuation and the effect of the geometry we fit a modified starburst law to the radiation transfer models of \cite{seon2016a} and we present in Fig.~\ref{fig:seon} the variation of $\delta$ as a function of the attenuation and the geometry.
The distribution of stars and dust is parametrized by the ratio of their radii, Rs/Rd. A value of 0 corresponds to a situation where the stars are centrally concentrated in the dust whereas a value of 1 corresponds to the stars and the dust following the same distribution. We see that at fixed attenuation, the more extended the stellar distribution is compared to the dust, the flatter the attenuation curve is. This is because an increasingly larger fraction of the stars is located near the edge of the cloud, and therefore an also larger fraction of the stars will face a lower optical depth at shorter wavelengths. We also see that even though the specific dust models behind the SMC extinction curves, \cite{weingartner2001a} and \cite{witt2000a}, do not yield exactly the same results, the general trend is driven by the geometry, with the steepest curves at the lowest attenuation.

It is also interesting to consider larger spatial scales. Displacements between the rest-frame UV and the rest-frame dust emission \citep[e.g.,][]{hodge2016a, tadaki2017a, gomez2018a, puglisi2019a} can lead to changes in the attenuation curves. These offsets do not appear to affect modeling negatively \citep{buat2019a}. However, as shown in \cite{fujimoto2020a} the offsets in ALPINE galaxies appear small or altogether absent, so displacements should not play a major role in the evolution of attenuation curves in our sample. This being said, a number of ALPINE galaxies are mergers \citep{lefevre2020a, jones2020a, jones2021a} and most of these sources are likely undergoing inflow and/or unresolved minor mergers. This is also true for our subsample. This is likely to have some degree of influence. However our sample is too small to split into subsamples based on the morpho-kinematic properties to investigate whether this would have a systematic influence.

Another mechanism that can lead to attenuation curves steeper than the intrinsic extinction curves is the effect of the differential reddening between different stellar populations \citep{inoue2005a}. This is the case because star-forming regions generally account for a much larger fraction of the luminosity in the UV than at longer wavelengths. Because star-forming regions are dustier than the general ISM, there is a transition from low attenuation populations in the near-infrared and in the optical to high-attenuation populations in the UV. Whereas the details of the dust properties and its spatial distribution described above are difficult to assess in this work, we can easily quantify the effect of the differential reddening, which fits very well within the framework of the \cite{charlot2000a} models, which we previously described \citep[see also][]{lin2021a}. To quantify the degree of steepening that differential reddening can bring, we have built a simple model with CIGALE: a fixed delayed SFH (age of 1.3~Gyr, $\tau$ of 500~Myr, and a burst fraction of 5\%, with the other parameters as described in Table~\ref{tab:cigale}), stellar populations with a metallicity $Z=0.008$, following a \cite{chabrier2003a} IMF, and the \cite{charlot2000a} model. The latter has four parameters: the indices of the attenuation curves of the BC and of the ISM, which are described by power laws and that we set to constant values of $n_{BC}=-1.3$ and $n_{ISM}=-0.7$ respectively, following \cite{charlot2000a}, the V-band attenuation of the ISM, and $\mu$, which is defined as the ratio of the ISM attenuation to the sum of the attenuation from the ISM and BC, all in the V band.

\begin{figure}[!htbp]
 \includegraphics[width=\columnwidth]{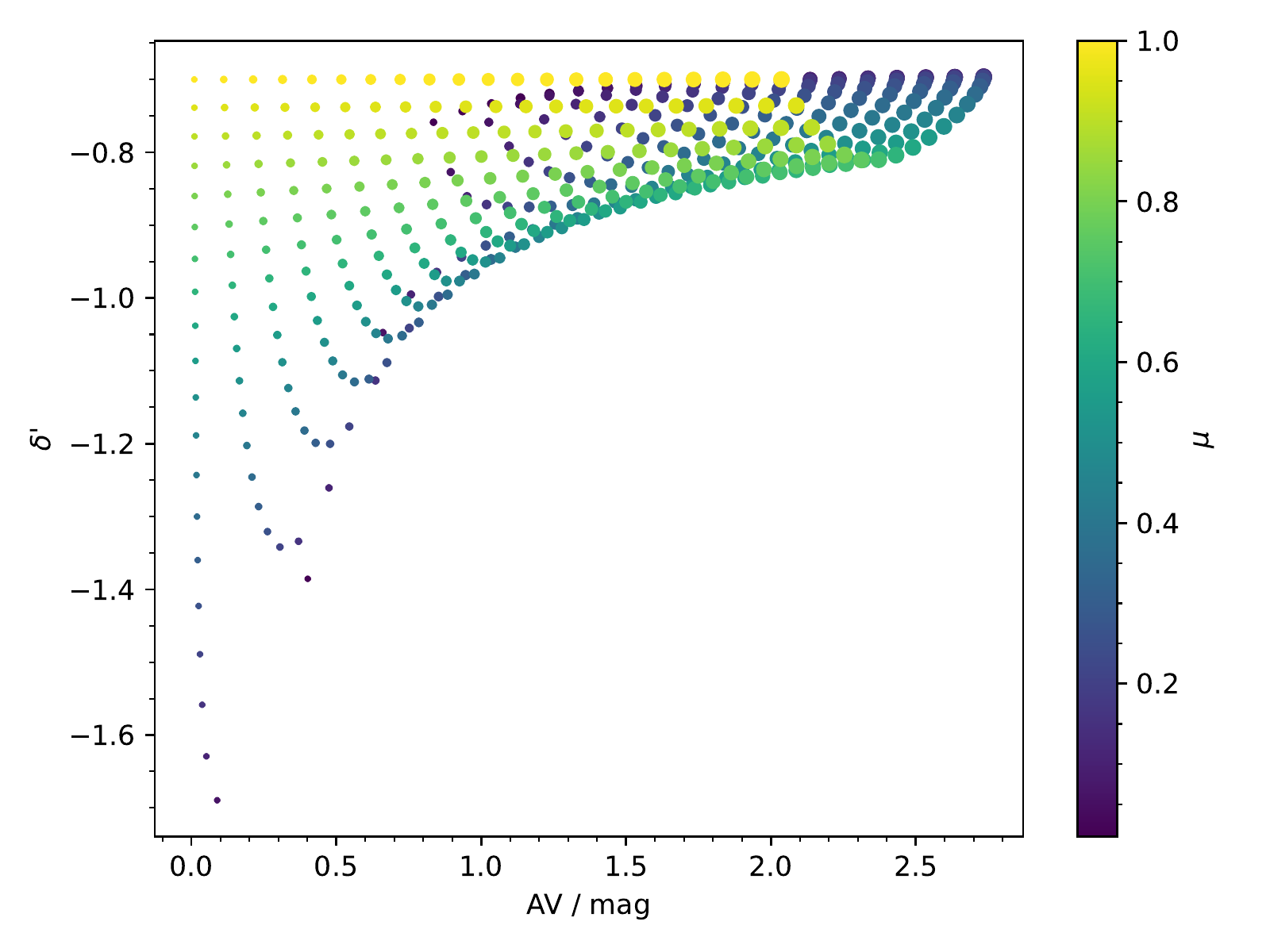}
 \caption{Index $\delta'$ of the power law describing the effective attenuation curve versus the attenuation in the V band for the \cite{charlot2000a} model. The size of the points is proportional to the V band ISM attenuation from 0.01~mag to 2~mag and the color gives $\mu$, the fraction of the dust in the ISM following the color bar to the right of the figure. For reference, a starburst curve corresponds to $\delta'\simeq-0.7$. We see that in the low $\mathrm{A_V}$ and low $\mu$ regime, the attenuation curve can become much steeper. \label{fig:delta-AV-mu}}
\end{figure}

In order to illustrate the effect of the steepening of the slope due to differential reddening, as a starting point, we consider a fairly attenuated galaxy for which we take $\mu=1$ (that is, all the dust is located in the ISM), with an ISM V-band attenuation of 1~mag. Fitting the resulting effective attenuation curve with a power law of index $\delta'$ between 150~nm and 800~nm, we retrieve as expected a power law of index $\delta'\simeq-0.7$, which is similar to the starburst curve (in this formulation, $\delta'\simeq\delta-0.7$). We show the effect of varying both the ISM attenuation and $\mu$ in Fig.~\ref{fig:delta-AV-mu}.
Quantitatively, if we reduce $\mu$ to 0.5 (half the dust is located in BC), the effective slope steepens to $\delta'=-1.07$ and the effective V-band attenuation increases to 1.35~mag. If we now reduce the ISM attenuation to 0.25~mag, the slope steepens to $\delta'=-1.22$, slightly steeper than an SMC curve. In turn, reducing $\mu$ to $1/3$ (that is, $2/3$ of the dust is located in BC) while keeping the ISM attenuation of 0.25~mag yields $\delta'=-1.38$. A more extreme combination of $\mu=0.1$ (90\% of the dust is located in BC) with an ISM attenuation of 0.1~mag corresponds to $\delta'=-1.59$ for an effective V-band attenuation of 0.43~mag. While the steepening reaches nearly 0.9~dex compared to a starburst curve, it is not clear whether such low $\mu$ are realistic, even at high redshift. However, given the young age of these galaxies and their gas-rich nature, it is probably reasonable to consider that dust may remain confined more efficiently in the vicinity of star-forming regions, and is dispersed much less efficiently than it would in more gas-poor galaxies, leaving the general ISM relatively dust free compared to star-forming regions.

Another aspect not to neglect in the previous paragraph is the choice of $n_{BC}$ and $n_{ISM}$ in the \cite{charlot2000a} models. Steeper $n_{BC}$ and $n_{ISM}$ naturally yield steeper effective attenuation curves. We also need to consider a possible variation of $n_{ISM}$ with the attenuation. Adopting the stellar populations described above, $\mu=1/3$ and the dependence of $n_{ISM}$ on the attenuation derived in \cite{chevallard2013a}, we see in Fig.~\ref{fig:delta-AV} that their model appears to broadly reproduce the general trend and amplitude of the variations. 

The SMC extinction curve, the steepest curve we can measure in the local universe, is not a single curve but rather a mean curve over a variety of regions that can be observed across the SMC. It is therefore reasonable to consider that there could be steeper curves, in particular at high redshift, where the physical conditions can differ from that of the SMC, yielding different composition and size distribution. Evidence towards the existence of extinction curves in distant galaxies at least as steep as that of the SMC can be found in observations of quasars and $\gamma$-ray bursts \citep{zafar2015a, zafar2018a}. They are specific cases where we can measure the extinction curve beyond the local universe thanks to their point-like nature that allows to eliminate geometrical effects.

Finally, changes in the properties of the attenuation may be related with the metallicity of the ISM. In a study of redshift $2.0$-$2.5$ galaxies, \cite{shivaei2020a} found that galaxies with a lower metallicity tend to have a steeper attenuation curve. While we do not know the metallicity of the galaxies in our sample, their young age and gas-rich nature suggest that they tend to be on the lower side, which would go towards having steeper curves.

From the above description, we see that there are multiple mechanisms at play that can result in the steepening of the slope at lower attenuation. However, it appears unlikely that any single process can explain by itself the particularly steep slopes we find at the lowest attenuation. In their recent study, \cite{lin2021a}, combined dust grain growth models with radiation transfer computations of one and two-zone geometries of galaxies at very high redshift, showing how different processes can act together, yielding attenuation curves considerably steeper than the extinction curve. More generally, hydrodynamical simulations of galaxies such as those presented in \cite{narayanan2018b} \citep[see in particular panel C of Fig.~6 in][]{salim2020a} or \cite{trayford2020a} (see in particular their Fig.~5) that combine radiation transfer computations with complex stellar populations and geometries are in excellent agreement with the overall trend we observe in our sample. Ultimately, the combination of steep extinction curves at high redshift, the effect of the geometry, and differential reddening probably results in the increasingly steep attenuation curves we measure in lower attenuation galaxies in our sample. 

\subsection{Impact of adopting a single attenuation curve\label{ssec:impact}}

\begin{figure*}[!htbp]
 \includegraphics[width=\columnwidth]{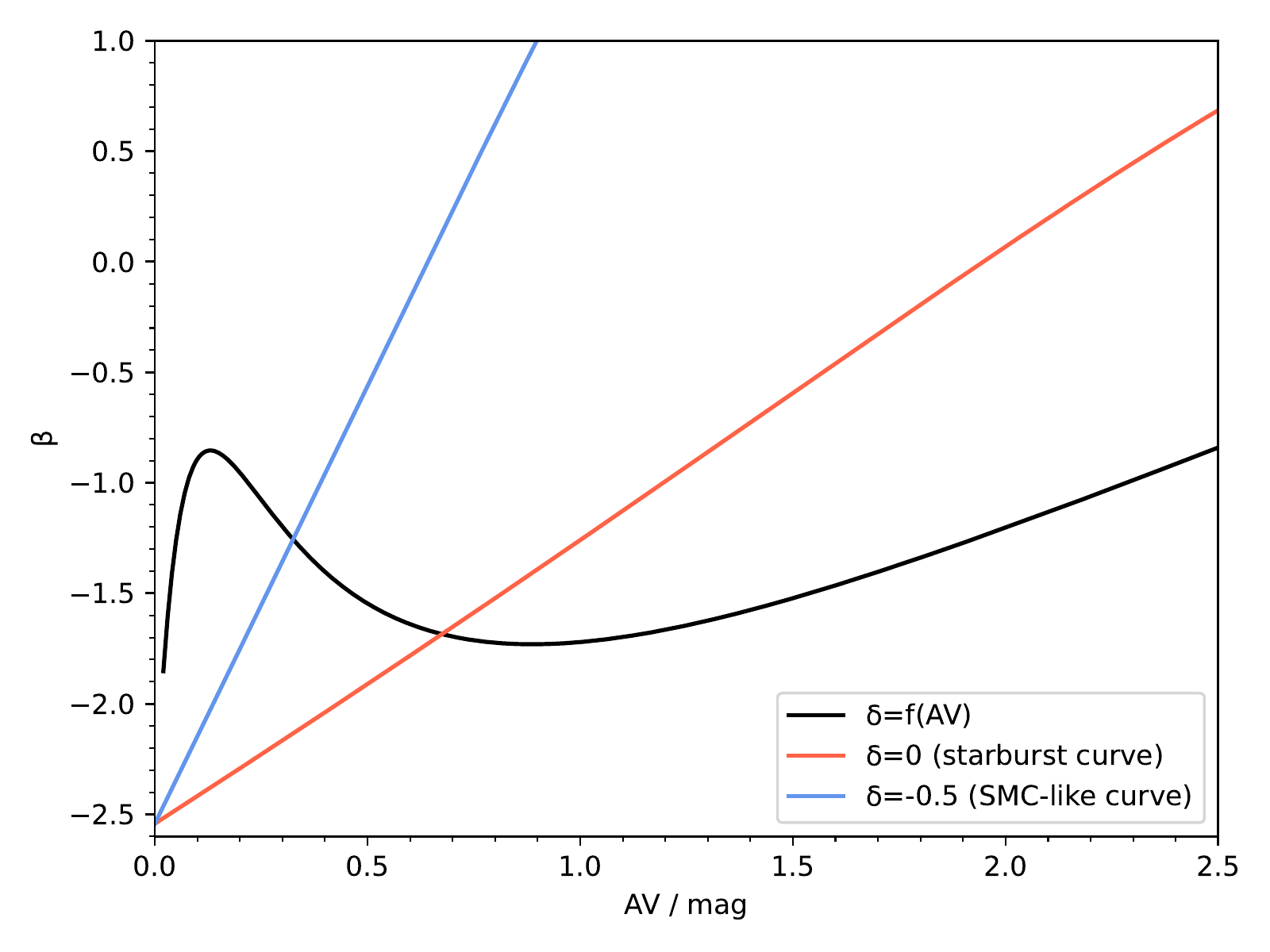}
 \includegraphics[width=\columnwidth]{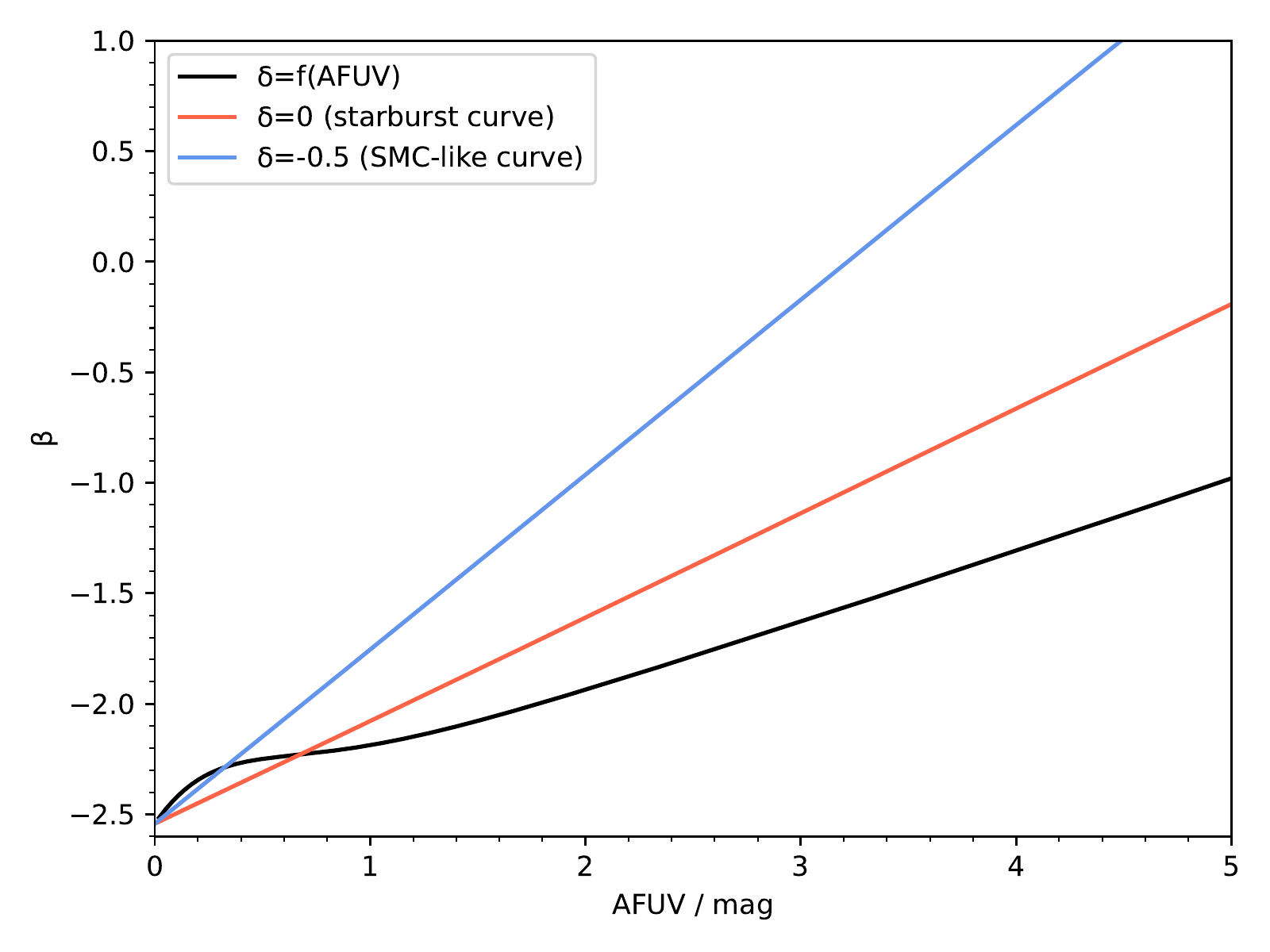}
 \caption{Observed UV slope $\beta$ as a function of the V-band (left) and FUV (right) attenuation (black), assuming the relation between the attenuation curve slope $\delta$ and the attenuation shown in Fig.~\ref{fig:delta-AV}. For comparison the evolution of $\beta$ for starburst and SMC-like attenuation curves are shown in red and blue, respectively. We see that the evolution of $\beta$ with the attenuation follows a very different track compared to the starburst and SMC curves. The consequence is that, in the case of main-sequence galaxies at $z \sim 5$, $\beta$ cannot easily be used to infer the attenuation of galaxies.\label{fig:beta-AV}}
\end{figure*}

The large range of attenuation curves necessarily has an impact on our capacity to reliably correct SFR estimators for the presence of dust. As mentioned earlier, the UV slope $\beta$ is one of the main tracers of the attenuation. Assuming sufficiently deep observations and a narrow range in the intrinsic slope $\beta_0$, the variation of the attenuation curve can be the main source of uncertainty. To quantify the effect of the dependence of the shape of the attenuation curve on $\mathrm{A_V}$, we first build a simple model with a fixed $\beta_0$, which we attenuate with a full range of attenuation curves from $\delta=-1.8$ to $\delta=0.5$. We then interpolate these models following the relation between $\delta$ and $\mathrm{A_V}$ shown in Fig.~\ref{fig:delta-AV}. We show the resulting variation of $\beta$ as a function of the attenuation in Fig.~\ref{fig:beta-AV}.
We immediately see that the evolution of $\beta$ with the attenuation follows a very different track compared to the starburst and SMC curves. If we consider a fixed attenuation curve, $\beta$ becomes monotonically redder, more (SMC) or less (starburst) rapidly. However, if $\delta$ depends on the attenuation, $\beta$ rapidly becomes redder, even at low attenuation, peaking a bit below $\beta=-0.8$ between $0.1$ and $0.2$~mag before slowly decreasing until $\mathrm{A_V}$ reaches slightly under 1~mag and then increases again but more slowly than in the case of a starburst curve, while always staying below $\beta=-0.5$ for an attenuation lower than 2.5~mag. The evolution of $\beta$ with the attenuation is not as dramatic in the FUV. It increases monotonically but not at the constant rate of fixed attenuation curves. We also observe a similar transition from curves steeper than the SMC at low attenuation to shallower than starburst at higher attenuation.

The direct consequence of the aforementioned relation is that, in the case of main-sequence galaxies at $z \sim 5$, $\beta$ cannot easily be used to infer the attenuation of galaxies. Naturally, the magnitude of this issue depends on the exact relation between $\delta$ and $\mathrm{A_V}$. An important question is its behavior at higher attenuation. We see in Fig.~\ref{fig:delta-AV} that this relation is constrained by only two galaxies with an attenuation larger than 1 magnitude that both have $\delta>0$. The starburst curve already being fairly gray, even a moderate flattening with a positive $\delta$ rapidly makes it that dust only has little reddening effect, leaving even dusty galaxies quite blue. \textit{In effect, this means that the blue color of high-redshift galaxies is not necessarily a proof that they would harbor little dust.} As we see, the question of the asymptotic convergence of $\delta$ bears important consequences. If in our case the reliance on only a couple of data points does not allow us to draw a firm conclusion, it is interesting to note that \cite{salmon2016a} also found $\delta>0$ at higher attenuation. This is in agreement with the radiation transfer models of \cite{witt2000a, seon2016a} as well as the simulations of \cite{narayanan2018b, trayford2020a}, which show that this possibility needs to be taken into account when interpreting observations of dusty galaxies.

At lower attenuation, a possible bias to consider is that, all things being equal, we preferentially select galaxies with steeper curves as they have a higher dust luminosity and the uncertainty on $\delta$ is smaller. In that sense, it is possible that the variation of $\delta$ with respect to the attenuation would not be as fast in the sense that there could be galaxies at low attenuation with larger $\delta$. However, even if that is the case, we still expect a variation (see Sect.~\ref{ssec:discussion-dust-properties}) and our sample shows that at least some main-sequence galaxies have a very steep $\delta$, shedding doubt on whether even for an homogeneous set of galaxies a single relation would be appropriate.

\begin{figure*}[!htbp]
 \includegraphics[width=\columnwidth]{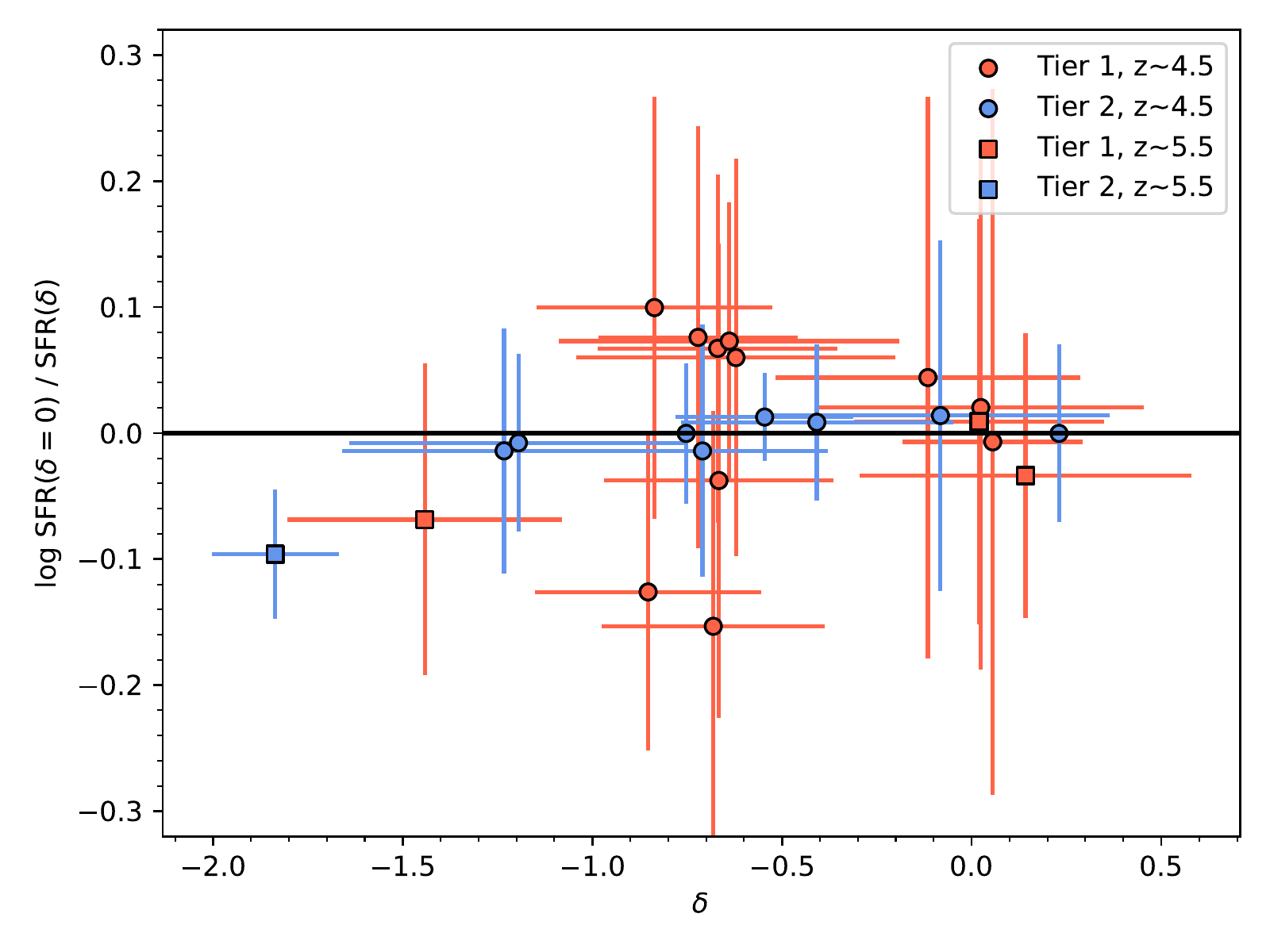}
 \includegraphics[width=\columnwidth]{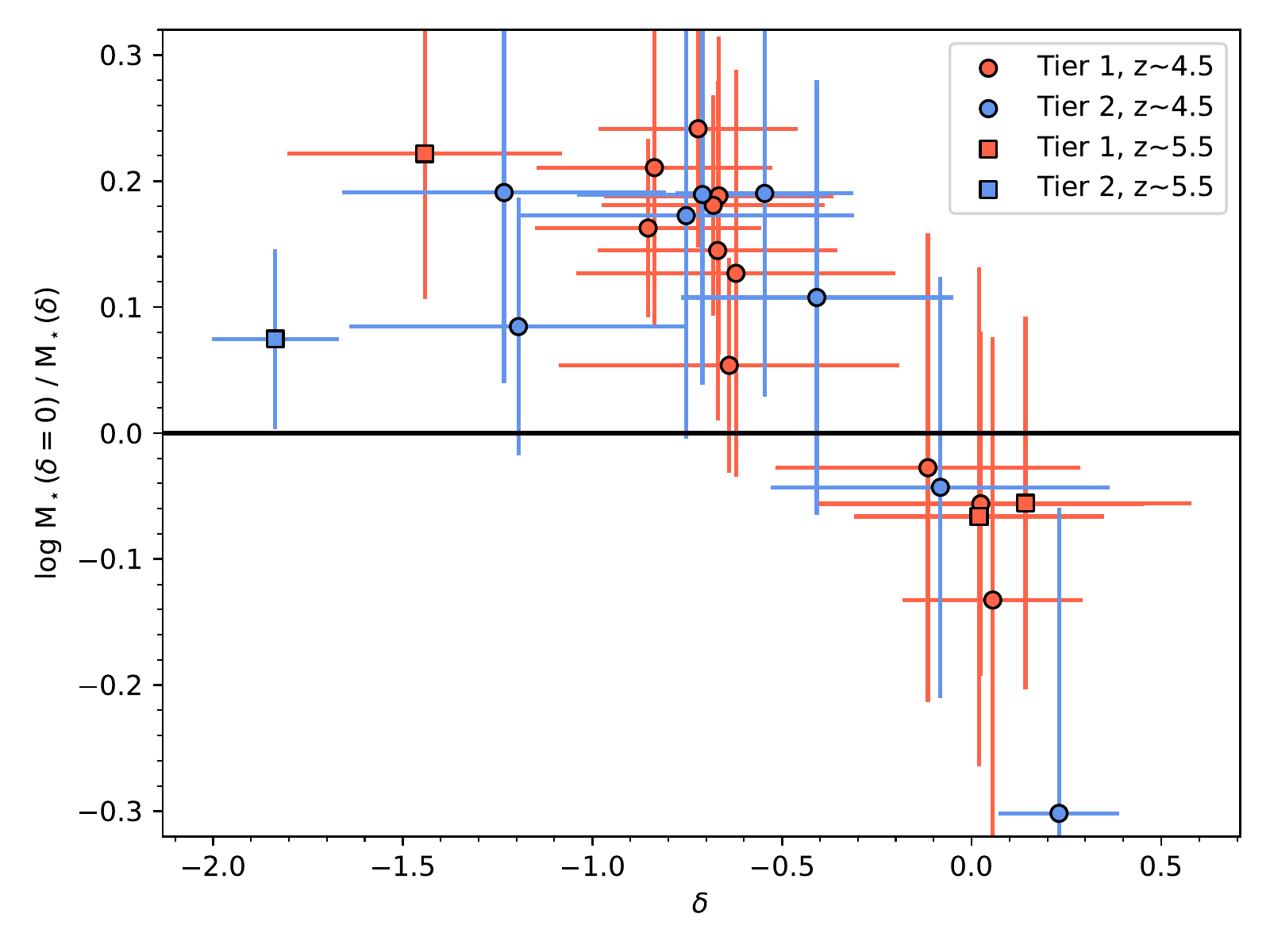}
 \caption{Ratio of the estimated SFR (left) and M$_\star$ (right) when setting $\delta=0$ and leaving $\delta$ free. The symbols are identical to those of Fig.~\ref{fig:mocks}.\label{fig:diff-SFR-Mstar}}
\end{figure*}

Finally, a fair question is the magnitude of the impact when adopting a fixed attenuation curve for estimating the SFR and M$_\star$ from SED modeling. To evaluate this, we have fitted again the sample but this time setting $\delta=0$. We find that the difference and the corresponding dispersion fairly small: $\Delta \log\mathrm{SFR / (M_{\odot} yr^{-1})}=0.00\pm0.06$ and $\Delta \log \mathrm{M_\star / M_\odot}=0.08 \pm 0.14$ (Fig.~\ref{fig:diff-SFR-Mstar}).
The reason the difference in the estimates is minute for the SFR is probably three-fold. First, we do not consider the full range of possible variations of the recent SFH as the data at our disposal are not sufficient to constrain such variations. For instance the shortest timescales would ideally be probed with an hydrogen recombination line. At the same time, the constraint from the [C~\textsc{ii}] emission through the SFR limits deviations from the input value. However, there is also probably a more physical reason to this. Most of the energy emitted by the dust is absorbed in the UV. This is especially the case for actively star-forming galaxies at high redshift for which the influence of dust heating by evolved stellar population is naturally more limited than at lower redshift. It comes that the key property to retrieve the SFR is the amplitude of the attenuation curve in the UV rather than its exact shape. For M$_\star$, the difference in the estimates follows a clear trend with $\delta$. As energy balance is a critical factor to set the amplitude of the UV attenuation, the shape of the curve will gain importance to determine other physical properties such as M$_\star$. Having a fixed attenuation curve means that the optical-to-NIR attenuation will be strongly determined by the UV attenuation, which even at these high redshifts is not appropriate.

Ultimately, the strong dependence of $\delta$ on the attenuation has a dramatic and complex effect on the emerging emission from galaxies. We therefore strongly recommend to use utmost caution when using $\beta$ to correct for dust attenuation at $z \sim 5$, as there are considerable uncertainties on the shape of attenuation curves.

This being said, as we have seen in Fig.~\ref{fig:diff-SFR-Mstar}, the impact of using a fixed attenuation curve remains fairly limited when using SED modeling. No clear effect is seen on the SFR, most likely because the determination of AFUV, which is the critical factor for this, is less sensitive than the exact shape of the attenuation curve than $\mathrm{A_V}$. In effect, the choice of the attenuation curve is more impactful for measuring the stellar mass, with a systematic trend. The mass offset remains limited to 0.3 dex, however. In conclusion, even though the physical cause of these variations requires further studies (for instance spatially resolved observations with ALMA and JWST will allow to explore whether the large scale geometries of these objects could explain such variations in their attenuation curves), the conclusion of our study is that practical effects remain sufficiently contained that using a single attenuation curve may be sufficient for many studies as long as SED modeling can be employed.

\section{Summary and conclusion\label{sec:conclusion}}

In this work we have studied 23 main-sequence $z=4.4$--$5.5$ galaxies from the ALPINE survey to measure the dust attenuation properties in general and the shape of the attenuation curves in particular. To do so, we have used the CIGALE code to model the SED in at least 6 rest-frame UV-to-optical bands in addition to the dust luminosity or the [C~\textsc{ii}] luminosity. These models are based on a delayed+burst SFH, the stellar populations of \cite{charlot2000a}, and a power law-modified starburst attenuation curve allowing a broad range of shapes. We find the following.

\begin{itemize}
  \item The relations between SFR, M$_\star$, IRX, and $\beta$ (see Figs. \ref{fig:MS}, \ref{fig:irx-beta}, and \ref{fig:irx-Mstar}) are broadly consistent with the findings of \cite{fudamoto2020a}, who used the LePhare code but without including constraints from the ALMA observations.
  \item These relations suggest, however, that there exists a variety of attenuation curves in main-sequence galaxies at $z \sim 5$.
  \item Estimating the index $\delta$ of the power law modifying the starburst attenuation curve, we find it ranges from $\delta=-1.84$ to $\delta=0.23$ (Table \ref{tab:properties}). For reference, $\delta=0$ corresponds to a bona fide starburst curve and $\delta=-0.5$ to an SMC-like extinction curve.
  \item The index $\delta$ is strongly dependent on the V-band attenuation, with steeper curves corresponding to lower attenuation (Fig.~\ref{fig:delta-AV}).
  \item It is likely that the steeper curves are the product of a combination of intrinsically steep extinction curves, the effect of the dust distribution, and differential reddening between star-forming regions and the ISM (see Figs.~\ref{fig:seon} and \ref{fig:delta-AV-mu}).
  \item Ultimately, there is no single attenuation curve that is appropriate for all high-redshift main-sequence galaxies. Caution is strongly advised when interpreting the UV color in terms of attenuation. However, when SED modeling is available, the impact of the choice of the attenuation curve on the SFR and M$_\star$ is limited (Figs.~\ref{fig:beta-AV} and \ref{fig:diff-SFR-Mstar}).
\end{itemize}

Over the next few years, the combination of high resolution ALMA and JWST data will be decisive to better understand the physical processes that give rise to steep attenuation curves in galaxies.

\begin{acknowledgements}
We would like to thank the referee for their very insightful comments that we believe have helped clarify and improve our article.

Médéric Boquien gratefully acknowledges support by the ANID BASAL project FB210003 and from the FONDECYT regular grants 1170618 and 1211000. DN acknowledges funding from the Space Telescope Science Institute via grant AR-15043.001. G.C.J. acknowledges ERC Advanced Grant 695671 ``QUENCH'' and support by the Science and Technology Facilities Council (STFC).

This research made use of Astropy,\footnote{\url{http://www.astropy.org}} a community-developed core Python package for Astronomy \citep{astropy2013a, astropy2018a}.
\end{acknowledgements}

\bibliographystyle{aa}
\bibliography{article}

\appendix

\section{Best-fits}

\begin{figure*}[!htbp]
  \centering
  \includegraphics[width=0.33\textwidth]{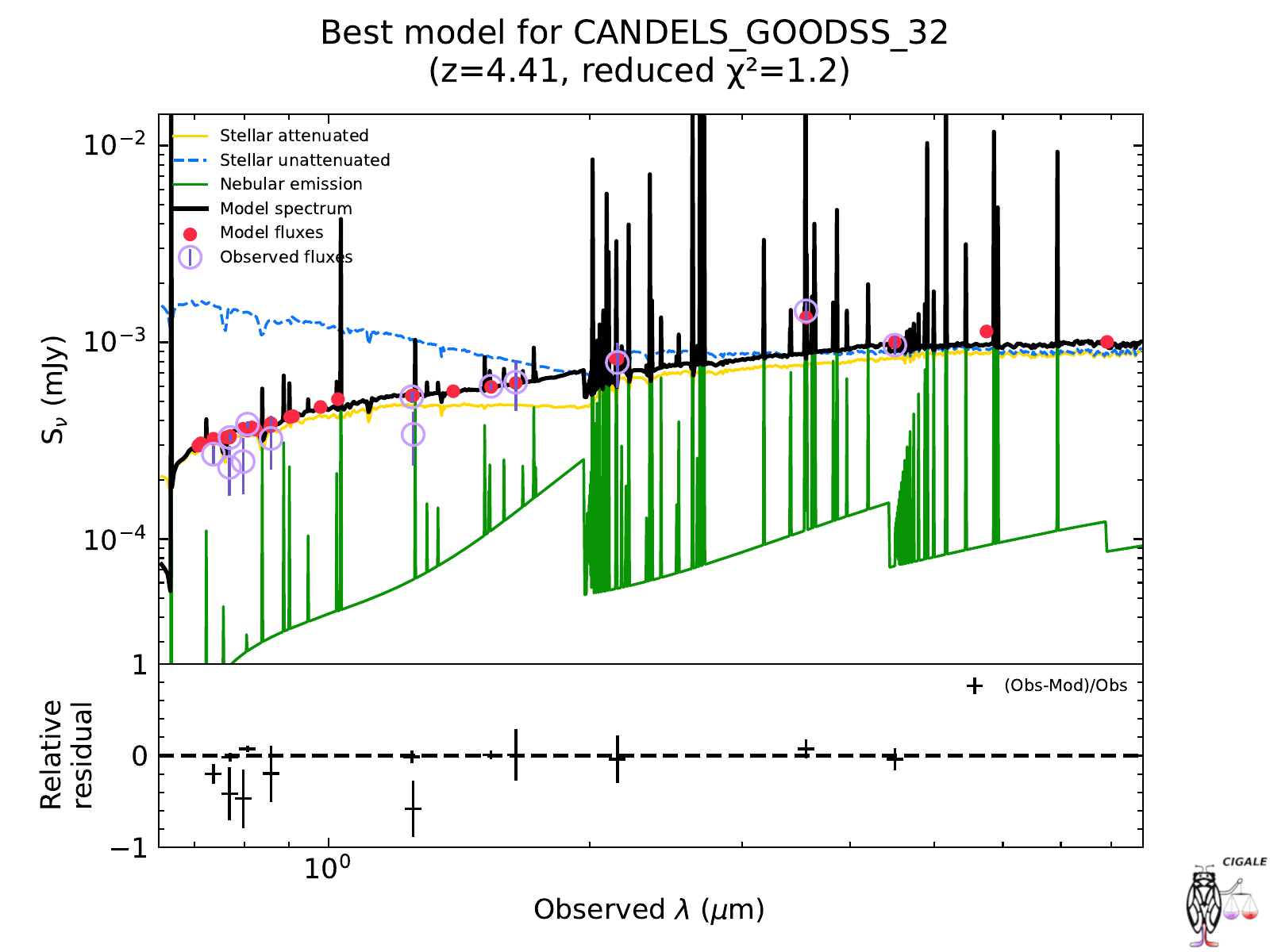}
  \includegraphics[width=0.33\textwidth]{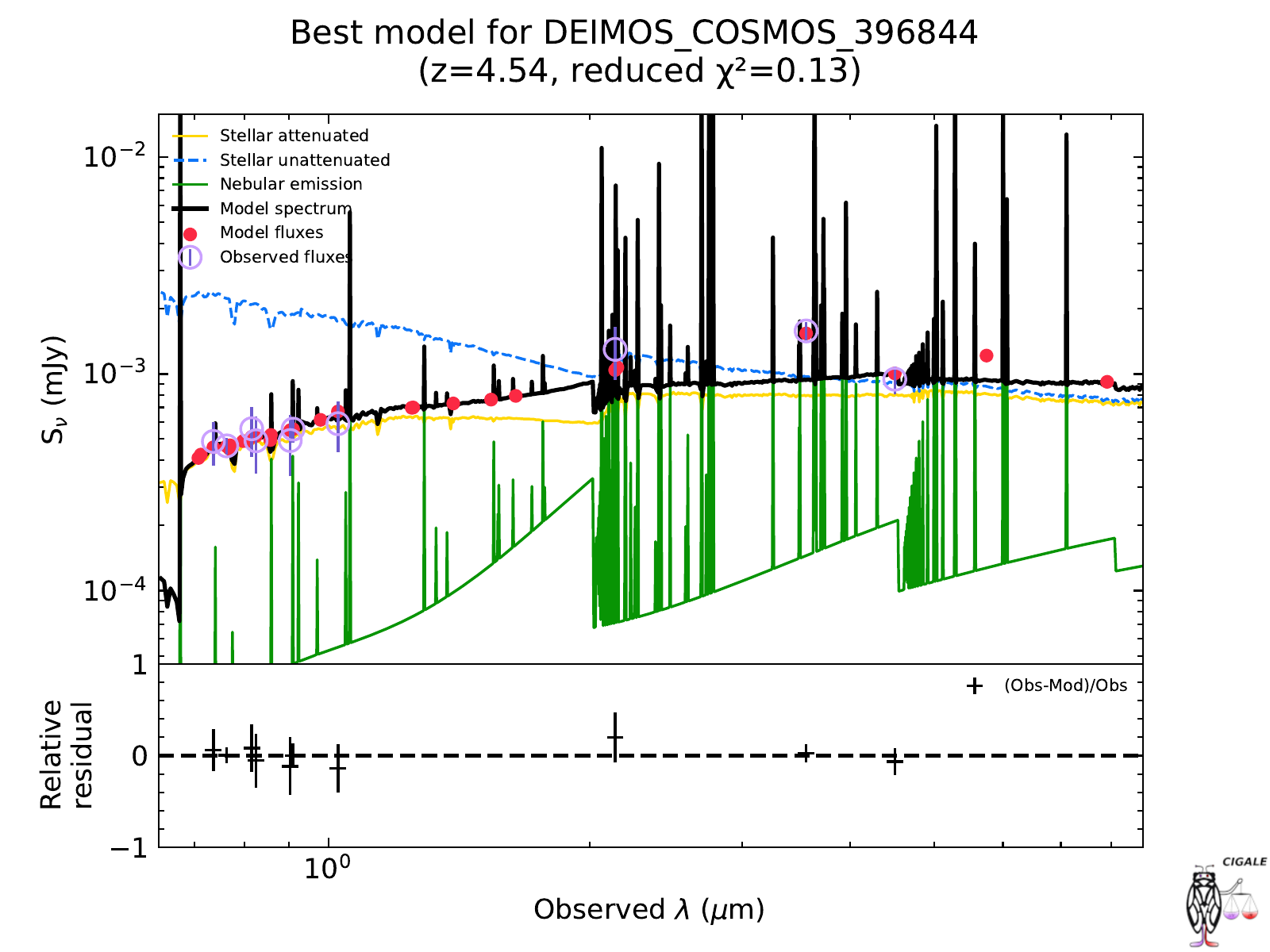}
  \includegraphics[width=0.33\textwidth]{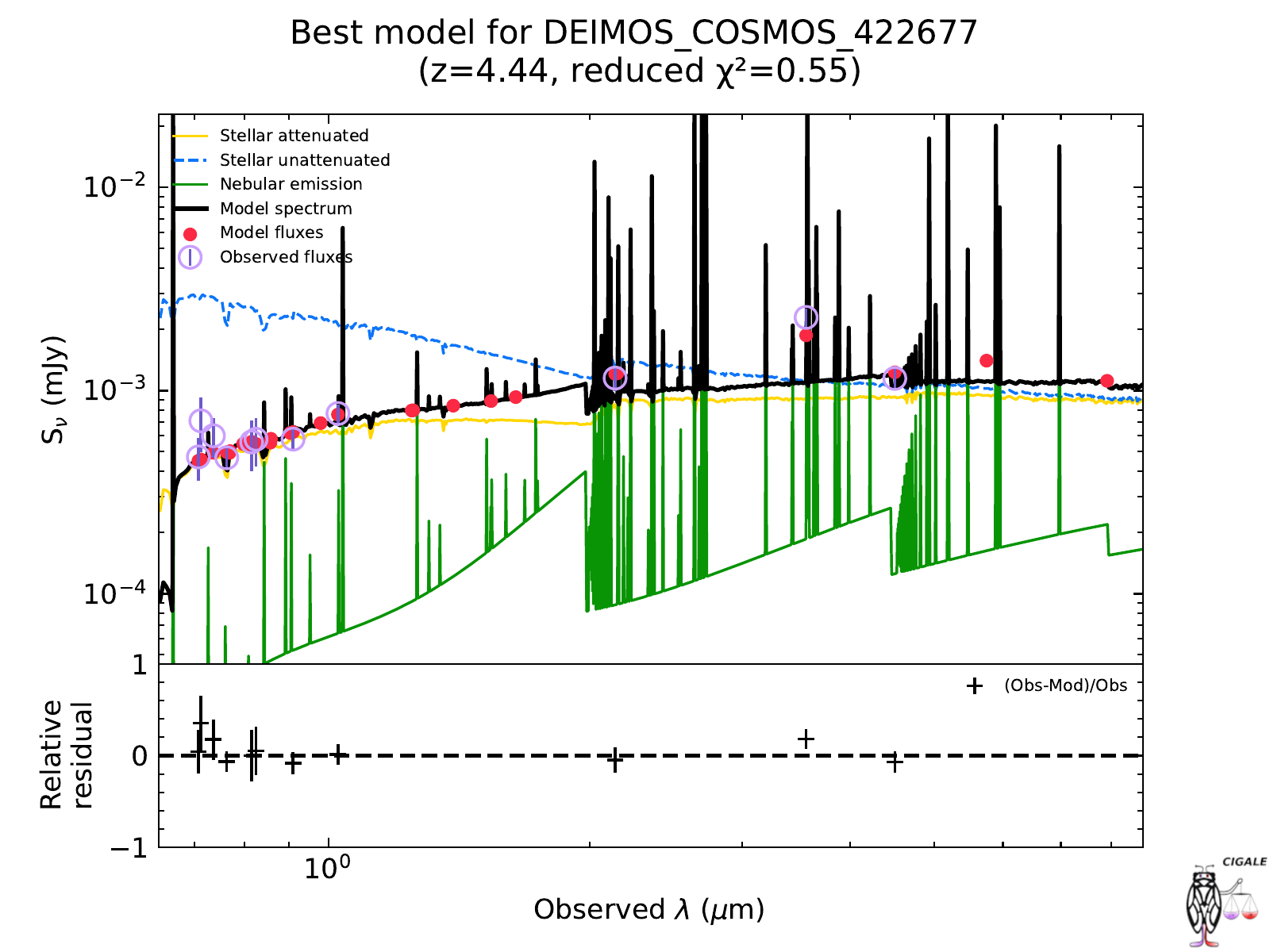}
  \includegraphics[width=0.33\textwidth]{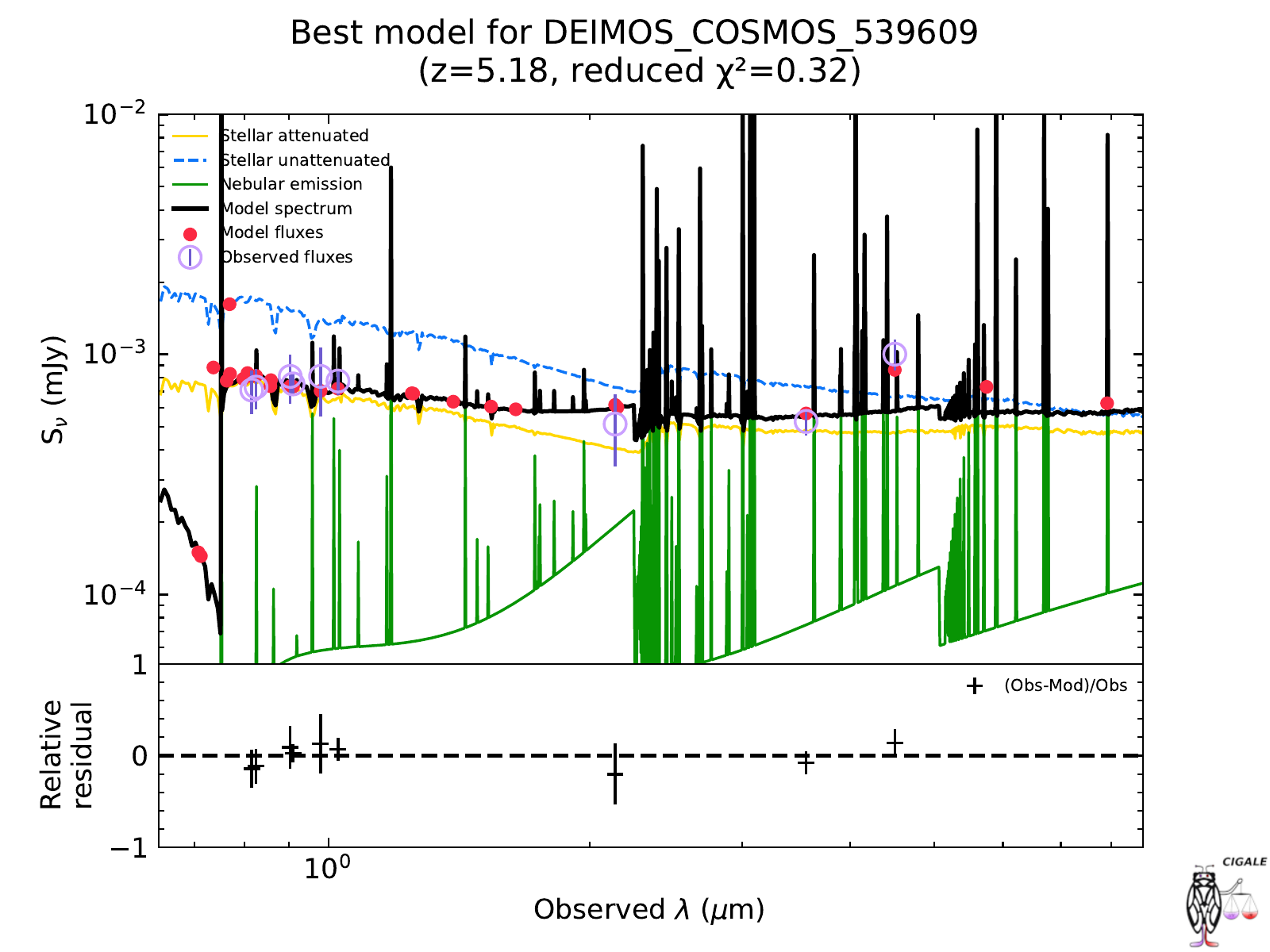}
  \includegraphics[width=0.33\textwidth]{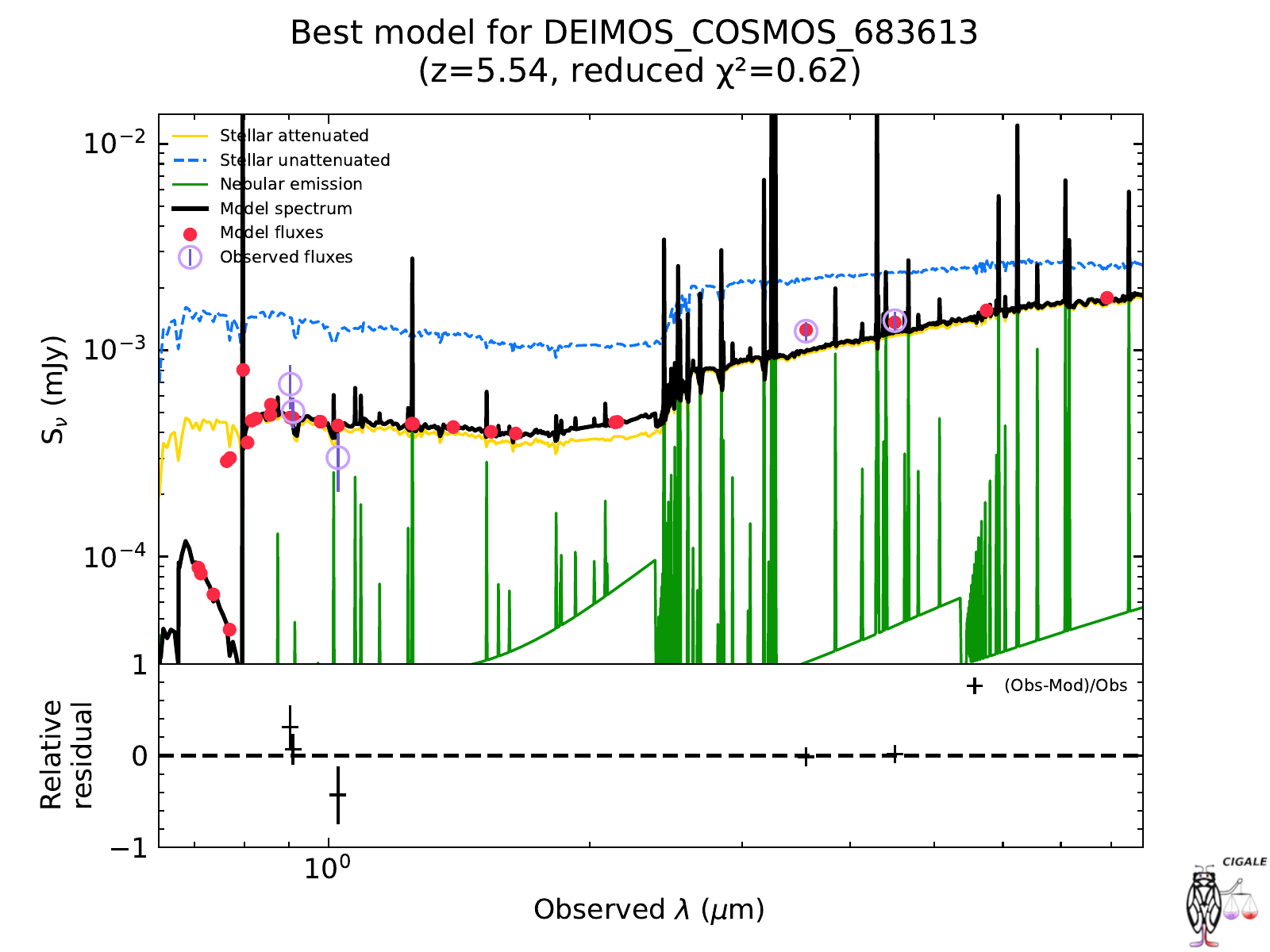}
  \includegraphics[width=0.33\textwidth]{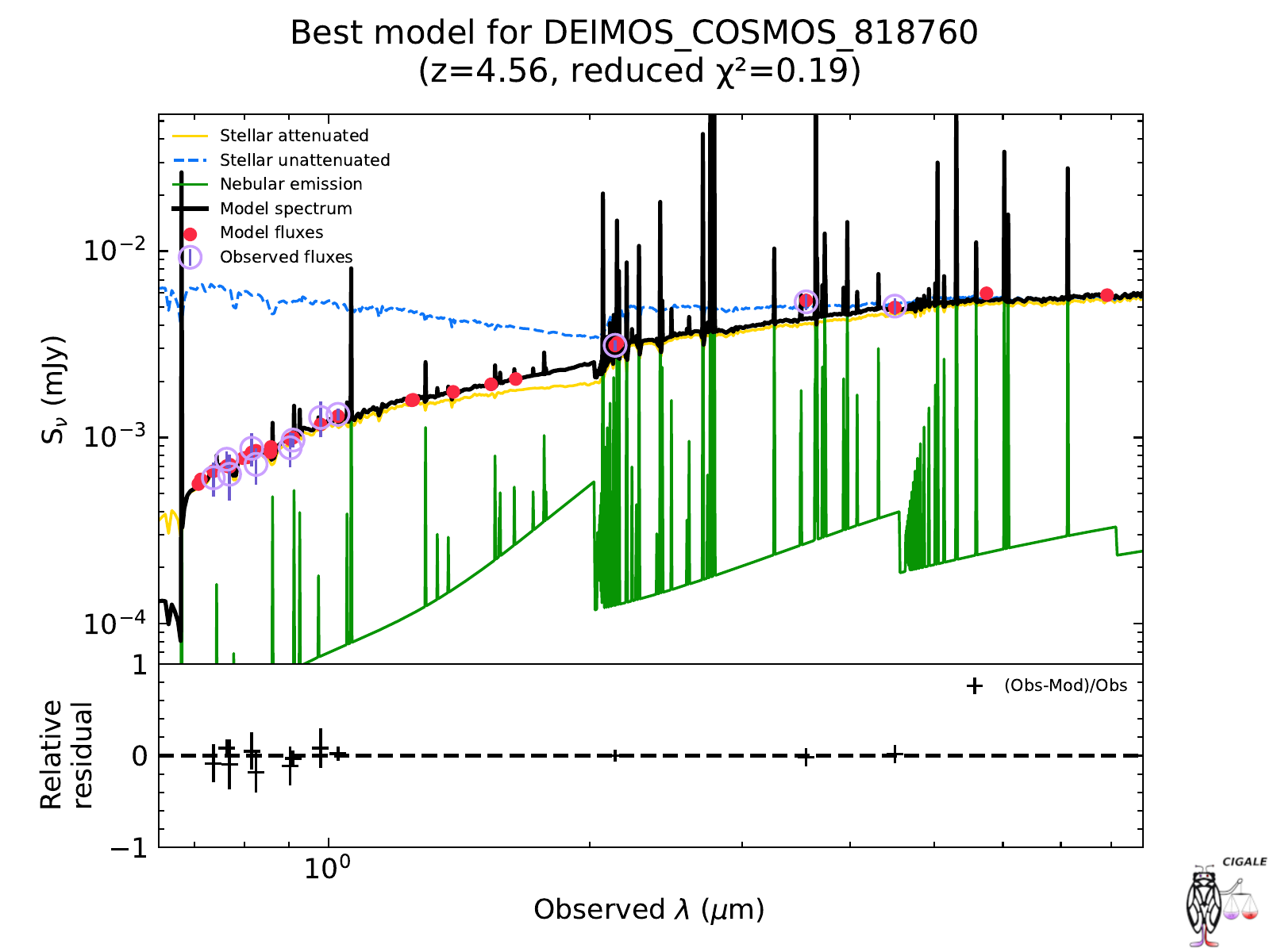}
  \includegraphics[width=0.33\textwidth]{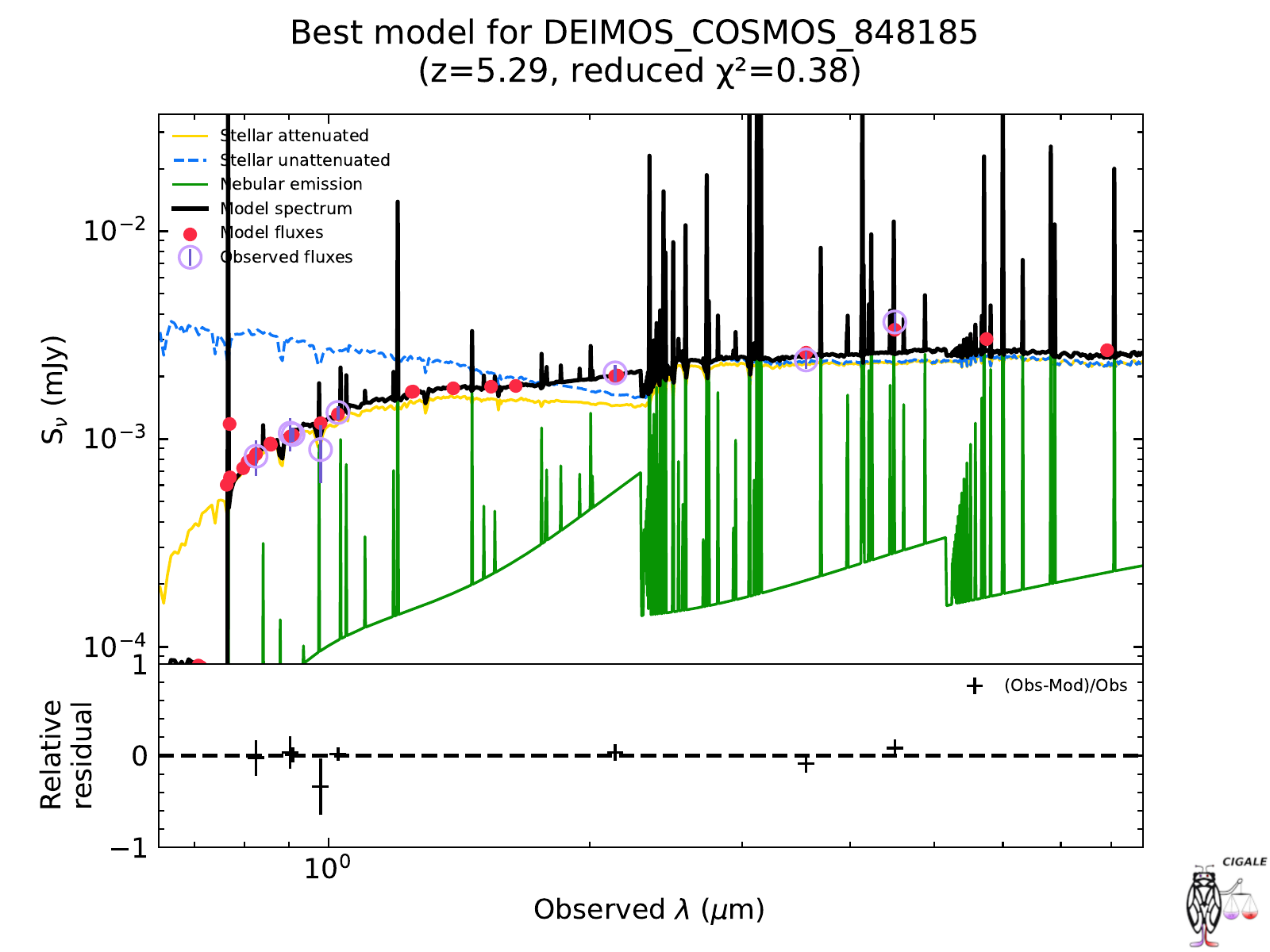}
  \includegraphics[width=0.33\textwidth]{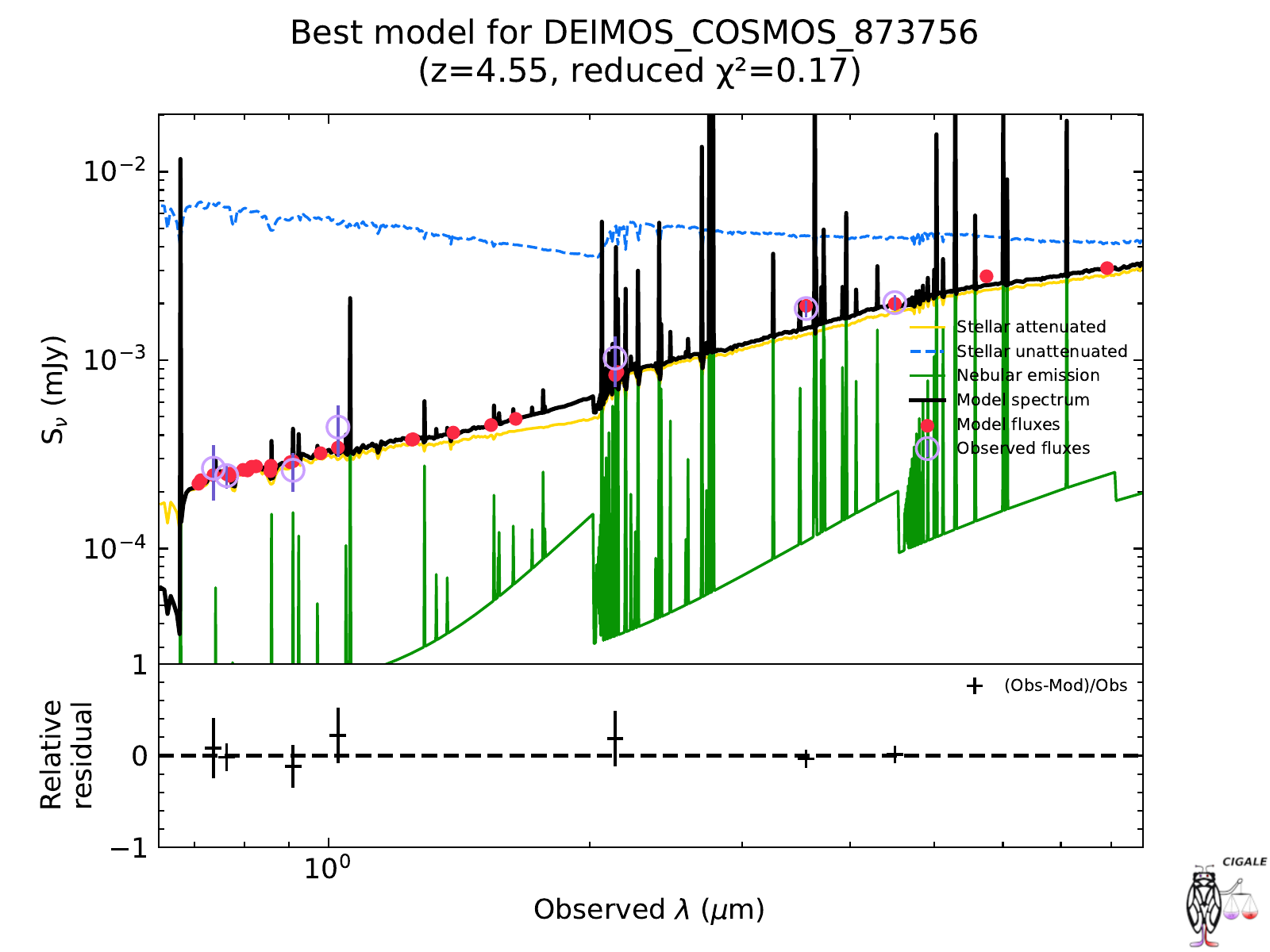}
  \includegraphics[width=0.33\textwidth]{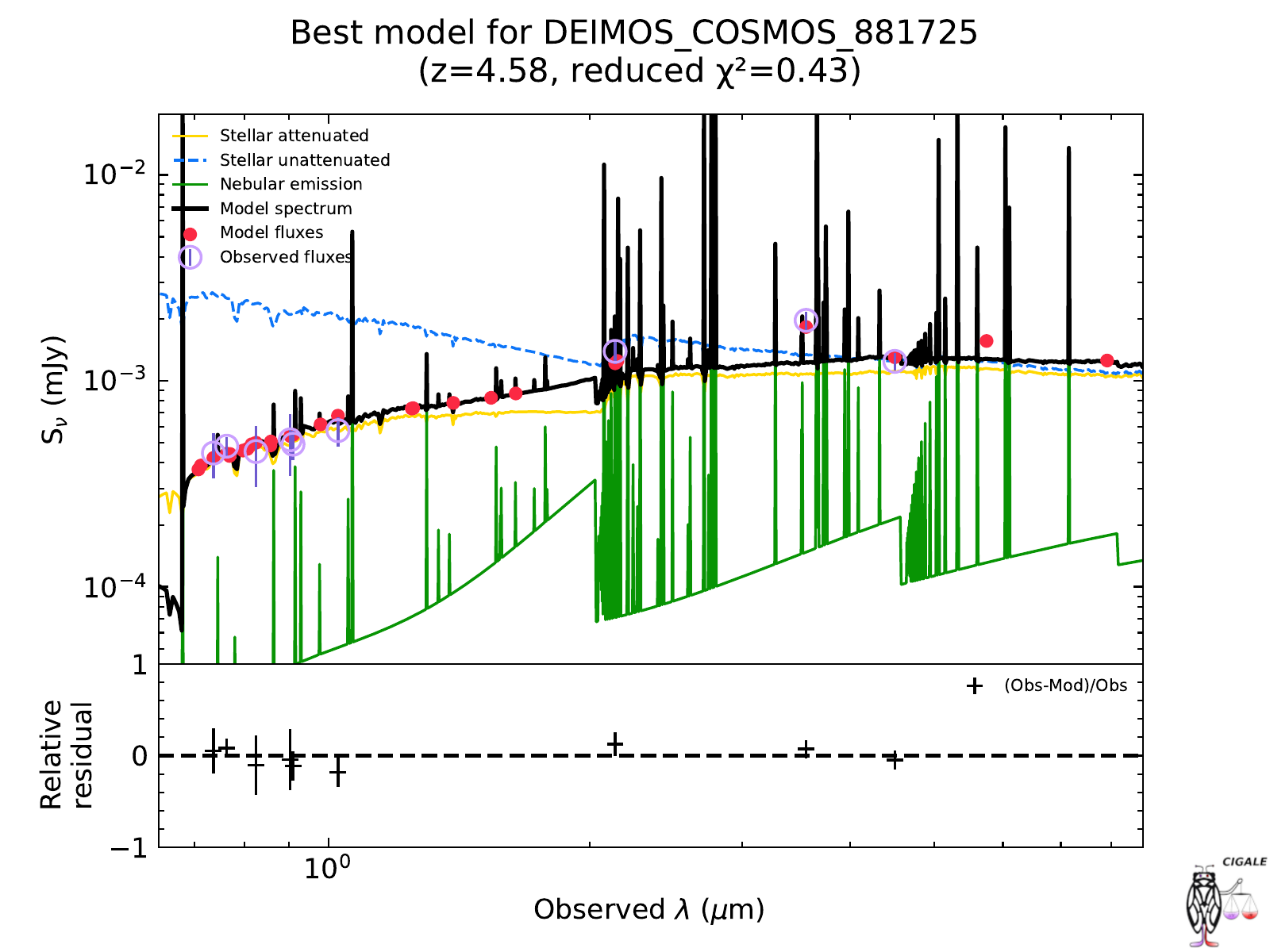}
  \includegraphics[width=0.33\textwidth]{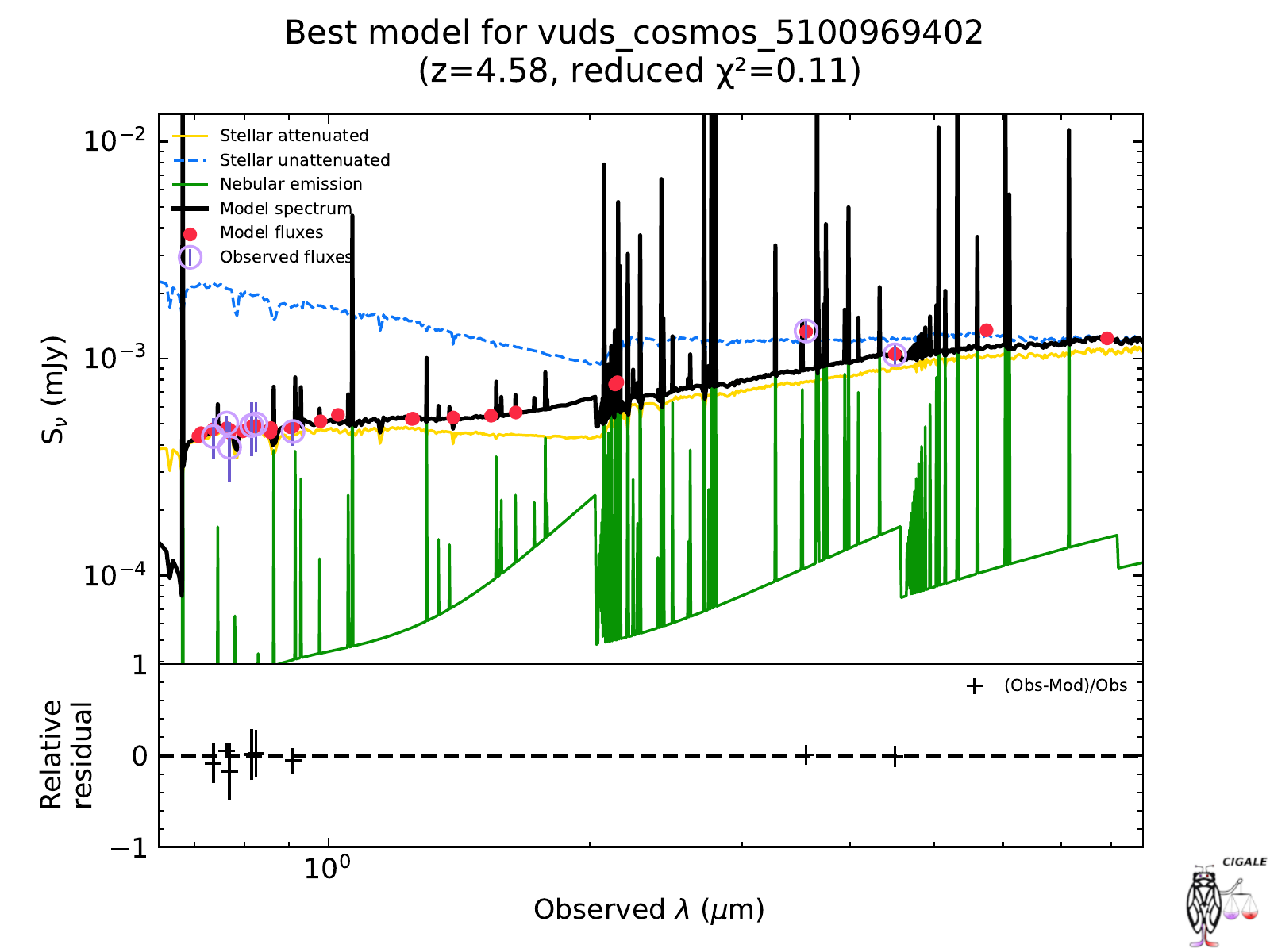}
  \includegraphics[width=0.33\textwidth]{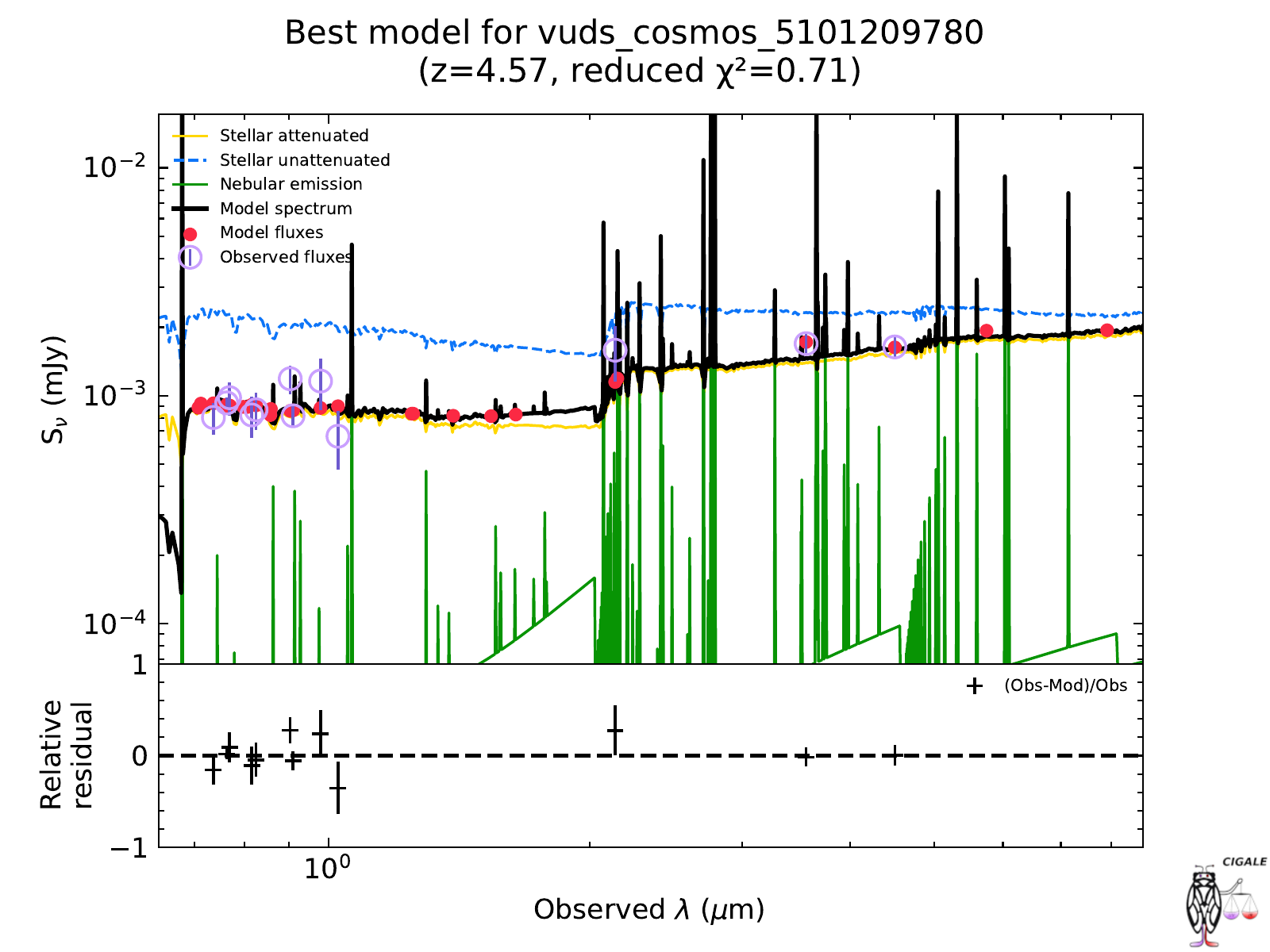}
  \includegraphics[width=0.33\textwidth]{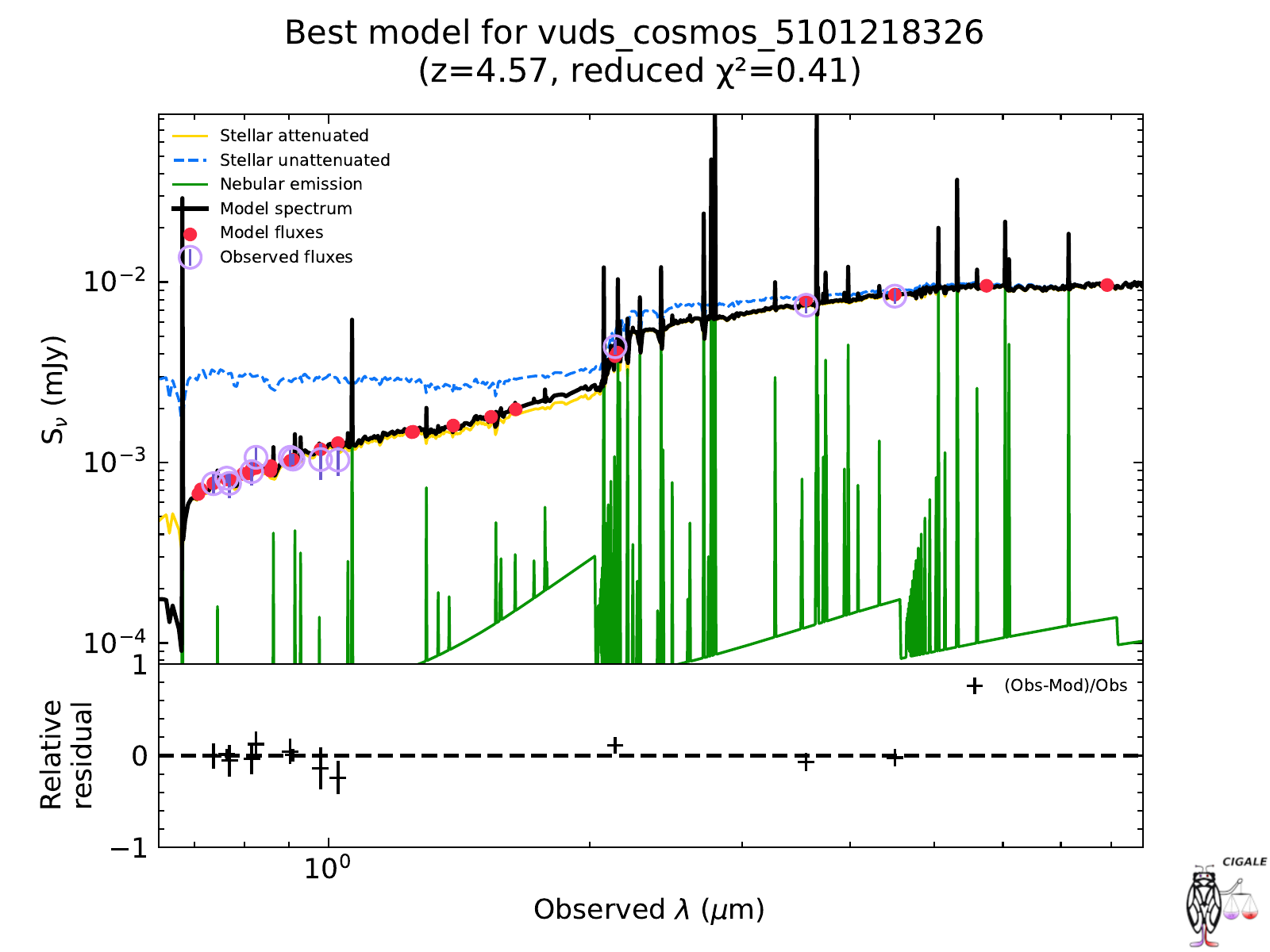}
  \caption{Best-fits for tier 1 galaxies (except for vuds\_cosmos\_5180966608 and vuds\_efdcs\_530029038, which are shown in Fig.~\ref{fig:best-fit}). The galaxy spectrum is shown in black, with the dust-attenuated stellar populations in yellow and the ionized gas in green. For comparison, the dust-free stellar populations are shown in blue. The model fluxes integrated in passbands are shown as red circles and the observations and 1-$\sigma$ uncertainties are indicated with purple circles with vertical lines. Finally the relative residuals are shown in the bottom panel of each plot.\label{fig:best-fits-1}}
\end{figure*}

\begin{figure*}[!htbp]
  \centering
  \includegraphics[width=0.33\textwidth]{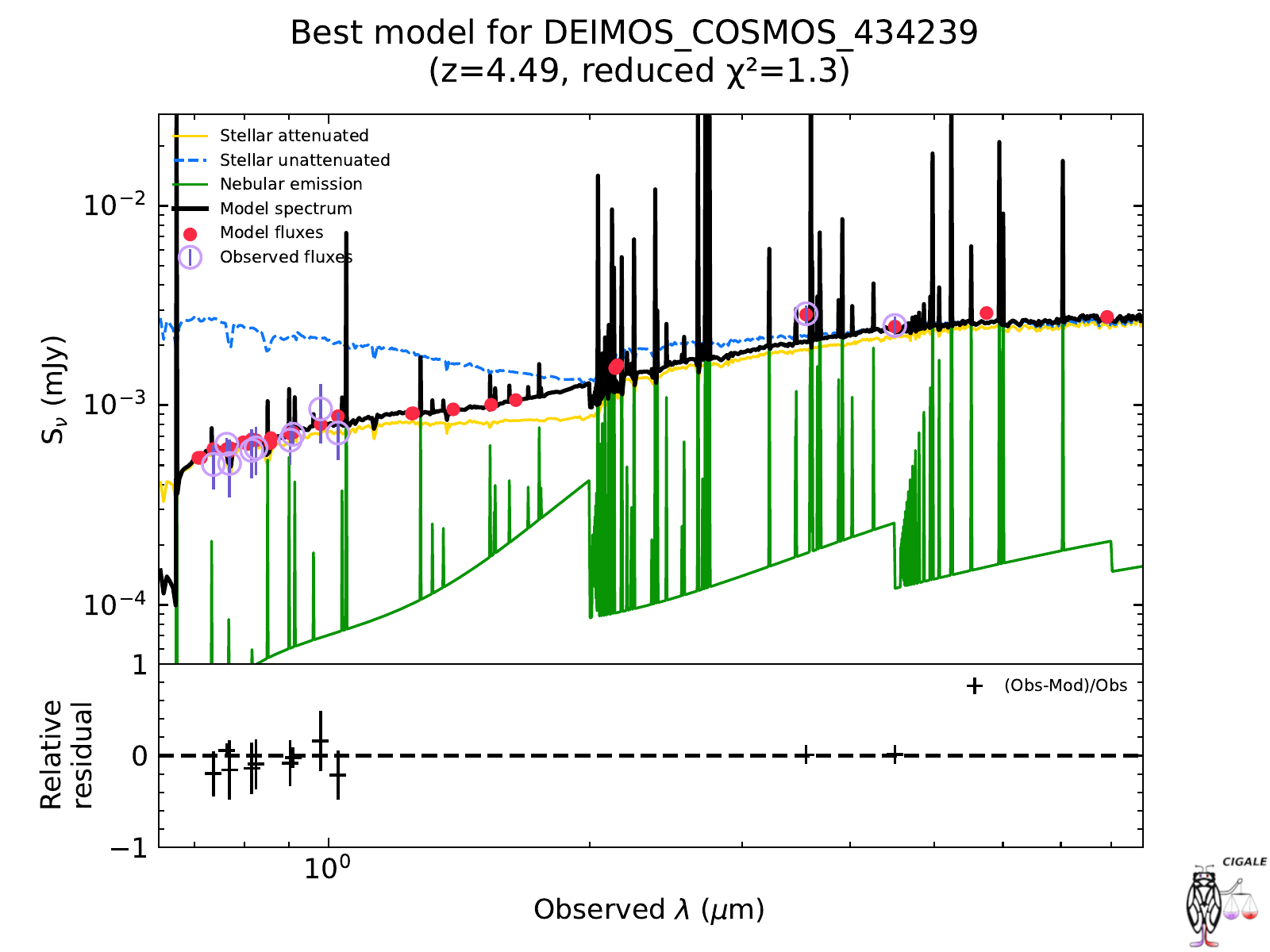}
  \includegraphics[width=0.33\textwidth]{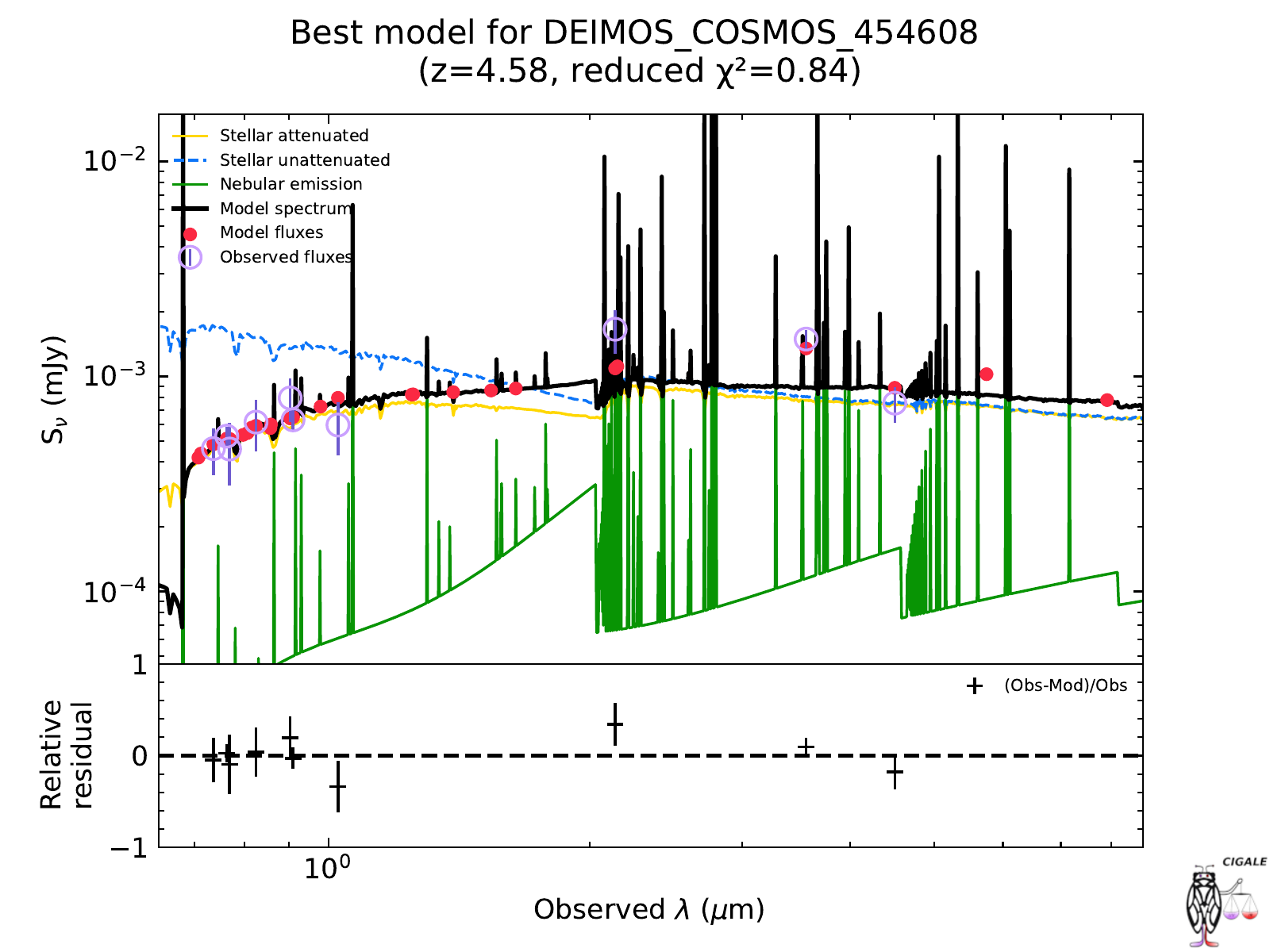}
  \includegraphics[width=0.33\textwidth]{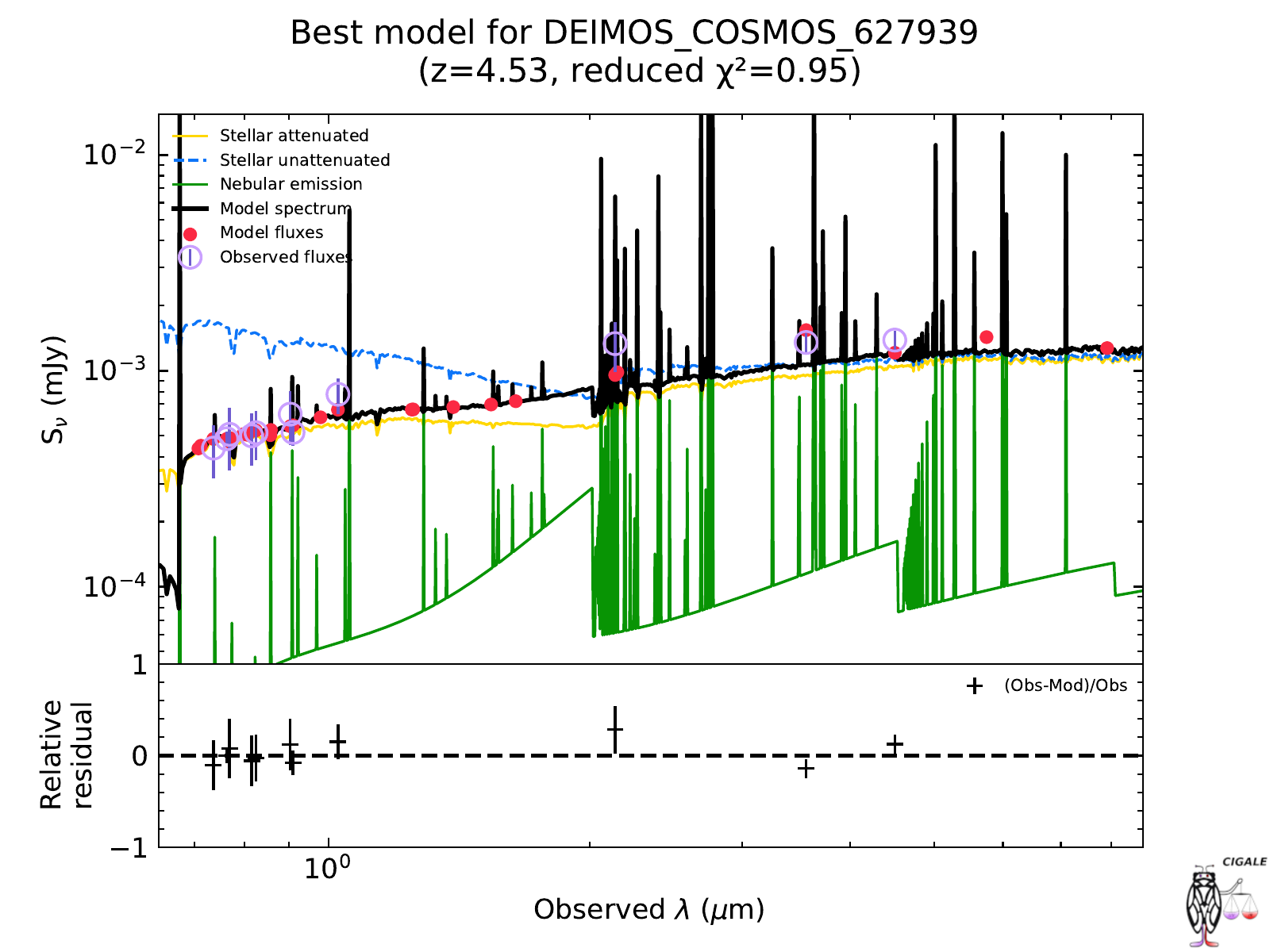}
  \includegraphics[width=0.33\textwidth]{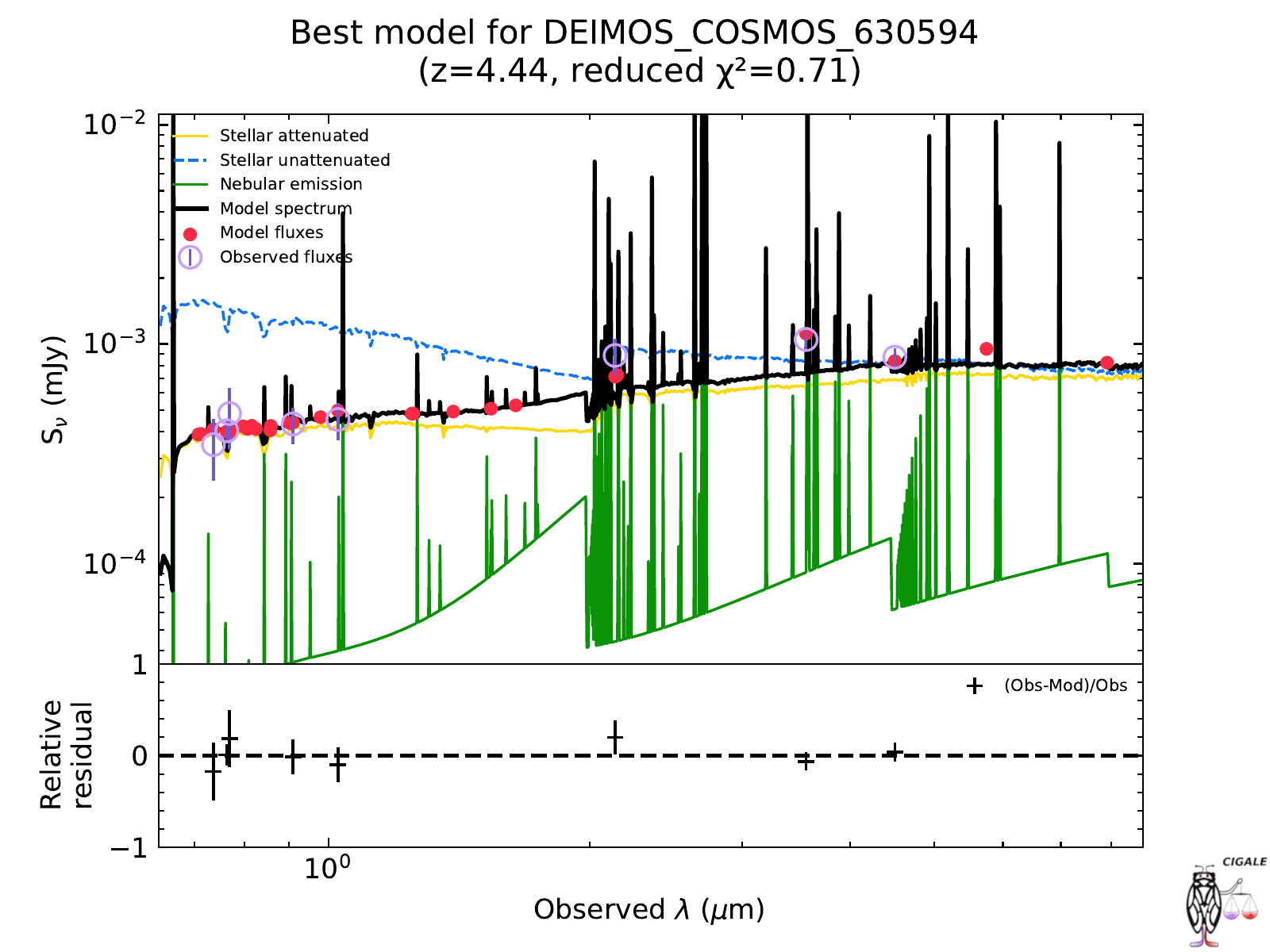}
  \includegraphics[width=0.33\textwidth]{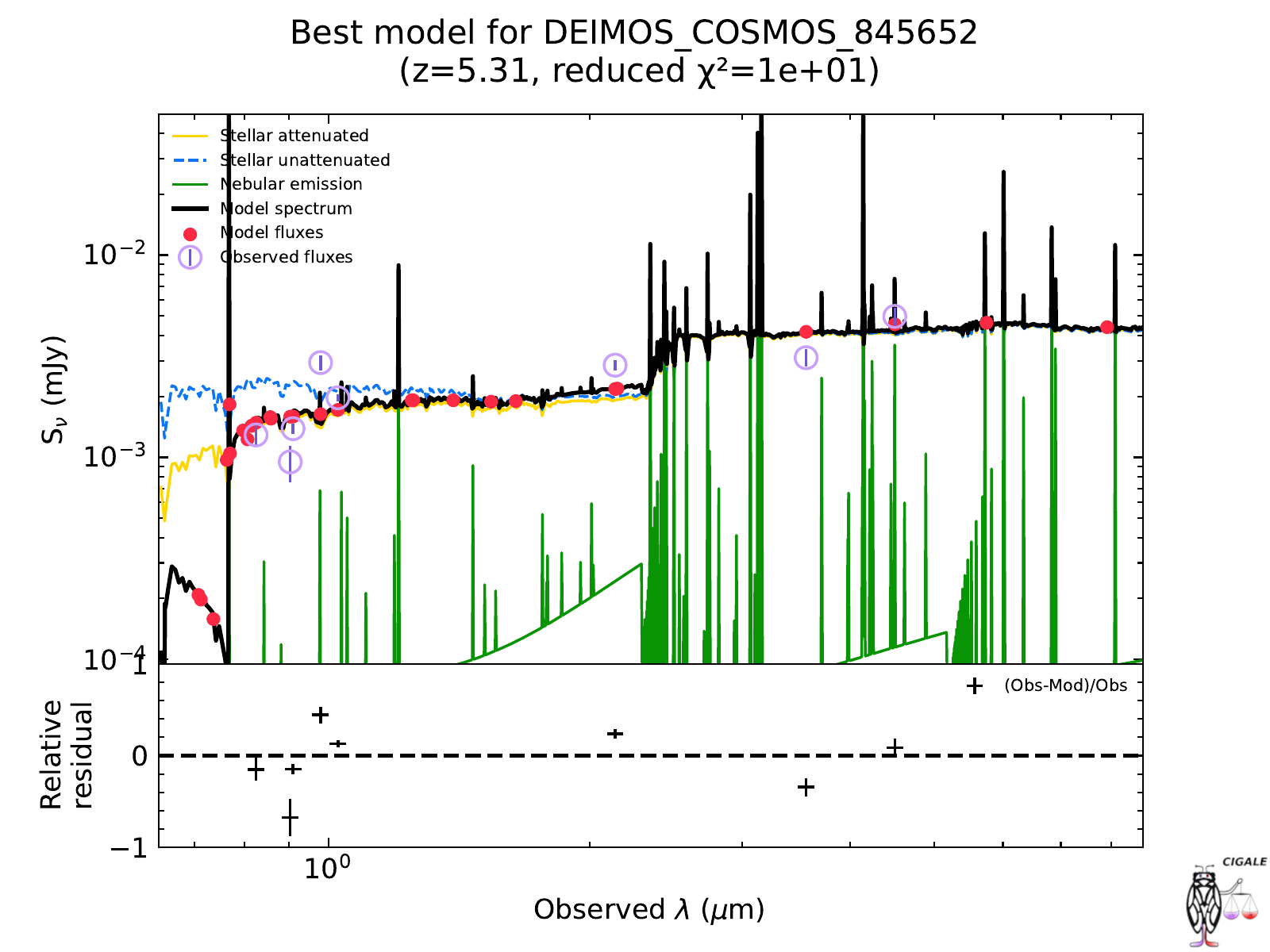}
  \includegraphics[width=0.33\textwidth]{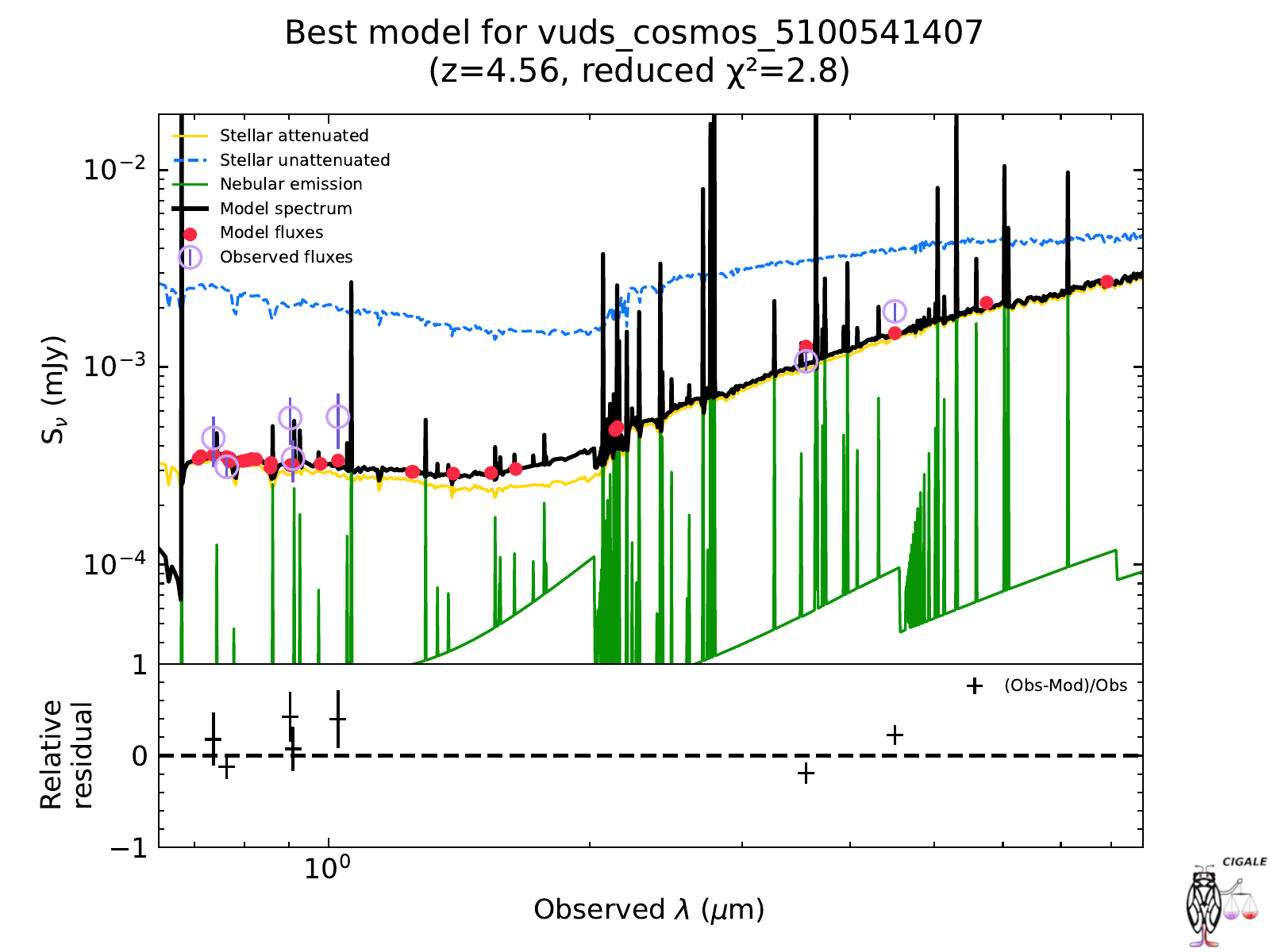}
  \includegraphics[width=0.33\textwidth]{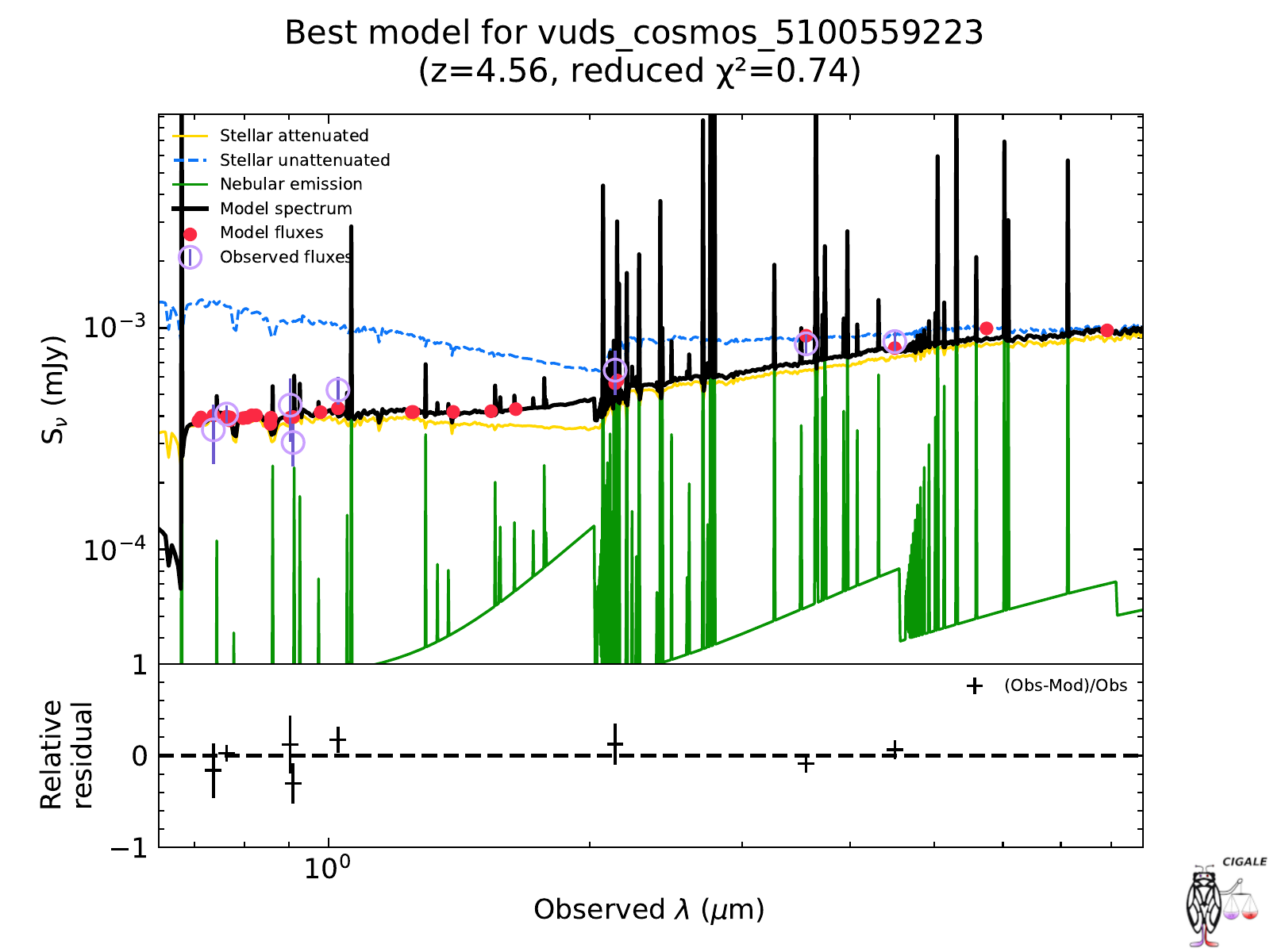}
  \includegraphics[width=0.33\textwidth]{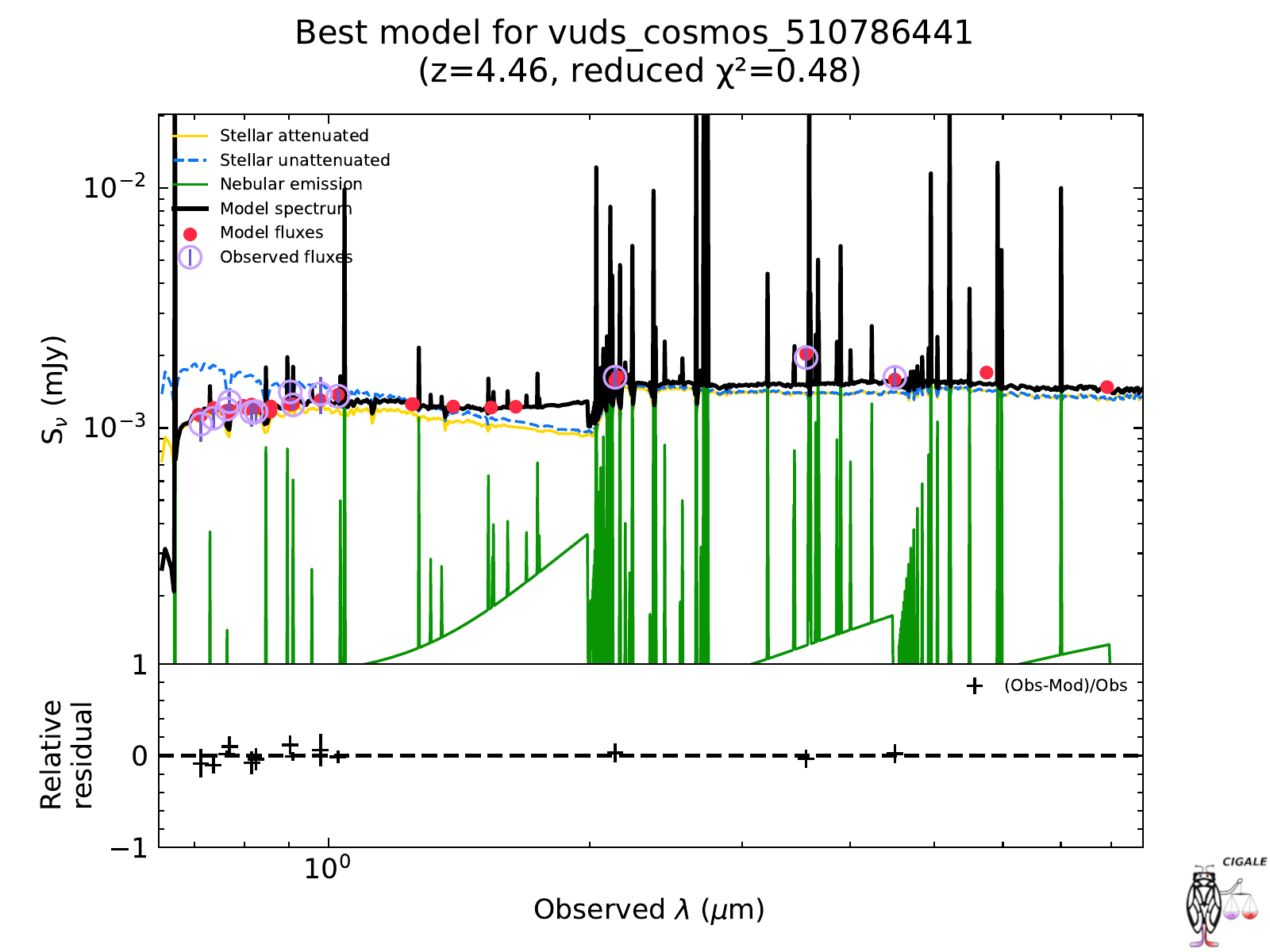}
  \includegraphics[width=0.33\textwidth]{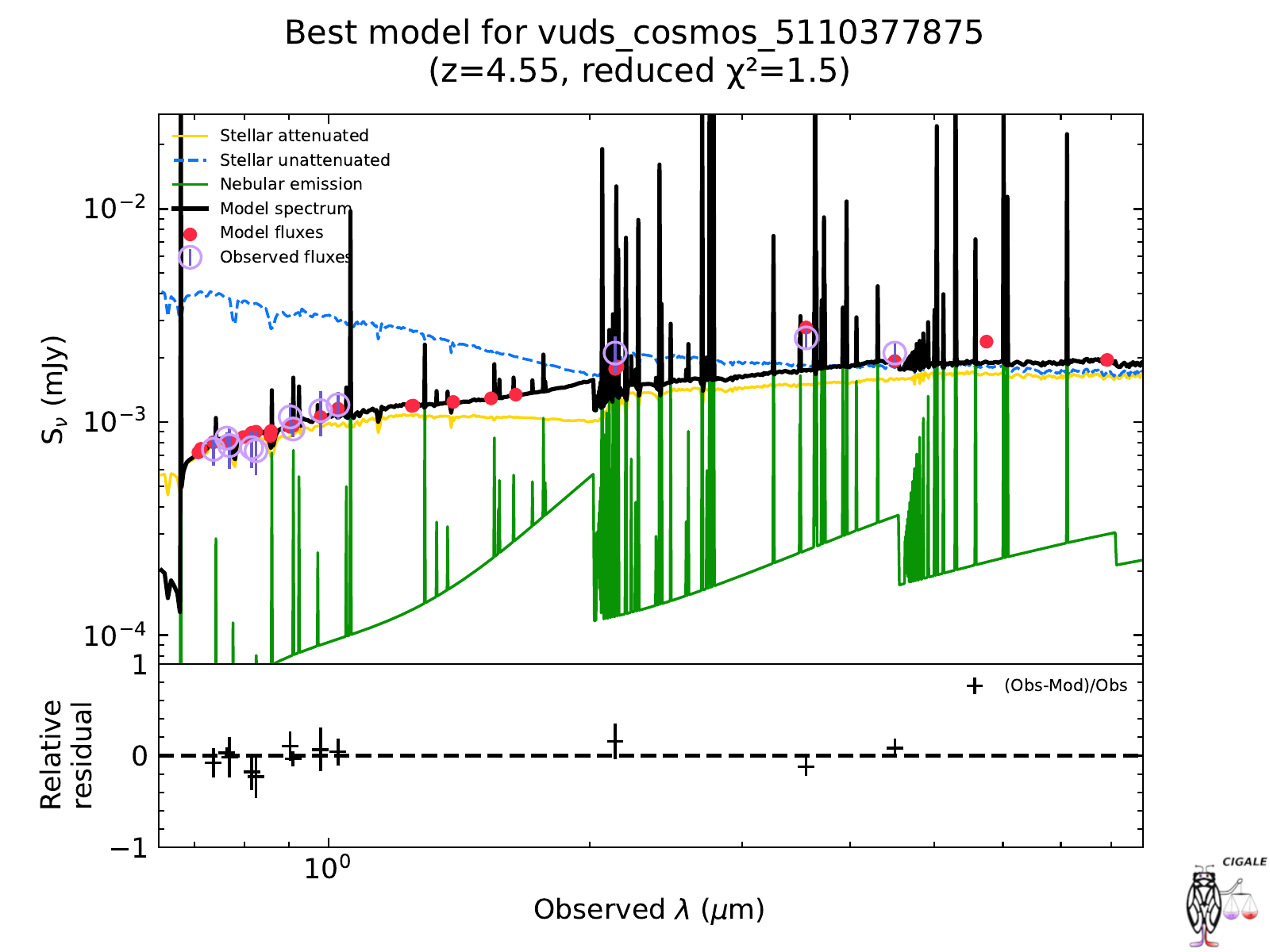}
  \caption{Same as Fig.~\ref{fig:best-fits-1} but for tier 2 galaxies.\label{fig:best-fits-2}}
\end{figure*}

We show in Fig.~\ref{fig:best-fits-1} and \ref{fig:best-fits-2} the best fits for each for the tier 1 and tier 2 galaxies in our final sample.

\clearpage
\section{Physical properties}

We present in Table \ref{tab:properties} the main physical properties of our sample.
\begin{sidewaystable*}
 \centering
 \begin{tabular}{lccccccccc}
  \hline\hline
  Name&Tier&$z$&IRX&$\beta$&$\delta$&AFUV&$\mathrm{A_V}$&SFR&M$_\star$\\
      &    &   &   &       &        &mag &mag&M$_\odot$ yr$^{-1}$&M$_\odot$\\\hline
CANDELS\_GOODSS\_32&1&4.41     &$ 0.57\pm 0.14$&$-1.20\pm 0.08$&$-0.84\pm 0.31$&$ 1.46\pm 0.24$&$ 0.24\pm 0.10$&$50.86\pm15.89$&$7.52\times10^9\pm1.83\times10^9$\\
DEIMOS\_COSMOS\_396844&1&4.54  &$ 0.57\pm 0.14$&$-1.44\pm 0.24$&$-0.62\pm 0.42$&$ 1.42\pm 0.23$&$ 0.32\pm 0.18$&$76.88\pm18.75$&$7.84\times10^9\pm2.27\times10^9$\\
DEIMOS\_COSMOS\_422677&1&4.44  &$ 0.63\pm 0.11$&$-1.29\pm 0.20$&$-0.67\pm 0.32$&$ 1.54\pm 0.20$&$ 0.31\pm 0.13$&$91.97\pm18.88$&$8.43\times10^9\pm1.92\times10^9$\\
DEIMOS\_COSMOS\_539609&1&5.18  &$ 0.07\pm 0.22$&$-2.42\pm 0.11$&$ 0.14\pm 0.44$&$ 0.67\pm 0.21$&$ 0.39\pm 0.20$&$66.52\pm13.38$&$5.41\times10^9\pm1.46\times10^9$\\
DEIMOS\_COSMOS\_683613&1&5.54  &$ 0.51\pm 0.17$&$-1.90\pm 0.22$&$ 0.02\pm 0.33$&$ 1.28\pm 0.26$&$ 0.58\pm 0.23$&$54.64\pm14.19$&$1.95\times10^{10}\pm7.70\times10^9$\\
DEIMOS\_COSMOS\_818760&1&4.56  &$ 0.81\pm 0.14$&$-0.74\pm 0.17$&$-0.72\pm 0.26$&$ 1.94\pm 0.26$&$ 0.35\pm 0.13$&$159.53\pm53.88$&$5.07\times10^{10}\pm9.63\times10^9$\\
DEIMOS\_COSMOS\_848185&1&5.29  &$ 0.32\pm 0.09$&$-1.20\pm 0.17$&$-1.44\pm 0.36$&$ 1.05\pm 0.14$&$ 0.08\pm 0.05$&$120.35\pm30.07$&$2.42\times10^{10}\pm5.32\times10^9$\\
DEIMOS\_COSMOS\_873756&1&4.55  &$ 1.36\pm 0.22$&$-1.31\pm 0.31$&$ 0.06\pm 0.24$&$ 3.03\pm 0.49$&$ 1.40\pm 0.47$&$142.62\pm66.46$&$4.26\times10^{10}\pm1.88\times10^{10}$\\
DEIMOS\_COSMOS\_881725&1&4.58  &$ 0.66\pm 0.13$&$-1.20\pm 0.22$&$-0.67\pm 0.30$&$ 1.60\pm 0.23$&$ 0.31\pm 0.13$&$87.12\pm23.58$&$1.04\times10^{10}\pm2.50\times10^9$\\
vuds\_cosmos\_5100969402&1&4.58&$ 0.54\pm 0.19$&$-1.85\pm 0.25$&$-0.12\pm 0.40$&$ 1.36\pm 0.30$&$ 0.55\pm 0.27$&$59.00\pm21.63$&$1.24\times10^{10}\pm4.58\times10^9$\\
vuds\_cosmos\_5101209780&1&4.57&$ 0.17\pm 0.26$&$-2.16\pm 0.15$&$ 0.02\pm 0.43$&$ 0.81\pm 0.28$&$ 0.40\pm 0.23$&$54.90\pm18.01$&$1.84\times10^{10}\pm4.73\times10^9$\\
vuds\_cosmos\_5101218326&1&4.57&$ 0.51\pm 0.14$&$-0.98\pm 0.18$&$-0.85\pm 0.30$&$ 1.34\pm 0.24$&$ 0.20\pm 0.09$&$82.33\pm20.71$&$1.22\times10^{11}\pm1.80\times10^{10}$\\
vuds\_cosmos\_5180966608&1&4.53&$ 0.76\pm 0.13$&$-0.80\pm 0.24$&$-0.68\pm 0.29$&$ 1.79\pm 0.26$&$ 0.33\pm 0.13$&$70.99\pm17.94$&$7.17\times10^{10}\pm1.32\times10^{10}$\\
vuds\_efdcs\_530029038&1&4.43  &$ 0.00\pm 0.16$&$-1.92\pm 0.07$&$-0.64\pm 0.45$&$ 0.62\pm 0.17$&$ 0.14\pm 0.09$&$32.50\pm 6.47$&$1.51\times10^{10}\pm2.35\times10^9$\\\hline
DEIMOS\_COSMOS\_434239&2&4.49  &$ 0.61\pm 0.08$&$-1.25\pm 0.26$&$-0.71\pm 0.33$&$ 1.49\pm 0.16$&$ 0.28\pm 0.12$&$103.13\pm15.75$&$2.74\times10^{10}\pm7.85\times10^9$\\
DEIMOS\_COSMOS\_454608&2&4.58  &$ 0.33\pm 0.08$&$-1.45\pm 0.20$&$-1.23\pm 0.43$&$ 1.02\pm 0.13$&$ 0.11\pm 0.07$&$62.19\pm 8.09$&$6.04\times10^9\pm1.55\times10^9$\\
DEIMOS\_COSMOS\_627939&2&4.53  &$ 0.48\pm 0.08$&$-1.46\pm 0.24$&$-0.75\pm 0.44$&$ 1.25\pm 0.13$&$ 0.24\pm 0.14$&$65.79\pm 5.93$&$1.03\times10^{10}\pm3.42\times10^9$\\
DEIMOS\_COSMOS\_630594&2&4.44  &$ 0.55\pm 0.09$&$-1.64\pm 0.25$&$-0.41\pm 0.36$&$ 1.37\pm 0.15$&$ 0.38\pm 0.17$&$57.27\pm 5.52$&$6.28\times10^9\pm2.09\times10^9$\\
DEIMOS\_COSMOS\_845652&2&5.31  &$-0.20\pm 0.10$&$-1.73\pm 0.10$&$-1.84\pm 0.17$&$ 0.43\pm 0.09$&$ 0.02\pm 0.01$&$72.76\pm 7.19$&$5.08\times10^{10}\pm7.70\times10^9$\\
vuds\_cosmos\_5100541407&2&4.56&$ 1.04\pm 0.09$&$-1.94\pm 0.26$&$ 0.23\pm 0.16$&$ 2.25\pm 0.19$&$ 1.25\pm 0.26$&$99.83\pm11.58$&$3.30\times10^{10}\pm1.56\times10^{10}$\\
vuds\_cosmos\_5100559223&2&4.56&$ 0.40\pm 0.18$&$-2.00\pm 0.22$&$-0.08\pm 0.45$&$ 1.11\pm 0.26$&$ 0.49\pm 0.27$&$34.92\pm 8.01$&$9.51\times10^9\pm3.16\times10^9$\\
vuds\_cosmos\_510786441&2&4.46 &$-0.10\pm 0.10$&$-2.00\pm 0.10$&$-1.19\pm 0.45$&$ 0.48\pm 0.09$&$ 0.05\pm 0.04$&$61.60\pm 6.62$&$1.18\times10^{10}\pm2.18\times10^9$\\
vuds\_cosmos\_5110377875&2&4.55&$ 0.65\pm 0.05$&$-1.38\pm 0.18$&$-0.55\pm 0.23$&$ 1.56\pm 0.09$&$ 0.34\pm 0.11$&$161.06\pm 9.09$&$1.41\times10^{10}\pm3.62\times10^9$\\\hline
\hline
\end{tabular}
\caption{Physical properties and corresponding uncertainties for the galaxies in our sample.\label{tab:properties}}
\end{sidewaystable*}

\end{document}